\pdfoutput=1

\documentclass[12pt,a4paper]{article}

\usepackage{ifthen}  
\newboolean{pdflatex}
\setboolean{pdflatex}{true}  

\newboolean{articletitles}
\setboolean{articletitles}{true} 

\newboolean{uprightparticles}
\setboolean{uprightparticles}{false} 

\newboolean{inbibliography}
\setboolean{inbibliography}{false} 

\def\paperauthors{LHCb Calorimeter group} 
\def\papertitle{Calibration and performance\\ of the LHCb calorimeters\\ in Run 1 and 2 at the LHC} 
\def\paperkeywords{{High Energy Physics}, {LHCb}, {Calorimetry}} 
\def\papercopyright{\the\year\ CERN for the benefit of the LHCb collaboration}  
\def\paperlicence{CC-BY-4.0 licence}
\def\paperlicenceurl{https://creativecommons.org/licenses/by/4.0/}

\usepackage[top=1in, bottom=1.25in, left=1in, right=1in]{geometry}

\usepackage{microtype}
\usepackage{lineno}  
\usepackage{xspace} 
\usepackage{caption} 

\usepackage{graphicx}  
\usepackage{color}
\usepackage{colortbl}
\graphicspath{{./figs/}} 
\DeclareGraphicsExtensions{.pdf,.PDF,png,.PNG}

\usepackage{amsmath} 
\usepackage{amssymb}
\usepackage{amsfonts}
\usepackage{upgreek} 

\newcommand*\patchAmsMathEnvironmentForLineno[1]{%
\expandafter\let\csname old#1\expandafter\endcsname\csname #1\endcsname
\expandafter\let\csname oldend#1\expandafter\endcsname\csname
end#1\endcsname
 \renewenvironment{#1}%
   {\linenomath\csname old#1\endcsname}%
   {\csname oldend#1\endcsname\endlinenomath}%
}
\newcommand*\patchBothAmsMathEnvironmentsForLineno[1]{%
  \patchAmsMathEnvironmentForLineno{#1}%
  \patchAmsMathEnvironmentForLineno{#1*}%
}
\AtBeginDocument{%
\patchBothAmsMathEnvironmentsForLineno{equation}%
\patchBothAmsMathEnvironmentsForLineno{align}%
\patchBothAmsMathEnvironmentsForLineno{flalign}%
\patchBothAmsMathEnvironmentsForLineno{alignat}%
\patchBothAmsMathEnvironmentsForLineno{gather}%
\patchBothAmsMathEnvironmentsForLineno{multline}%
\patchBothAmsMathEnvironmentsForLineno{eqnarray}%
}

\usepackage{hyperxmp}

\usepackage[pdftex,
            pdfauthor={\paperauthors},
            pdfkeywords={\paperkeywords},
            pdfcopyright={Copyright (C) \papercopyright},
            pdflicenseurl={\paperlicenceurl}]{hyperref}

\usepackage[all]{hypcap} 

\usepackage{xspace} 
\usepackage{upgreek}

\def\lhcb   {\mbox{LHCb}\xspace}

\def\lhc    {\mbox{LHC}\xspace}

\def\spd    {SPD\xspace}
\def\presh  {PS\xspace}
\def\ecal   {ECAL\xspace}
\def\hcal   {HCAL\xspace}
\def\MagUp {\mbox{\em Mag\kern -0.05em Up}\xspace}

\def\lone   {L0\xspace}
\def\hlt    {HLT\xspace}

\ifthenelse{\boolean{uprightparticles}}%
{

 \def\Pmu         {\ensuremath{\upmu}\xspace}

 \def\Ppi         {\ensuremath{\uppi}\xspace}

 \def\Pchi        {\ensuremath{\upchi}\xspace}                 
 \def\Ppsi        {\ensuremath{\uppsi}\xspace}

 \def\PDelta      {\ensuremath{\Delta}\xspace}                 
 \def\PXi         {\ensuremath{\Xi}\xspace}                 
 \def\PLambda     {\ensuremath{\Lambda}\xspace}                 
 \def\PSigma      {\ensuremath{\Sigma}\xspace}                 
 \def\POmega      {\ensuremath{\Omega}\xspace}                 
 \def\PUpsilon    {\ensuremath{\Upsilon}\xspace}

 \def\PB      {\ensuremath{\mathrm{B}}\xspace}                 
                  
 \def\PD      {\ensuremath{\mathrm{D}}\xspace}

 \def\PJ      {\ensuremath{\mathrm{J}}\xspace}                 
 \def\PK      {\ensuremath{\mathrm{K}}\xspace}

 \def\Pc      {\ensuremath{\mathrm{c}}\xspace}                 
                  
 \def\Pe      {\ensuremath{\mathrm{e}}\xspace}

 \def\Pi      {\ensuremath{\mathrm{i}}\xspace}

 \def\Ps      {\ensuremath{\mathrm{s}}\xspace}

}
{

 \def\Pmu         {\ensuremath{\mu}\xspace}

 \def\Ppi         {\ensuremath{\pi}\xspace}

 \def\Pchi        {\ensuremath{\chi}\xspace}                 
 \def\Ppsi        {\ensuremath{\psi}\xspace}                 
                  
 \mathchardef\PDelta="7101
 \mathchardef\PXi="7104
 \mathchardef\PLambda="7103
 \mathchardef\PSigma="7106
 \mathchardef\POmega="710A
 \mathchardef\PUpsilon="7107
                  
 \def\PB      {\ensuremath{B}\xspace}                 
                  
 \def\PD      {\ensuremath{D}\xspace}

 \def\PJ      {\ensuremath{J}\xspace}                 
 \def\PK      {\ensuremath{K}\xspace}

 \def\Pc      {\ensuremath{c}\xspace}                 
                  
 \def\Pe      {\ensuremath{e}\xspace}

 \def\Pi      {\ensuremath{i}\xspace}

 \def\Ps      {\ensuremath{s}\xspace}

}

\makeatletter
\ifcase \@ptsize \relax
  \newcommand{\miniscule}{\@setfontsize\miniscule{4}{5}}
\or
  \newcommand{\miniscule}{\@setfontsize\miniscule{5}{6}}
\or
  \newcommand{\miniscule}{\@setfontsize\miniscule{5}{6}}
\fi
\makeatother

\DeclareRobustCommand{\optbar}[1]{\shortstack{{\miniscule (\rule[.5ex]{1.25em}{.18mm})}
  \\ [-.7ex] $#1$}}

\def\en         {{\ensuremath{\Pe^-}}\xspace}   
\def\ep         {{\ensuremath{\Pe^+}}\xspace}

\def\mup        {{\ensuremath{\Pmu^+}}\xspace}
\def\mun        {{\ensuremath{\Pmu^-}}\xspace} 

\def\squark    {{\ensuremath{\Ps}}\xspace}

\def\cquark    {{\ensuremath{\Pc}}\xspace}

\def\pion   {{\ensuremath{\Ppi}}\xspace}
\def\piz    {{\ensuremath{\pion^0}}\xspace}
\def\pip    {{\ensuremath{\pion^+}}\xspace}
\def\pim    {{\ensuremath{\pion^-}}\xspace}

\def\kaon    {{\ensuremath{\PK}}\xspace}
  \def\Kbar    {{\kern 0.2em\overline{\kern -0.2em \PK}{}}\xspace}

\def\KorKbar    {\kern 0.18em\optbar{\kern -0.18em K}{}\xspace}

\def\Kp      {{\ensuremath{\kaon^+}}\xspace}
\def\Km      {{\ensuremath{\kaon^-}}\xspace}
\def\Kpm     {{\ensuremath{\kaon^\pm}}\xspace}

\def\Kstarz  {{\ensuremath{\kaon^{*0}}}\xspace}

\def\Kstarp  {{\ensuremath{\kaon^{*+}}}\xspace}

  \def\Dbar    {{\kern 0.2em\overline{\kern -0.2em \PD}{}}\xspace}
\def\D       {{\ensuremath{\PD}}\xspace}

\def\DorDbar    {\kern 0.18em\optbar{\kern -0.18em D}{}\xspace}
\def\Dz      {{\ensuremath{\D^0}}\xspace}

\def\theDstarp{{\ensuremath{\D^{*}(2010)^{+}}}\xspace}

\def\B       {{\ensuremath{\PB}}\xspace}
\def\Bbar    {{\ensuremath{\kern 0.18em\overline{\kern -0.18em \PB}{}}}\xspace}

\def\BorBbar    {\kern 0.18em\optbar{\kern -0.18em B}{}\xspace}
\def\Bz      {{\ensuremath{\B^0}}\xspace}

\def\Bu      {{\ensuremath{\B^+}}\xspace}
\def\Bub     {{\ensuremath{\B^-}}\xspace}

\def\Bm      {{\ensuremath{\Bub}}\xspace}
\def\Bpm     {{\ensuremath{\B^\pm}}\xspace}

\def\Bs      {{\ensuremath{\B^0_\squark}}\xspace}

\def\jpsi     {{\ensuremath{{\PJ\mskip -3mu/\mskip -2mu\Ppsi\mskip 2mu}}}\xspace}

\def\chiczero {{\ensuremath{\Pchi_{\cquark 0}}}\xspace}
\def\chicone  {{\ensuremath{\Pchi_{\cquark 1}}}\xspace}
\def\chictwo  {{\ensuremath{\Pchi_{\cquark 2}}}\xspace}
\def\Y#1S{\ensuremath{\PUpsilon{(#1S)}}\xspace}

\def\chic  {{\ensuremath{\Pchi_{c}}}\xspace}
\def\Upsilonres  {{\ensuremath{\PUpsilon}}\xspace}

\def\Lbar        {{\ensuremath{\kern 0.1em\overline{\kern -0.1em\PLambda}}}\xspace}
\def\LorLbar     {\kern 0.18em\optbar{\kern -0.18em \PLambda}{}\xspace}

\def\to                 {\ensuremath{\rightarrow}\xspace}

\def\AT#1     {\ensuremath{A_{\mathrm{T}}^{#1}}\xspace}           

\def\C#1      {\ensuremath{\mathcal{C}_{#1}}\xspace}                       
\def\Cp#1     {\ensuremath{\mathcal{C}_{#1}^{'}}\xspace}                    
\def\Ceff#1   {\ensuremath{\mathcal{C}_{#1}^{\mathrm{(eff)}}}\xspace}        
\def\Cpeff#1  {\ensuremath{\mathcal{C}_{#1}^{'\mathrm{(eff)}}}\xspace}       
\def\Ope#1    {\ensuremath{\mathcal{O}_{#1}}\xspace}                       
\def\Opep#1   {\ensuremath{\mathcal{O}_{#1}^{'}}\xspace}                    

\newcommand{\tev}{\ifthenelse{\boolean{inbibliography}}{\ensuremath{~T\kern -0.05em eV}}{\ensuremath{\mathrm{\,Te\kern -0.1em V}}}\xspace}
\newcommand{\gev}{\ensuremath{\mathrm{\,Ge\kern -0.1em V}}\xspace}
\newcommand{\mev}{\ensuremath{\mathrm{\,Me\kern -0.1em V}}\xspace}
\newcommand{\kev}{\ensuremath{\mathrm{\,ke\kern -0.1em V}}\xspace}
\newcommand{\ev}{\ensuremath{\mathrm{\,e\kern -0.1em V}}\xspace}
\newcommand{\gevc}{\ensuremath{{\mathrm{\,Ge\kern -0.1em V\!/}c}}\xspace}
\newcommand{\mevc}{\ensuremath{{\mathrm{\,Me\kern -0.1em V\!/}c}}\xspace}
\newcommand{\gevcc}{\ensuremath{{\mathrm{\,Ge\kern -0.1em V\!/}c^2}}\xspace}
\newcommand{\gevgevcccc}{\ensuremath{{\mathrm{\,Ge\kern -0.1em V^2\!/}c^4}}\xspace}
\newcommand{\mevcc}{\ensuremath{{\mathrm{\,Me\kern -0.1em V\!/}c^2}}\xspace}

\def\m    {\ensuremath{\mathrm{ \,m}}\xspace}

\def\mm   {\ensuremath{\mathrm{ \,mm}}\xspace}

\def\mum  {\ensuremath{{\,\upmu\mathrm{m}}}\xspace}

\def\invpb {\ensuremath{\mbox{\,pb}^{-1}}\xspace}

\def\invfb   {\ensuremath{\mbox{\,fb}^{-1}}\xspace}

\def\ns   {\ensuremath{{\mathrm{ \,ns}}}\xspace}

\def\mhz  {\ensuremath{{\mathrm{ \,MHz}}}\xspace}
\def\khz  {\ensuremath{{\mathrm{ \,kHz}}}\xspace}
\def\hz   {\ensuremath{{\mathrm{ \,Hz}}}\xspace}

\def\Xrad {\ensuremath{X_0}\xspace}
\def\NIL{\ensuremath{\lambda_{int}}\xspace}
\def\mip {MIP\xspace}

\newcommand{\chisqndf}{\ensuremath{\chi^2/\mathrm{ndf}}\xspace}
\newcommand{\chisqip}{\ensuremath{\chi^2_{\text{IP}}}\xspace}

\newcommand{\chisqvtxndf}{\ensuremath{\chi^2_{\text{vtx}}/\mathrm{ndf}}\xspace}

\def\gsim{{~\raise.15em\hbox{$>$}\kern-.85em
          \lower.35em\hbox{$\sim$}~}\xspace}
\def\lsim{{~\raise.15em\hbox{$<$}\kern-.85em
          \lower.35em\hbox{$\sim$}~}\xspace}

\def\sPlot{\mbox{\em sPlot}\xspace}

\def\sqs   {\ensuremath{\protect\sqrt{s}}\xspace}

\def\pt         {\ensuremath{p_{\mathrm{T}}}\xspace}

\def\et         {\ensuremath{E_{\mathrm{T}}}\xspace}

\def\mrad{\ensuremath{\mathrm{ \,mrad}}\xspace}

\def\tell1  {TELL1\xspace}
\def\ukl1   {UKL1\xspace}

\def\pmt    {PMT\xspace}

\def\adc    {ADC\xspace}
\def\led    {LED\xspace}

\def\hv     {HV\xspace}

\newcommand{\ie}{\mbox{\itshape i.e.}\xspace}

\usepackage{cite} 
\usepackage{mciteplus}

\usepackage{longtable}  

\begin{document}
\renewcommand{\thefootnote}{\fnsymbol{footnote}}
\setcounter{footnote}{1}



\begin{titlepage}
\pagenumbering{roman}

\vspace*{-1.5cm}
\centerline{\large EUROPEAN ORGANIZATION FOR NUCLEAR RESEARCH (CERN)}
\vspace*{1.5cm}
\noindent
\begin{tabular*}{\linewidth}{lc@{\extracolsep{\fill}}r@{\extracolsep{0pt}}}
\ifthenelse{\boolean{pdflatex}}
{\vspace*{-1.5cm}\mbox{\!\!\!\includegraphics[width=.14\textwidth]{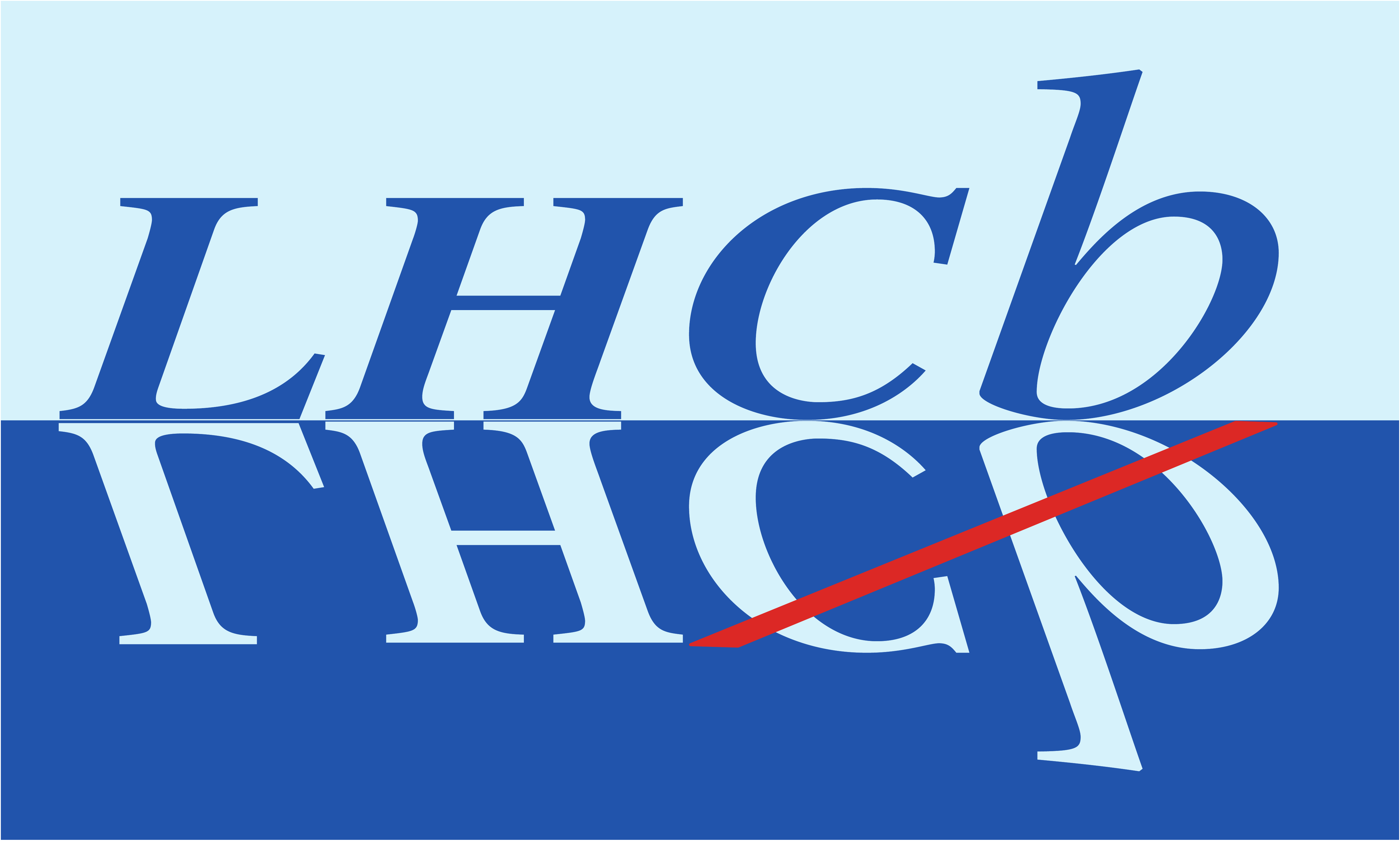}} & &}
{\vspace*{-1.2cm}\mbox{\!\!\!\includegraphics[width=.12\textwidth]{figs/lhcb-logo.eps}} & &}
\\
 & & CERN-LHCb-DP-2020-001 \\  
 & & August 26, 2020 \\ 
 & & \\
\end{tabular*}

\vspace*{4.0cm}

{\normalfont\bfseries\boldmath\huge
\begin{center}
  \papertitle 
\end{center}
}

\vspace*{2.0cm}

\begin{center}
\paperauthors\footnote{Authors are listed at the end of this paper.}
\end{center}

\vspace{\fill}

\begin{abstract}
  \noindent 
The calibration and performance of the LHCb Calorimeter system in Run~1 and 2 at the LHC are described.
After a brief description of the sub-detectors and of their role in the trigger, the calibration methods used for each part 
of the system are reviewed. The changes which occurred with the increase of beam energy 
in Run~2 are explained. The performances of the calorimetry for $\gamma$ and $\piz$  are detailed.
A few results from collisions recorded at $\sqrt {s}$ = 7, 8 and 13 TeV are shown.
\end{abstract}

\vspace*{2.0cm}

\begin{center}
  Submitted to JINST
\end{center}

\vspace{\fill}

{\footnotesize 
\centerline{\copyright~\papercopyright. \href{\paperlicenceurl}{\paperlicence}.}}
\vspace*{2mm}

\end{titlepage}


\newpage
\setcounter{page}{2}
\mbox{~}
%
%
%
%

\cleardoublepage


\renewcommand{\thefootnote}{\arabic{footnote}}
\setcounter{footnote}{0}

\tableofcontents
\cleardoublepage

\pagestyle{plain} 
\pagenumbering{arabic}
\setcounter{page}{1}


\section{Introduction}

\begin{figure}[!ht]
\centering
\includegraphics[width=0.85\textwidth]{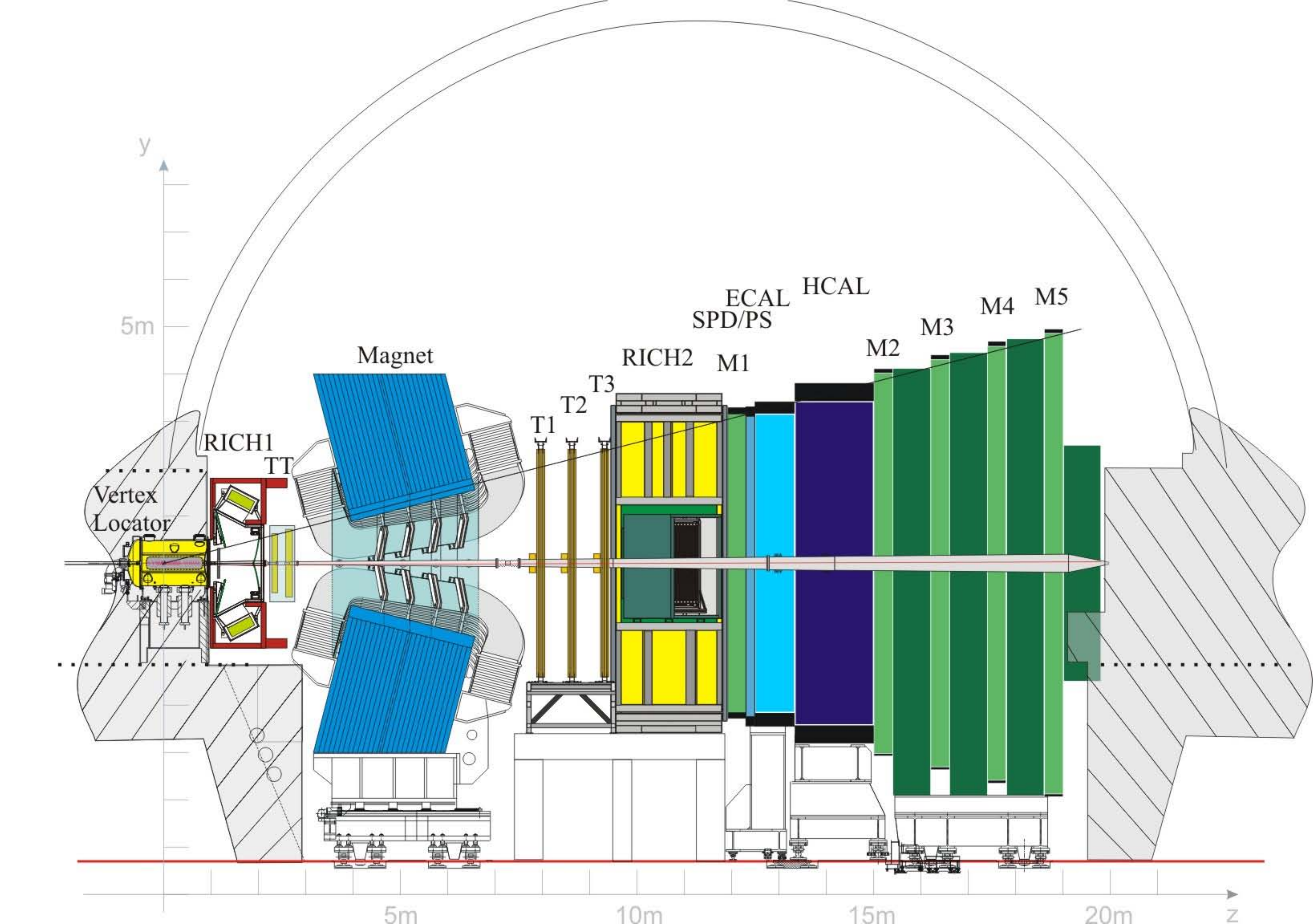}
\caption{Layout of the \lhcb experiment.}
\label{lhcb_setup}
\end{figure}

The \lhcb detector~\cite{LHCb-DP-2008-001} (Figure~\ref{lhcb_setup}) is a single-arm forward spectrometer covering the \mbox{pseudorapidity} range $2<\eta <5$, designed for the study of particles containing $b$ or $c$ quarks. The detector includes a high-precision tracking system consisting of a silicon-strip vertex detector surrounding the $pp$ interaction region, a large-area silicon-strip detector located upstream of a dipole magnet with a bending power of about $4~\textrm{Tm}$, and three stations of silicon-strip detectors and straw drift tubes placed downstream. The polarity of the magnet is reversed repeatedly during data taking, which causes all detection asymmetries that are induced by the left--right separation of charged particles due to the magnetic field to change sign. The combined tracking system has a momentum resolution $\Delta p/p$ that varies from 0.4\% at 5\gevc to 0.6\% at 100\gevc, and an impact parameter resolution of 20\mum for tracks with high transverse momentum. Charged hadrons are identified using two ring-imaging Cherenkov detectors. Muons are identified by a system composed of alternating layers of iron and multiwire proportional chambers. The identification of hadrons, electrons\footnote{In all what follows, except obvious exceptions, "electron" will stand for electron or positron.} and photons and the measurement of their energies and positions is performed by a system of 4 sub-detectors: the \lhcb calorimeter system~\cite{LHCb-TDR-002}. The trigger~\cite{LHCb-TDR-010} consists of a hardware stage, based on information collected from the calorimeter and muon systems at a rate of 40\mhz, called "\lone trigger", followed by a software stage, which applies a full event reconstruction, known as the "High Level Trigger" (\hlt).

The information provided by the calorimeter system is also used by the \lone trigger (see Sec.~\ref{sec:l0trigger}). The 4 sub-detectors are located between 12.3 and 15.0 m along the beam axis downstream of the interaction point (light and dark blue in Figure~\ref{lhcb_setup}): a Scintillator Pad Detector (\spd), a PreShower (\presh), an Electromagnetic CALorimeter (\ecal) and a Hadronic CALorimeter (\hcal), all placed perpendicular to the beam axis. A 2.5~\Xrad lead foil\footnote{\Xrad is the radiation length.} is interleaved between the \spd and the \presh. A signal in the \spd marks the presence of a charged particle. Energy deposited in the \presh indicates the start of an electromagnetic shower. \ecal and \hcal determine the electromagnetic or hadronic nature of the particles reaching them. Minimum ionizing particles are also detected in all four sub-detectors.

After briefly recalling the main characteristics of the 4 sub-detectors of the calorimetric system, this paper describes the various methods developed to calibrate the \lhcb calorimeters and the evolution of these methods over the course of the two distinct data taking periods (Run 1 and Run 2) at the LHC. The performance of the calorimeters is then presented.

\section{The \lhcb Calorimeters}

\subsection{Detector layout}

\begin{figure}[!htb]
\centering 
\includegraphics[width=0.7\textwidth]{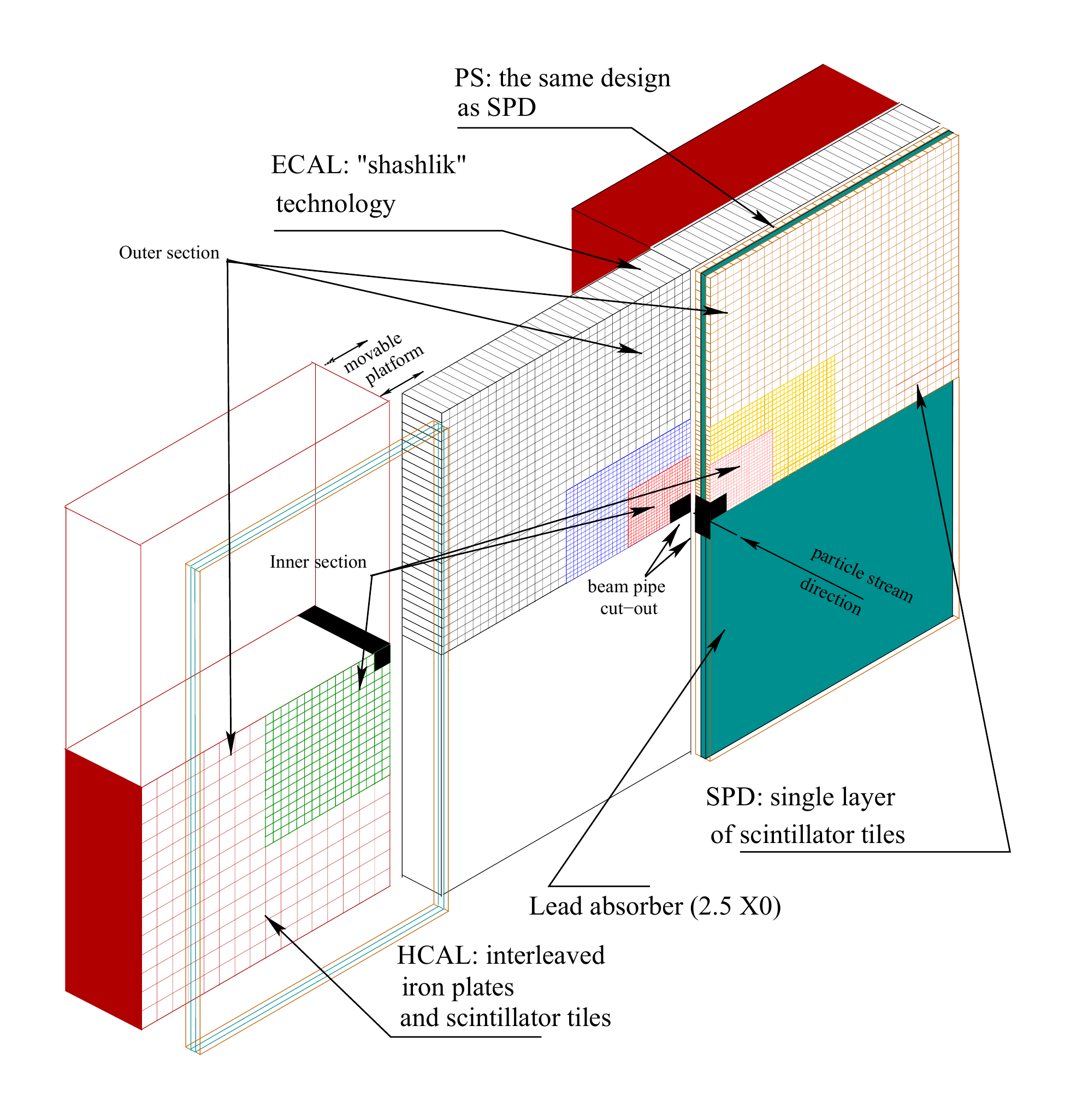} 
\caption{Layout of the calorimeter system.}
\label{calo_layout}
\end{figure}

As most detectors of LHCb, the calorimeters are assembled in two halves (A and C sides), which can be moved out horizontally for assembly and maintenance purposes, as well as to provide access to the beam-pipe. The \spd, \presh, \ecal and \hcal (Figure~\ref{calo_layout}) detectors are segmented in the plane perpendicular to the beam axis into square cells.

\begin{figure}[!htb]
\centering
\begin{minipage}{0.49\textwidth}
\includegraphics[width=1.0\textwidth]{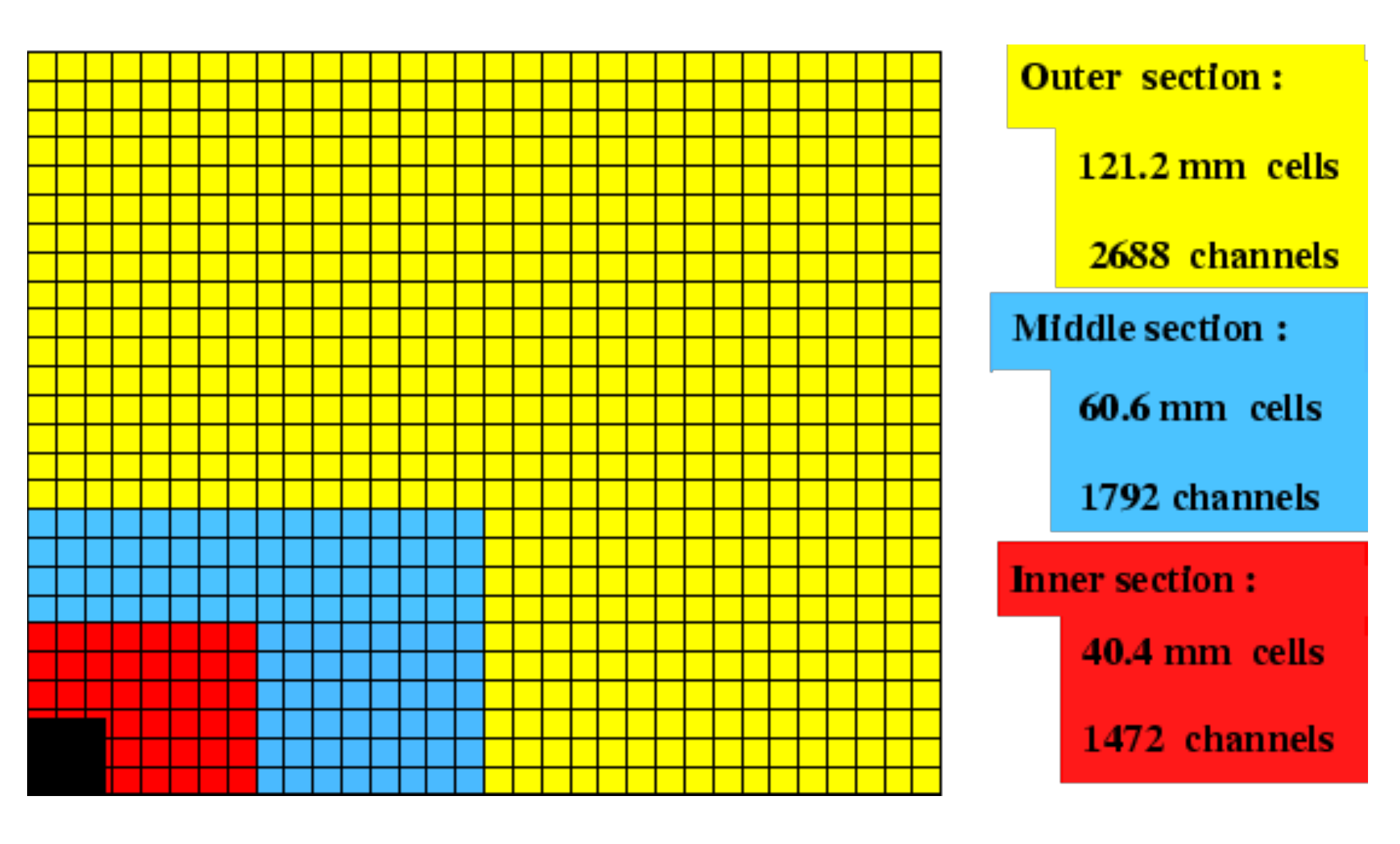}
\end{minipage}
\begin{minipage}{0.49\textwidth}
\includegraphics[width=1.0\textwidth]{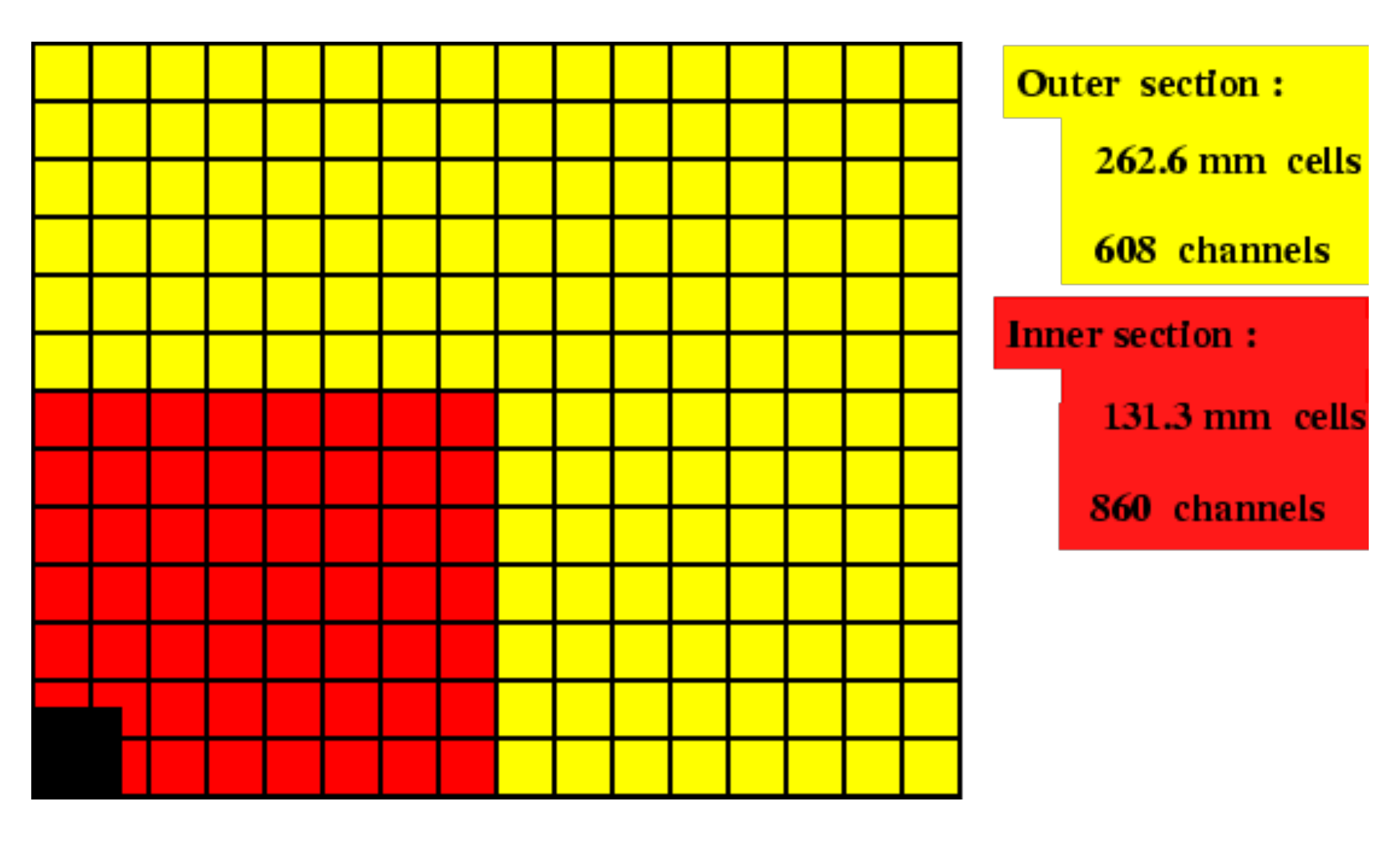}
\end{minipage}
\caption{Calorimeter cells segmentation of the \spd, \presh, and \ecal (left), and the \hcal (right).}
\label{calo_segment} 
\end{figure}

At the LHC, the density of the particles hitting the calorimeters varies by two orders of magnitude over the calorimeter surface, the inner part close to the \lhc beam pipe being subject to the highest density. In order to follow these variations the calorimeters are divided into regions with cells of different sizes, smaller cells being placed close to the beam pipe and the cell size increasing with increasing distance to the beam. There are three regions for the \spd, \presh and \ecal and two regions for the \hcal (Figure~\ref{calo_segment}, Table~\ref{geomcalo}).

\begin{table}
\begin{center}
\begin{tabular}{l|l|l|l}
sub-detector & \spd/\presh & \ecal & \hcal \\
\hline
number of channels & 2$\times$6016 & 6016 & 1488 \\
overall lateral & 6.2 m $\times$ 7.6 m & 6.3 m $\times$ 7.8 m & 6.8 m $\times$ 8.4 m \\
dimension in x, y & & & \\
cell size  (mm) Inner & 39.7 (\spd), 39,8 (\presh) & 40.4 & 131.3\\
cell size  (mm) Middle & 59.5  (\spd), 59.76 (\presh) & 60.6 & \\
cell size  (mm) Outer & 119 (\spd), 119.5 (\presh) & 121.2 & 262.6\\
depth in z & 180 mm, & 835 mm, & 1655 mm, \\
           & 2.5~\Xrad, 0.1 \NIL & 25~\Xrad, 1.1 \NIL & 5.6 \NIL \\
light yield & $\sim$20 p.e./MIP & $\sim$3000 p.e./GeV &$\sim$105 p.e./GeV \\
dynamic range & 0 - 100 \mip   & 0 - 10 \gev \et & 0 - 20 \gev \et \\
            & 10 bits (\presh), 1 bit (\spd) & 12 bits & 12 bits \\
\end{tabular}
\caption{\label{geomcalo} Main parameters of the \lhcb calorimeter sub-detectors.}
\end{center}
\end{table}

All four sub-detectors consist of successions of absorbers (lead or iron) and scintillator plates. The scintillation light resulting from the showers of the particles going through the detector is collected by wavelength-shifting (WLS) fibres. The fibres transport the light to photomultiplier tubes (\pmt) or multi-anode photomultiplier tubes (MA-\pmt) that convert the light into electrical signals which are readout by electronic Front-End Boards (FEB). Each cell of the \ecal and \hcal is equipped with a photomultiplier tube and the corresponding FEB converts the signal into an energy value, $E$. In order to obtain a constant dynamic range in transverse energy over the calorimeter plane, the gain of the electronic channels are set as a function of the position of the cell, namely its distance to the beam. The transverse energy, \et, is defined as $\et = E \sin\theta$, where $\theta$ is the angle between the beam axis and a line from the interaction point to the centre of the cell. 

The \spd and \presh provide information for particle identification of electrons and photons, which are used in particular by the \lone trigger. The \presh information is used to separate electrons, photons and pions while the \spd measurements contribute to the separation of neutral particles from charged ones. In addition, the overall \spd hit multiplicity provides an estimation of the charged particles multiplicity in each beam crossing. This information can be used to veto complex events that would be extremely difficult to analyse offline. The \ecal measures the transverse energy of electrons, photons and \piz for the \lone trigger, which is relevant to select $B$ decays with an electron or a photon in the final state. The offline reconstruction and energy computation of \piz, electrons and photons make also use of the information from the \presh detector. The \hcal provides a measurement of the transverse energy of hadrons for the \lone trigger in order to select a large variety of $D$ and $B$ decays with a charged hadron (pion, kaon or proton) in the final state. The \spd, \presh, \ecal and \hcal are used for offline particle identification.

The \lhcb calorimeter system was installed between years 2004 and 2008. The commissioning phase took place between 2005 and 2009 using cosmic rays and secondary particles resulting from the first tests of injection of proton beams into the \lhc rings. The first trace of a cosmic ray was detected in January 2008 and the system was fully operational for the first collisions in September 2008. 

\subsection{Calorimeter \lone Trigger}\label{sec:l0trigger}

The performance of the calorimeters in the \lhcb trigger is discussed in Ref.~\cite{LHCb-DP-2012-004}. The main aspects are summarised below. The calorimeter part of the \lone trigger, called in the following \lone-Calorimeter, computes the transverse energy (\et) deposited in clusters of $2\times 2$ cells of the same size (or equivalently of the same region). Three different types of clusters, or "trigger candidates", are built by the \lone-Calorimeter. The information from the different calorimeter sub-detectors allows to build through a trigger validation board the electron/photon/hadron candidate, within the 25\ns timing. The first type is the "hadron candidate", which is an \hcal cluster, with \et equal to the \et of the \hcal cluster plus the \et of the \ecal cluster in front of the \hcal one. The second type is the "photon candidate" which is an \ecal cluster with 1 or 2 \presh cells hit in front of the \ecal cluster, and no hit in the \spd cells corresponding to the \presh ones. In the central zone of \ecal, due to the higher multiplicity, an \ecal cluster with 3 or 4 \presh cells hit in front of it is also identified as photon, in order to increase the trigger efficiency. The last type of candidate is the "electron candidate" which is built with the same requirements concerning \presh hits as the photon candidate requiring in addition at least one \spd cell hit in front of the \presh ones. The transverse energy of the photon and electron candidates is the \et of the \ecal cluster. The \et of these candidates is compared with a fixed threshold, 3.5\gev for hadrons and 2.5\gev for electrons and photons at the start of Run 1.These two thresholds have slightly evolved (mainly increased) over time. Only events containing at least one candidate above these thresholds are sent to the software trigger for further processing. The performances of the \lone-Calorimeter are computed from data recorded by the \lhcb detector, using trigger-unbiased samples of well-identified decay modes. The efficiency of the hadron trigger on $\Bm\to\Dz(\Km\pip)\pim$, is $(11\pm 1)\%$ for \Bm with \pt of 3.5\gevc, $(52\pm 1)\%$ at 10\gevc and $(87\pm 1)\%$ at 15\gevc. The efficiency of the electron trigger is equal to $(86\pm 1)\%$ for the mode $\jpsi\to \ep\en$ with a $\jpsi$ \pt larger than 10\gevc. The performance of the photon trigger is measured using $\Bz\to \Kstarz \gamma$ decays that allow to extract an efficiency of $(50\pm 4)\%$ for selecting the events over the full \pt spectrum. 

\subsection{\spd and \presh}

\begin{figure}[!ht]
\centering
\includegraphics[width=0.7\textwidth]{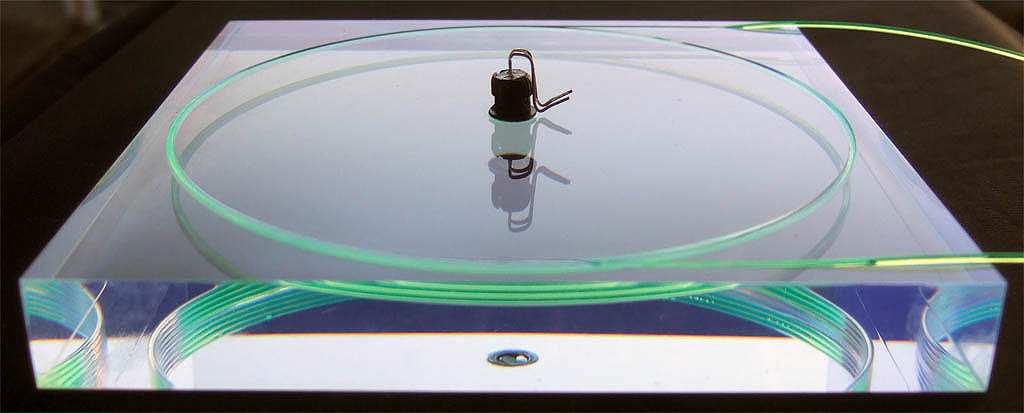}
\caption{\spd cell.}
\label{spd_cell}
\end{figure}

The \spd and \presh are walls of scintillator pads (cells) with a WLS fibre coil grooved inside for better light collection as seen in Figure~\ref{spd_cell}. The two walls are separated by a lead curtain with a thickness corresponding to 2.5 radiation lengths. The \spd and \presh have in total 6016 cells each. Their readout is multi-anode photomultipliers (MA-\pmt) of type "Hamamatsu 5900 M64".  

The \spd delivers a binary information per cell, depending on the comparison of the energy deposited in the cell with a threshold. This is used to distinguish charged particles from neutral ones and, in association with the energy measured in the corresponding \presh cells, helps photon and electron identifications. 

\subsection{\ecal}

\begin{figure}[!ht]
\centering
\includegraphics[width=0.7\textwidth]{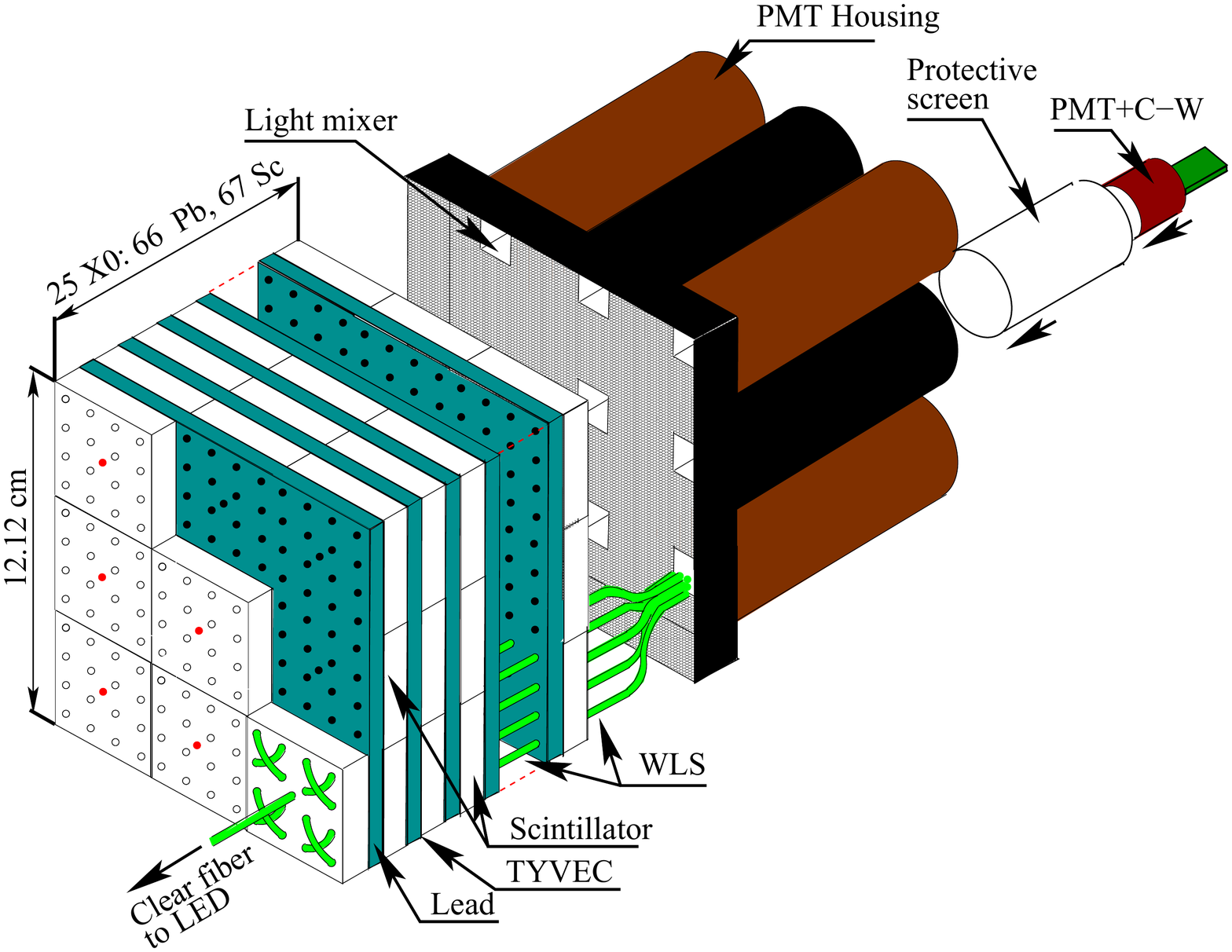}
\caption{\ecal cell.}
\label{ecal_cell}
\end{figure}

The \ecal thickness is of 25 radiation lengths to ensure the full containment of the high energy electromagnetic showers and to get an optimal energy resolution. The \ecal cells have a shashlik structure, as shown in Figure~\ref{ecal_cell}, with scintillator (4\mm) and lead (2\mm) layers. The scintillation light readout is performed by Hamamatsu R7899-20 photomultipliers. The total number of cells is 6016. 

As shown in Figure~\ref{calo_segment}, three regions (inner, middle and outer) have been defined according to the distance of the cells to the beam-pipe. The cell size is such that the \spd-\presh-\ecal system is projective, as seen from the interaction point. In addition, the outer dimensions of the \ecal match projectively those of the tracking system, $\theta_x < 300\mrad$ and $\theta_y < 250 \mrad$, while the inner angular acceptance of \ecal is limited to $\theta_{x,y} >25 \mrad$ around the beam pipe, where $\theta_x$ and $\theta_y$ are the polar angles in the \lhcb frame in which $(x,y)$ is the plane perpendicular to the beam axis. The \ecal front surface is located at about $12.5\m$ from the interaction point. The energy resolution of \ecal for a given cell, measured with test-beam electrons is parameterised~\cite{Arefev:1103500} as
\begin{equation}
\frac{\sigma(E)}{E} =  \frac{(9.0\pm 0.5)\%}{\sqrt{E}}\, \oplus \, (0.8 \pm 0.2)\%\, \oplus \frac{0.003}{E \sin\theta},
\label{ecal_resolution}
\end{equation}
where $E$ is the particle energy in \gev, $\theta$ is the angle between the beam axis and a line from the \lhcb interaction point and the centre of the \ecal cell. The second contribution is the constant term (corresponding to mis-calibrations, non-linearities, leakage, ...) while the third one is due to the noise of the electronics which is evaluated on average to 1.2 \adc counts~\cite{LHCb-DP-2008-001}.

\subsection{\hcal}

\begin{figure}[!ht]
\centering
\includegraphics[width=0.7\textwidth]{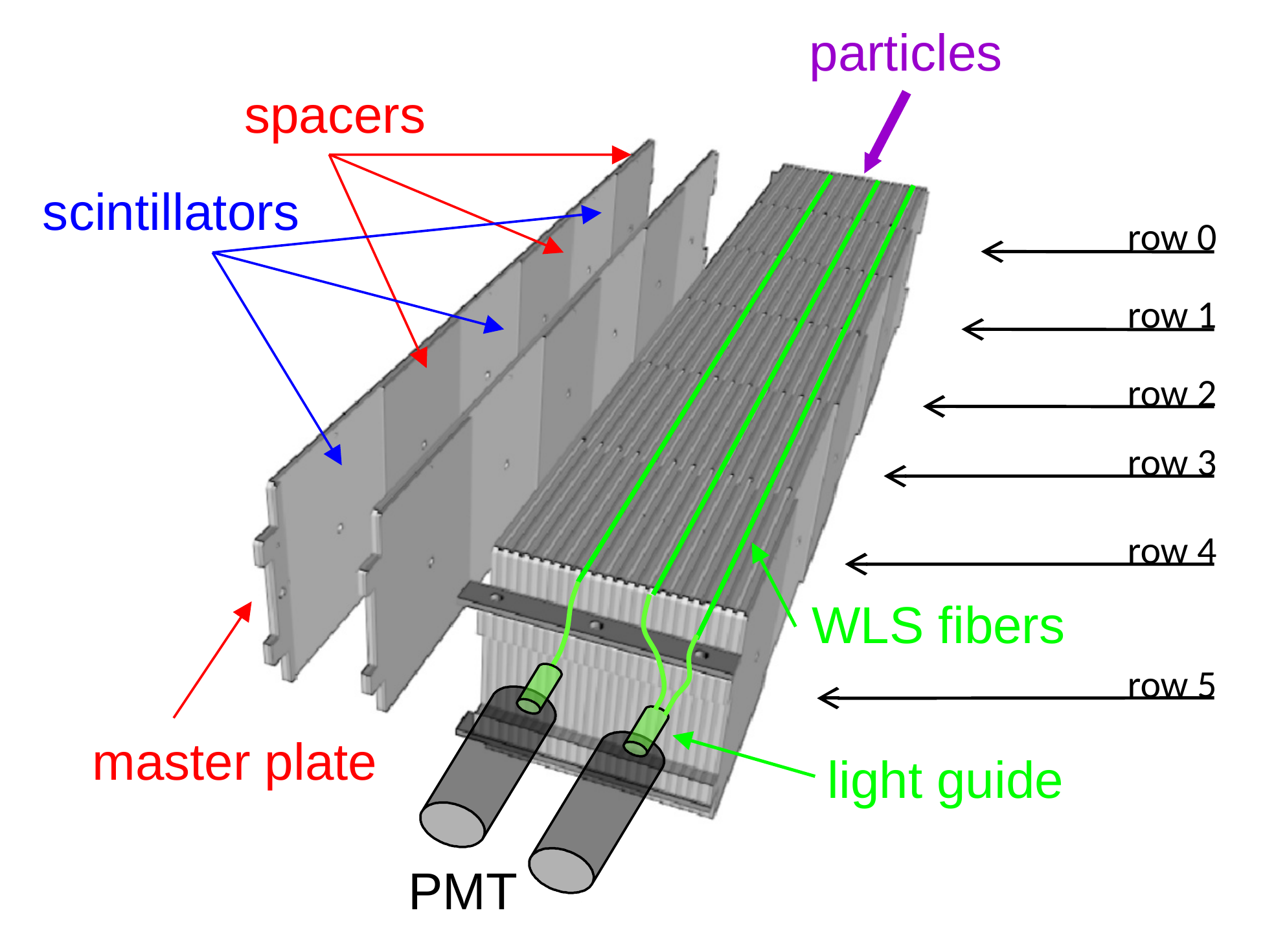}
\caption{\hcal cell.}
\label{hcal_cell} 
\end{figure}

The \hcal thickness is 5.6 interaction lengths due to space limitations. A sampling structure was chosen, made from iron and scintillating tiles, as absorber and active material, respectively. The special feature of this sampling structure is the orientation of the scintillating tiles that are placed parallel to the beam axis (Figure~\ref{hcal_cell}) thus enhancing the light collection compared to a perpendicular orientation of the scintillating tiles. The same photomultiplier type as in \ecal (Hamamatsu R7899-20) is used for the readout. The \hcal has in total 1488 cells all of the same dimension located in two regions (inner and outer), depending on their distance to the beam-pipe.

The energy resolution measured in test beams with pions is~\cite{Coca:691519}
\begin{equation}
\frac{\sigma(E)}{E} =  \frac{(67\pm 5)\%}{\sqrt{E}}\, \oplus \, (9\pm 2)\%,
\end{equation}
where $E$ is the deposited energy in \gev.

\subsection{Electronics, readout and monitoring}

The common approach used in all four detectors to process the photomultiplier signal is to make first a shaping and an analogue integration of the signal then sample and finally digitise it. The reduced digitised information (8 bits) is sent to the trigger validation board which builds signal candidates (hadron, electron, photon). The full digitised information (12 bits) is available to data acquisition for higher level trigger selections and offline analysis. The design of the calorimeters causes the MA-\pmt or \pmt output signal to be wider than 25\ns. Different solutions are adopted to face this problem. 

The \spd and \presh subtract, in turn, a fraction of the signal measured in the previous clock cycle to solve the problem of the width of the MA-\pmt pulse. The integration is performed by 2 interleaved integrators running at 20\mhz. One channel integrates while the other discharges. Since the readout uses 64 channel MA-\pmt, small boards, called Very Front-End (VFE) boards, located close to the detector, host the MA-\pmt and the associated signal-processing electronics. The \spd output is reduced to a single bit of data, comparing the integrated signal value with a threshold. This threshold is tuned by measuring the \spd efficiency for different thresholds, looking for the best efficiency while keeping the noise rate low. The analogue signal of the \presh and the \spd bit are sent to the FEB in the crates located near the \ecal and \hcal readout crates. In these FEB, the \presh signal is digitised by a 10-bit analogue to digital converter (\adc).

The \ecal and \hcal electronics performs a clipping of the signal prior to integration so that it almost fully fits in one clock cycle. The readout is performed by an analogue chip hosted in the FEB located in crates installed in the \lhcb cavern, directly on top of the calorimeter structure. The integrator discharge is done by injecting altogether with the current \pmt signal an inverted copy of the signal delayed by 25\ns~\cite{LHCb-TDR-002,LHCb-DP-2008-001}. Hence, the signal at the integrator output reaches its maximum on a plateau, which is stable (within 1\%) during 4\ns where it is sampled and digitised on a 12-bit precision scale. 

High Voltage (\hv) is provided for the four subsystems by full custom Cockroft-Walton voltage multipliers. The \spd and \presh share a common design for their 64 channel MA-\pmt, just as \ecal and \hcal do for their \pmt.

The first adjustment necessary to take data efficiently is the time-alignment of the electronics. This allows to integrate the maximum of the signal pulse in the 25\ns bunch spacing. The principle for the time alignment uses an asymmetry method. It consists in delaying the overall integration clock by half a cycle. In a well tuned system, the same amount of signal is expected in the two consecutive clock cycles containing the energy deposit. This procedure gives a timing per cell for the \ecal and \hcal and a timing per MA-\pmt for the \spd and \presh. Full details of the method, its implementation and results can be found in Ref.~\cite{AbellanBeteta:2012dd} and references therein. 

To monitor the functionality of the readout chain described above and the stability of its characteristics, the \presh, \ecal and \hcal are equipped with monitoring systems based on light emitting diodes (\led). Each photomultiplier is illuminated by \led flashes of known intensity. The light emitted is transported to the cell \pmt by clear fibres and altogether to a stable PIN-diode photodiode. The intensity of the signal of the \pmt is normalised by the PIN-diode response, so that the value of the average \pmt response can be used to follow up the behaviour of each readout channel. The system was aimed at using these \led flashes to follow the \pmt gain variation over large periods. However, due to non-uniform radiation damage of the clear fibres taking the \led light inside the detector cells, this system was not used for the fine calibration of \ecal, but performed as expected for \hcal. The monitoring of the \ecal and \hcal response is performed online, in parallel with data taking, and accuracy on the gain variation measurement reached by this calibration method is better than 1\%. The LEDs flash with a frequency of 11\khz during empty packets intervals of the \lhc beam sequence. A subset corresponding to a flash frequency of 50\hz is sent to a monitoring CPU farm for analysis. 

\section{Calibration of the \lhcb Calorimeters}

The \lhc operation started at the end of 2009 for a few weeks of $pp$ collisions at \sqs = 0.9\tev. The \lhc Run 1 period of data taking started in 2010 and ended in 2012 with an increase of \sqs energy from 7 (2010-2011)  to 8 (2012) \tev. During this period, the various monitoring and calibration methods of the calorimeters did not change although ageing effects were already developing. The \ecal fibres were damaged by radiations in a non-uniform way, preventing to use the \led system to follow the ageing during Run 1, but the response of individual cells could be followed by the \led system. Taking advantage of the Long Shutdown 1 (LS1 --- 2013-2014), the fibres of the \ecal, suffering from radiation/ageing effects were replaced. \lhc production resumed in 2015. The beam energy was increased to reach \sqs = 13\tev and the bunch space was reduced from 50 to 25\ns. Along the Run 2 period (2015-2018) several techniques were implemented to follow online the evolution of the ageing and to correct its effects. These various operations are described in detail in this section. 

\subsection{\spd}

\subsubsection{Expected efficiency}

\begin{figure}[!ht]
\centering
\includegraphics[width=0.7\textwidth]{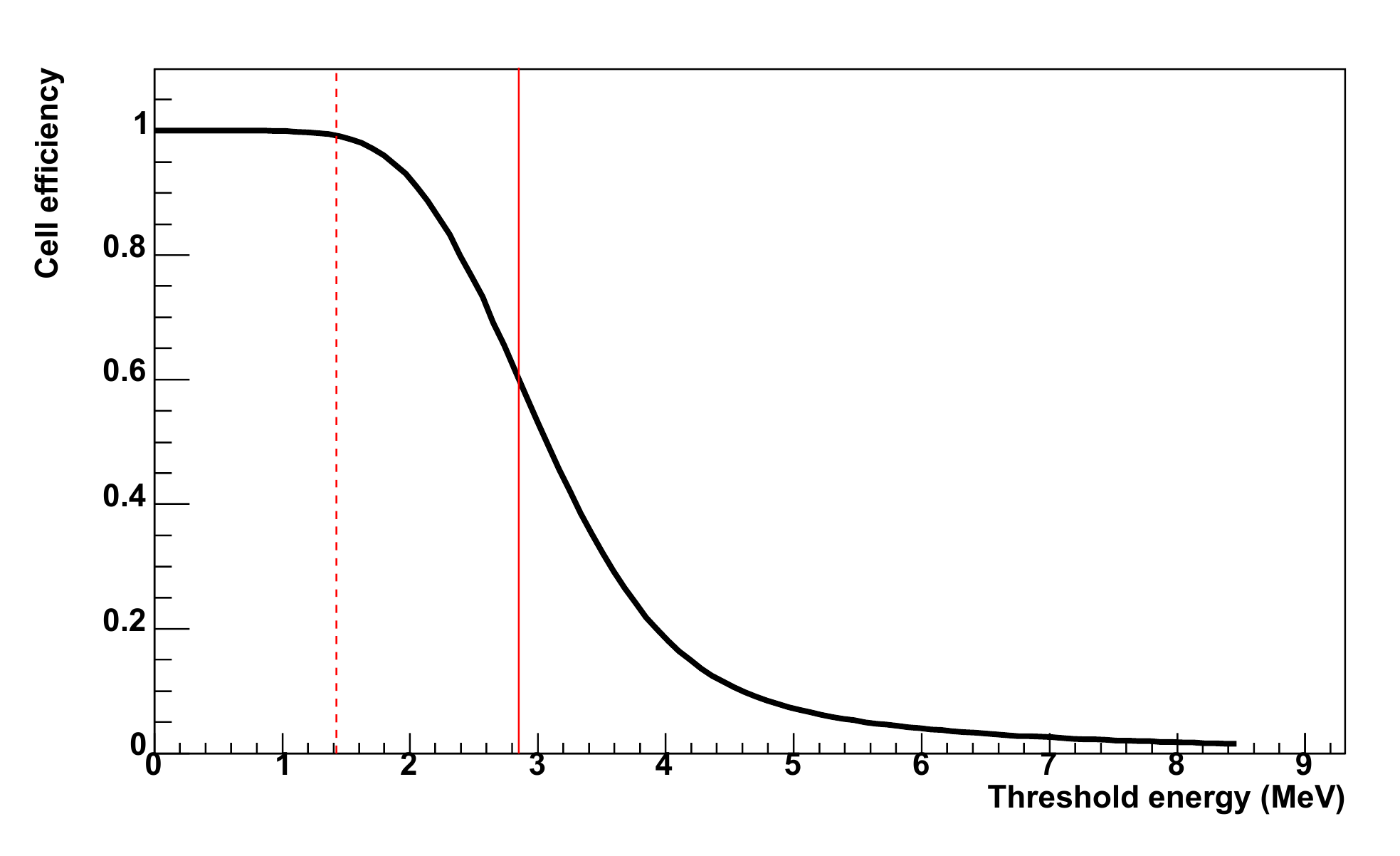}
\caption{Expected efficiency of the \spd response to particles as a function of the threshold value, obtained from the integration of convoluted Poisson and Landau distributions. The solid line corresponds to the position of the averaged minimum energy ionisation (MIP), the dashed one to the threshold set at $0.5\,E_{\rm MIP}$.}
\label{spd_expected_eff}
\end{figure}

The calibration of the \spd uses charged particles pointing to its cells after extrapolating their trajectories to the \spd plane. The tracks reconstructed by the tracking system of \lhcb should coincide with signals or "hits" recorded in the \spd. The efficiency is defined as the fraction of tracks effectively giving a hit in the detector. The energy deposited by charged particles traversing a thin material is parameterised by a Landau distribution  whose average is the MIP energy, $E_{\rm MIP} = 2.85\mev$. Taking into account that the number of photoelectrons, $N_{\rm phe}$, generated in the \pmt photocathode follows a Poisson distribution, the probability density function of the deposited energy is given by the convolution of both distributions. The theoretical efficiency curve can be derived and is shown in Figure~\ref{spd_expected_eff}. 

\subsubsection{Pre-calibration with cosmic rays}

After tests and measurements in test benches and laboratories, pre-calibration gains are obtained for each cell~\cite{VazquezGomez:1399119}. Cosmic ray data are then taken with a threshold energy value per cell corresponding to 1 MIP. Events are triggered by a coincidence of \ecal and \hcal signals and charged tracks are reconstructed. Several cuts are applied to ensure that in each selected event, a well measured single track hits a \spd cell. With these conditions horizontal cosmic rays are selected giving a poor statistic to perform a cell calibration; therefore the 64 cells belonging to the same VFE card were globally adjusted and a correction of 15\% was applied to the pre-calibrated gains.

\subsubsection{Run 1 calibration}

\begin{figure}[!ht]
\centering
\includegraphics[width=1.0\textwidth]{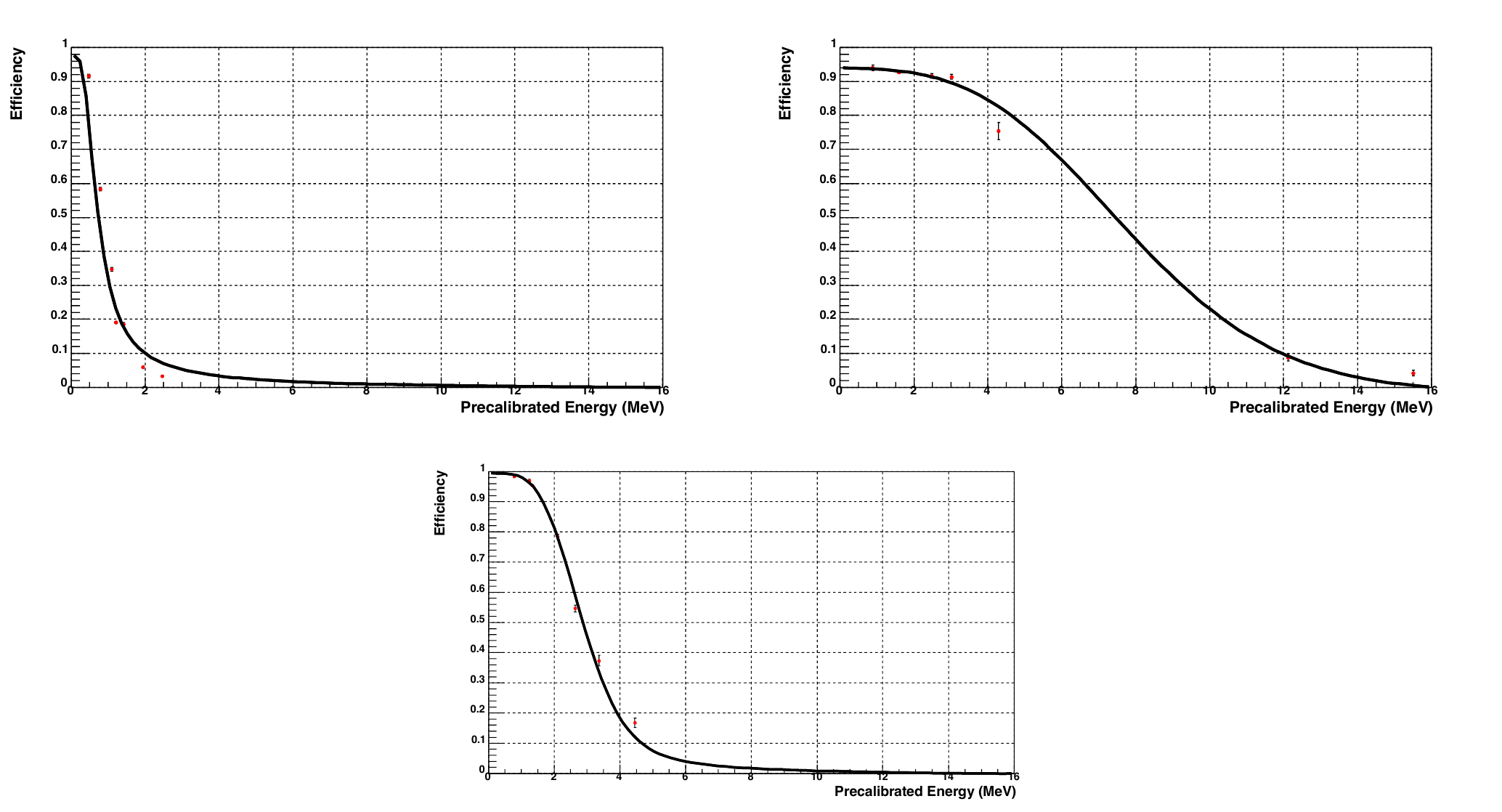}
\caption{Examples of efficiency curves for individual \spd cell fits covering the whole correction factor range.}
\label{spd_eff}
\end{figure}

In March 2010, the \lhc reached a centre of mass of \sqs = 7\tev. Data were taken at seven different thresholds values, from 0.3 to 2.1 $E_{\rm MIP}$. Only 3\% of the statistics was available for the highest threshold value (2.1 $E_{\rm MIP}$). A specific event reconstruction using only the T1, T2, T3 tracking stations was performed and several selections were applied on the reconstructed tracks. The theoretical expected efficiency (Figure~\ref{spd_expected_eff}) was then fitted for each cell with the data taken at each threshold value. Examples of these fits are shown in Figure~\ref{spd_eff}.

\begin{figure}[!ht]
\centering
\includegraphics[width=0.7\textwidth]{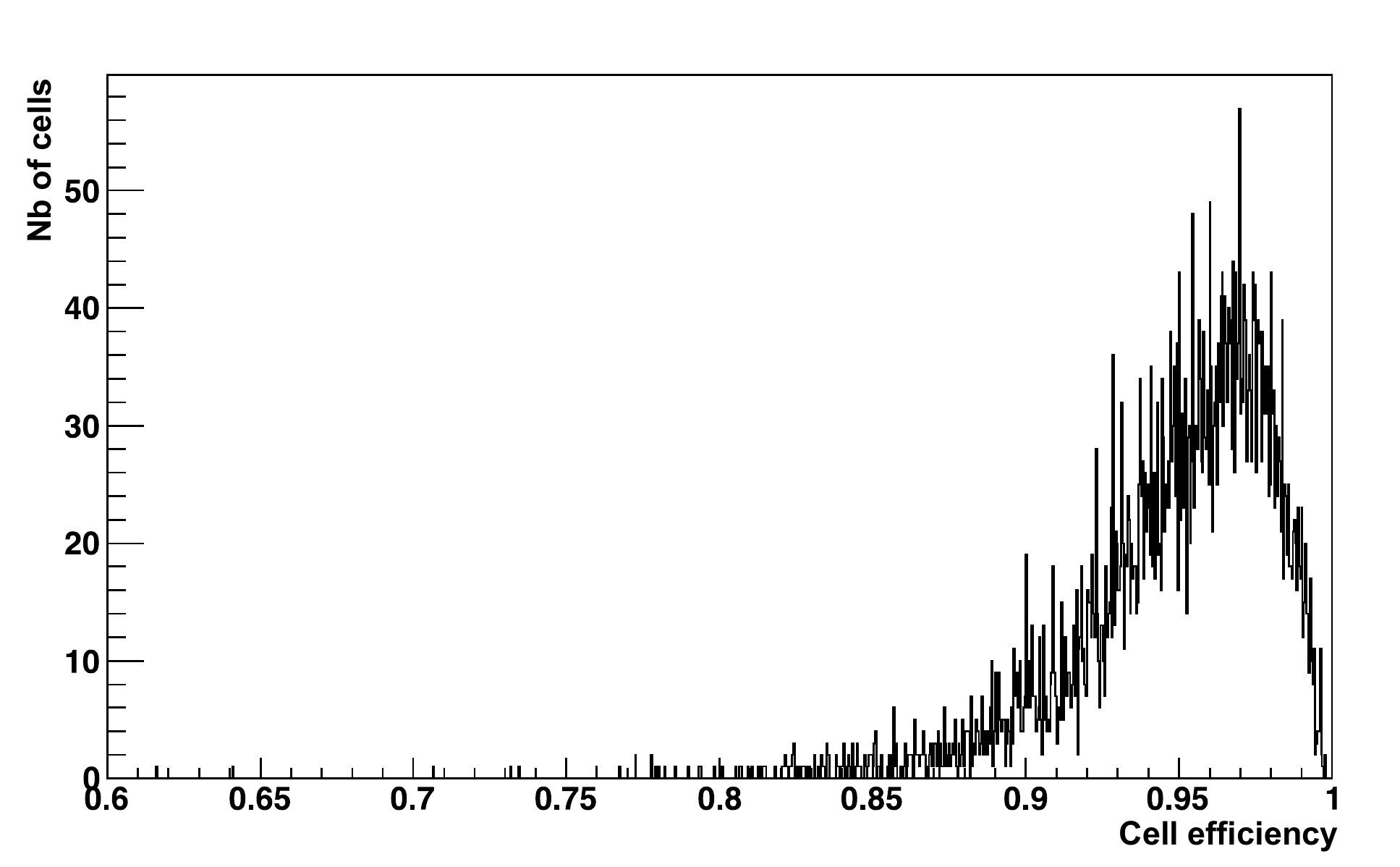}
\caption{Distribution of the cell efficiencies in the \spd in 2011. }
\label{spd_eff_2011}
\end{figure}

Cell per cell correction factors are extracted from these fits and applied into the \spd electronics at the start of the 2011 data taking period. Early 2011 data were used to measure the average efficiency, which is found to be 95\%, with a RMS of 3\% as shown in Figure~\ref{spd_eff_2011}. Later on, the evolution of the efficiency has been measured several times during the \lhc Run 1 period: June 2011, April 2012 and November 2012. The expected degradation of the efficiency with time and accumulated luminosity is visible and reaches $\approx$ 15\% in the Inner region (largest occupancy) of the \spd at the end of Run 1. This degradation is due to the radiations causing ageing effects on the PMTs, specially the photocathodes,  and on the scintillator material. At the end of Run 1 and an accumulated luminosity of 3.1\invfb, the average efficiency of the \spd decreased to 88\%, going down to 84\% in the case of the Inner region.

\subsubsection{Run 2 calibration}

At the beginning of Run 2 (October 2015), a partial recovery of the efficiency loss, of the order of 3\%, is observed. This recovery is due to annealing effects during LS1. A complete re-calibration is performed with the first collisions of 2017 in order to obtain new efficiency curves and new corrections to take into account the actual ageing of the detector components. The procedure is generally the same as for Run 1 with 6 threshold values (0.3 to 2.0 MIP) and 15 million minimum bias events per threshold value. The average efficiencies obtained range from 97.1\% at 0.3 MIP to 6.6\% at 2.0 MIP. The average MIP position is 0.89 giving about 10 photoelectrons per MIP. The correction to the pre-calibrated values has decreased by 24\% on average compared to the 2010 calibration. After recalibration, the efficiency of the \spd has been regularly monitored taking advantage of the implementation of new fast streams of calibration data~\cite{Aaij:2016rxn}. The average cell efficiency after recalibration is 94\%. From the monitoring of the \spd efficiency performed along all these years, an approximate ageing effect of 1.5\% efficiency decrease per collected \invfb is observed.

\subsection{\presh}

\subsubsection{Introduction}

The \presh uses charged particles pointing to its cells for calibration, by extrapolating their trajectories to the \presh. Several calibrations of the \presh occurred along the following timeline. Cosmic data making use of calorimeter tracks~\cite{AbellanBeteta:2012dd} were used prior to the first beam circulations. The raw data from proton-proton collisions without a complete reconstruction were then considered to refine the calibration corrections. Once the full reconstruction of charged tracks was available, the MIP pointing to a \presh cell were used to equalise the responses for each individual cell. This was cross-checked by a calibration method based on the energy flow in the detector~\cite{Martens:1391730}. This section focuses solely on the calibration with MIPs which was performed once the detector was properly aligned geometrically and in time with the tracking system. Run 1 calibration and Run 2 re-calibration are described. 

\subsubsection{Initial calibration and monitoring}

The 10-bit dynamic range of the \presh detector was designed to cover energy deposits from 0.1 to 100 times that of a MIP. It allows to measure simultaneously the high amount of energy deposited by the electromagnetic particles (photons and electrons) and the MIP energy deposits on the lower part of the dynamic scale, around 10 \adc counts. The linearity of the energy response recorded by the apparatus~\cite{LHCb-DP-2008-001} allows to use MIP energy deposits to calibrate each cell of the detector. 

The basic readout unit of the \presh is the FEB which gathers 64 scintillating detector cells read out through a MA-\pmt. The \presh calibration strategy with particles at minimum ionisation hence follows a two step procedure:  
\begin{itemize}
\item The cells belonging to a FEB are inter-calibrated such that they produce the same response to the passage of a MIP. The two sub-channels corresponding to the alternated integration at the level of the VFE boards receive an individual correction accounting for the gain differences between the two VFE amplification channels.
\item All the FEB responses are equalised by adjusting the \hv of the corresponding MA-\pmt.   
\end{itemize}

\subsubsection{MIP sample and fit model of the \presh response}

\begin{figure}[!ht]
\centering
\includegraphics[width=0.7\textwidth]{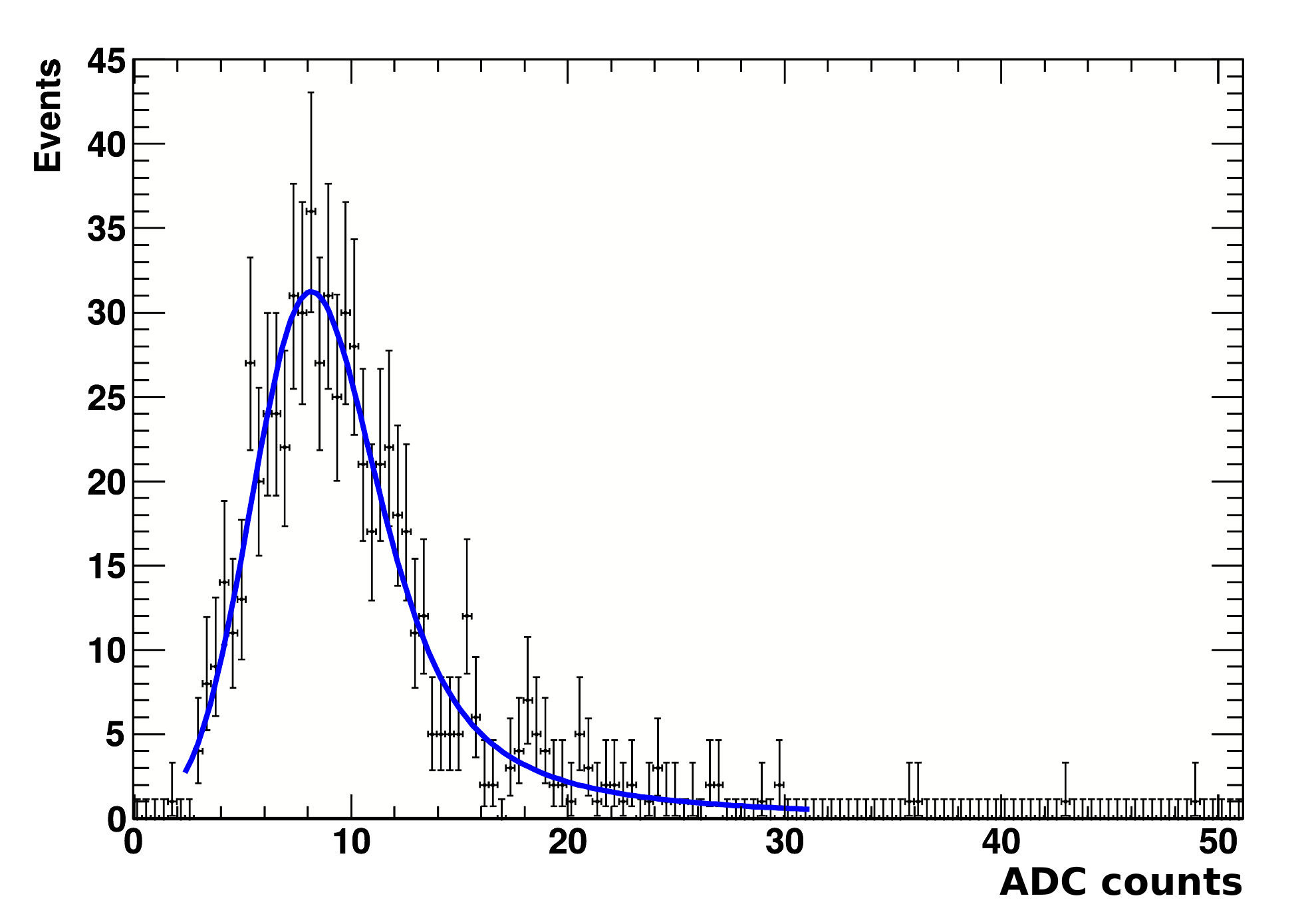}
\caption{Distribution of the energies, expressed in \adc counts, recorded in a \presh cell with a charged track pointing to it. 
The result of the fit with the model discussed in the text is superimposed.}
\label{ps_cell_adc}
\centering
\end{figure}

The MIP sample is composed of reconstructed tracks of charged particles in the \lhcb tracking system, requiring their momentum to be larger than 2\gevc in order to reach the \presh. The response of each cell with a charged track pointing to it is recorded. Figure~\ref{ps_cell_adc} shows a typical distribution of the measured energy of such cells. The energy is corrected for the track trajectory length in the scintillator; the electronics offsets of each sub-channel are as well corrected using the values measured in the calibration sequences taken during the proton collision runs. The response is modelled by a convolution of a Landau distribution with a Poisson distribution, the latter accounting for the fluctuation in the photo-statistics of the MA-\pmt channels. The estimator considered to represent the response of the cell is the most probable value of the Landau distribution, as determined from the fits.

\subsubsection{Run 1 calibration coefficients and performances}

A set of 128 coefficients for each board is derived (one for each sub-channel) and fed into the FEB electronics components processing the events at 40\mhz rate. These multiplicative corrective factors, denoted $g_{\rm ch}$, ensure a uniform response of the cells within a FEB. The next step consists in equalising the responses of the 100 FEB by setting accordingly the high voltage $V$ of each MA-\pmt. This is realised by comparing the current to previous responses of the cells. The response for each channel can be written (in \adc counts) as 
\begin{equation}
R_{\rm ch} = \alpha_{\rm ch} \cdot g_{\rm ch} \cdot V^{\beta},
\end{equation}
where $V$ is the applied high voltage and $\beta$ the high voltage power factor assumed to be the same for all channels of a given tube. $\alpha_{\rm ch}$ is a constant multiplicative factor characterising each channel, irrelevant for the new calibration since the latter is performed relative to the current calibration.   

For each channel,  the newly calibrated high voltage $V^{\prime}$ is derived from the following computation,
\begin{equation}
\label{Eq:hvgain}
V^{\prime}= V \left ( \frac{R^{\prime}_{\rm ch} \, g_{\rm ch} }{R_{\rm ch} \, g^{\prime}_{\rm ch}}\right )^{\frac{1}{\beta}},
\end{equation}
where $R^{\prime}_{\rm ch}$ is the required response of the channel (10 \adc counts here),  $g^{\prime}_{\rm ch}$ the newly measured  corrective factor to be fed in the electronics. The actual derivation of $V^{\prime}$ is computed from average values over all channels of the corresponding FEB unit.        

\begin{figure}[!ht]
\centering
\includegraphics[width=0.7\textwidth]{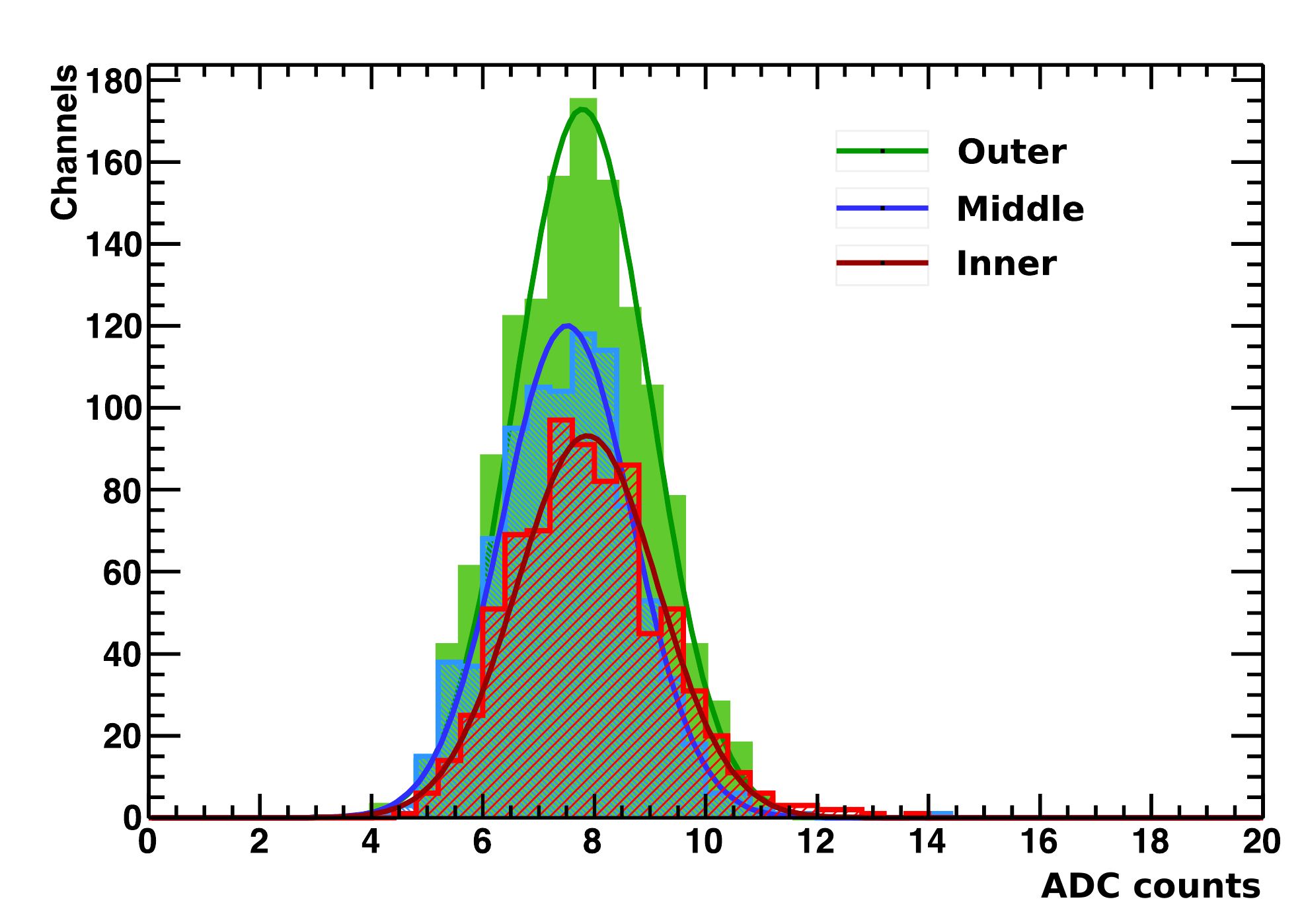}
\caption{Measurement of the MIP-calibrated energy deposit in 2011 data. The distributions in the three \presh regions are superimposed as well as the fit of a Gaussian function to each of these data distributions.}
\label{ps_mip_e_2011}
\end{figure}

The calibration performance is checked a posteriori by adjusting with a Gaussian function the overall response of all channels recorded in subsequent runs. The variance determined by the fit is used as the estimator of the performance, e.g. a calibration at the 10\% level means that 68\% of the channels are comprised in an interval of $\pm10\%$ around the central value.  A convenient way to estimate the typical performance of the calibration with MIP is to derive the response of each cell from the energy flow method described later. This comparison allows to establish that a typical calibration at a level better than 5\% is achieved. As an illustration, Figure~\ref{ps_mip_e_2011} shows the performance of the calibration made with the late 2010 data split by regions of the \presh and observed in the first data recorded in the 2011 run. Similar performance is achieved for 2012 data.        

\subsubsection{Run 2 recalibration}

The \presh has been recalibrated using 2016 data to correct the settings from ageing factors. The procedure has been the same one used for Run 1 and consists on the following steps. 
\begin{enumerate} 
\item Select clean muons from di-muons stream stripping lines~\cite{Aaij:1384386}.
\item Study the pre-shower response.
\item Correct for the track length in the scintillator and build up the \adc count distribution of each channel.
\item Correct for pedestals variations in the relevant periods.
\item Fit a Landau distribution function convoluted with a Gaussian to the \adc distribution.
\item Build the gains for each channel to make uniform the 64 responses within a board.
\item Adjust High Voltages to make all boards responses uniform (within 10 \adc counts).
\end{enumerate}
The relative change in the most probable value for the MIP responses (${\rm MPV}$) for 2016 compared to 2011,
\begin{equation}
{\cal A} = \frac{{\rm MPV}^{\rm 2016} - {\rm MPV}^{\rm 2011}}{{\rm MPV}^{\rm 2011}},
\end{equation}
is averaged over the three \presh detector areas for sides A and C. It is negative, lower on side C than side A and decreases from Outer to Inner. Lowest value is -26\% (side C, Inner) and largest value is -6\% (side A, Outer).

\subsection{\ecal}

\subsubsection{Introduction}

The different operations leading to the calibration of the \ecal are described: initial adjustment of \ecal energy scale, energy flow calibration and fine calibration of the \ecal cells. The corrections and methods introduced in Run 1 to compensate for detector ageing are detailed. In Run 2 new methods allowing to correct for ageing effects and recalibration on a daily basis are described.

\subsubsection{Initial adjustment of \ecal energy scale}

{\noindent\bf General principles}

The dynamic range of \ecal signals is defined by the following relation
\begin{equation}
\adc_{\rm max}= E_{\rm max} \, e  \, k \frac{Y \, G_{\rm nominal}}{s_{\rm \adc}},
\label{ecal_e_to_adc}
\end{equation}
where:
\begin{itemize}
\item $\adc_{\rm max}$ is the upper limit of the readout \adc dynamic range (coded on 12 bits),
\item $E_{\rm max}$ is the upper limit (saturation) for the deposited energies measured by individual \ecal cells. In order to ensure an optimal detector and trigger performance, the value of $E_{max}$ for a cell with coordinates $(x,y,z)$ is chosen as $E_{\rm max}(\gev) = 10/\sin\theta$ with $\sin\theta=\sqrt{x^2+y^2}/\sqrt{x^2+y^2+z^2}$,
\item $e$ is the electron charge,
\item $k=1/3$ is a factor introduced by the clipping circuit corresponding to the fraction of the signal remaining after filtering,
\item $Y$ is the photoelectron yield, which is the average number of photoelectrons produced per \gev of energy deposited by incoming particles. It is the product of the light yield of the \ecal module by the quantum efficiency of the \pmt photocathode,
\item $G_{\rm nominal}$ is the \pmt operational gain,
\item $s_{\rm \adc}$ is the sensitivity of the readout \adc, defining the value of integrated charge per \adc count.
\end{itemize}
The energy scale of an \ecal cell is set by adjusting the \pmt gain, $G$, to the value $G_{\rm nominal}$ defined by Eq.~(\ref{ecal_e_to_adc}). The gain itself is regulated by means of the \hv supplied to the \pmt. Thus, the parameters required for the energy scale adjustment are the \adc sensitivity, the photoelectron yield and the dependence of $G$ to the \hv, $G({\rm \hv})$ which will be referred as the {\it \pmt regulation curve} in the following.

All readout \adc chips passed a pre-installation selection in the laboratory. The permissible variation of $s_{\adc}$ was set to $\pm$5\% around the central value $s_{\adc}=0.0195~{\rm pC/\adc}$ count. The channel-to-channel dispersion of the sensitivity values among accepted \adc counts has a RMS 2.5\%.

\vspace{0.5\baselineskip}
{\noindent\bf Calibration of \pmt gains}

All photomultipliers were pre-calibrated {\it in situ} with the LEDs of the \ecal monitoring system~\cite{Machikhiliyan:1557395}. From the fixed amplitude of the \led light pulses, the absolute value of the nominal \pmt gain $G$ could be extracted from the mean $A$ and the width $\sigma_{A}$ of the Gaussian distribution of the output signal,
\begin{equation}
G = K \frac{\sigma_{A}^{2} - \sigma_{P}^{2}}{A - P},
\label{ecal_pmt_gain}
\end{equation}
where $P=0$ and $\sigma_{P}$ are the mean and the RMS of the pedestal distribution (typically $\sigma_{P} = 1.2$ \adc counts), while $K$ is a known factor defined by the hardware properties. It is proportional to the \adc sensitivity $K \sim s_{\rm \adc}$ and therefore the procedure of gain calibration allows to cancel the dispersion in \adc sensitivities as well.

The formula in Eq.~(\ref{ecal_pmt_gain}) is valid under the condition that the main contribution to the width $\sigma_{A}$ comes from photo-statistics, i.e. $\sigma_{A}\sim\sqrt{N_{\rm p.e.}}$, where $N_{\rm p.e.}$ is the number of photoelectrons emitted by the \pmt photocathode. Moreover, $\sigma_{A}$ should be large enough in comparison with the pedestal RMS, $\sigma_{P}$, and with the \adc resolution. In order to fulfil these requirements, the absolute gain values were measured in a reference point with specific hardware settings, when the intensity of the \led light was set as low as possible and, in compensation, the \pmt high voltages were significantly increased. 

\begin{figure}[!ht]
\centering
\includegraphics[width=0.7\textwidth]{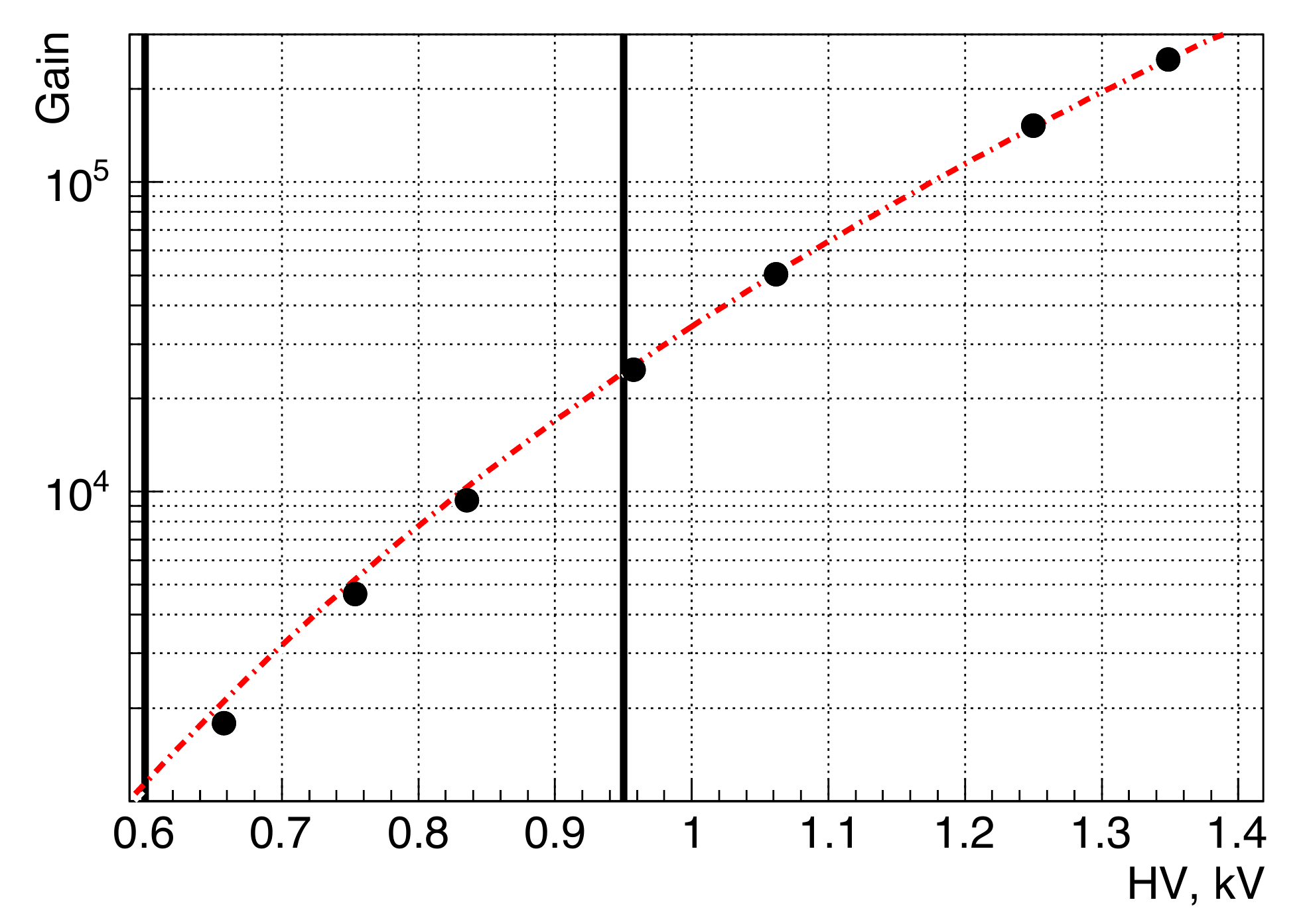}
\caption{Example of a \ecal \pmt regulation curve. The two bold vertical lines mark the working range of \ecal high voltages.}
\label{ecal_pmt_curve}
\end{figure}

The second step of the calibration procedure aims to obtain the individual profiles of regulation curves in wide \hv ranges, which were extracted from the relative change of \pmt response with \hv to fixed light intensities. The final dependence $G({\rm \hv})$ for each readout channel was parameterised in the form $G=G_{0} \, ({\rm \hv}/{\rm \hv}_{0})^\alpha$, where ${\rm \hv}_{0} = 1\,{\rm kV}$ is the \hv normalisation constant and $G_{0}$ and $\alpha$ are two parameters describing the \pmt regulation curve. A typical example of a measured $G({\rm \hv})$ function is shown in Figure~\ref{ecal_pmt_curve}, presenting both the set of experimental points and the fitted curve. 

\vspace{0.5\baselineskip}
{\noindent\bf Determination of the photoelectron yield}

All \ecal modules passed pre-installation checks on a dedicated test bench, where cosmic rays were used to measure the response on MIPs~\cite{Arefev:1080559}. The light readout system employed a \pmt of the same type as used in the \ecal. Each readout channel was equipped with a \led, which allowed to monitor its stability and to determine the gain and the amount of photoelectrons by the same method as described above. The absolute energy scale was determined by a complementary test-beam measurement of the equivalent energy deposition of MIPs (0.33\gev). The resulting numbers for the average photoelectron yield $Y$ were ~3100, 3500 and 2600 photoelectrons per \gev for inner, middle and outer modules, respectively, while the cell-to-cell variation of $Y$ was found to be within 7\%~\cite{Arefev:1080559}.

\vspace{0.5\baselineskip}
{\noindent\bf Energy scale adjustment}

Finally the individual \pmt regulation curves, the central value $s_{\rm \adc}$ and the average values of $Y$ for each \ecal region were used to calculate the initial \hv settings of \ecal photomultipliers. One additional correction was applied at this stage to compensate for the dispersion in the quantum efficiency of \pmt photocathodes (6\% RMS). 

The precision of the initial cell-to-cell inter-calibration was estimated to be within 10\%. Two main contributing factors were the dispersion in photoelectron yields and the accuracy of gain determination. Such a precision was sufficient to observe a clear \piz signal from $\piz \to \gamma \gamma$ decays as soon as the \lhc started to deliver the first proton collisions. The relative width of the \piz peak was at the level of 10\%~\cite{Machikhiliyan:1557395}.

\subsubsection{Energy flow calibration}

\begin{figure}[!ht]
\centering
\begin{minipage}{0.49\textwidth}
\includegraphics[width=1.0\linewidth]{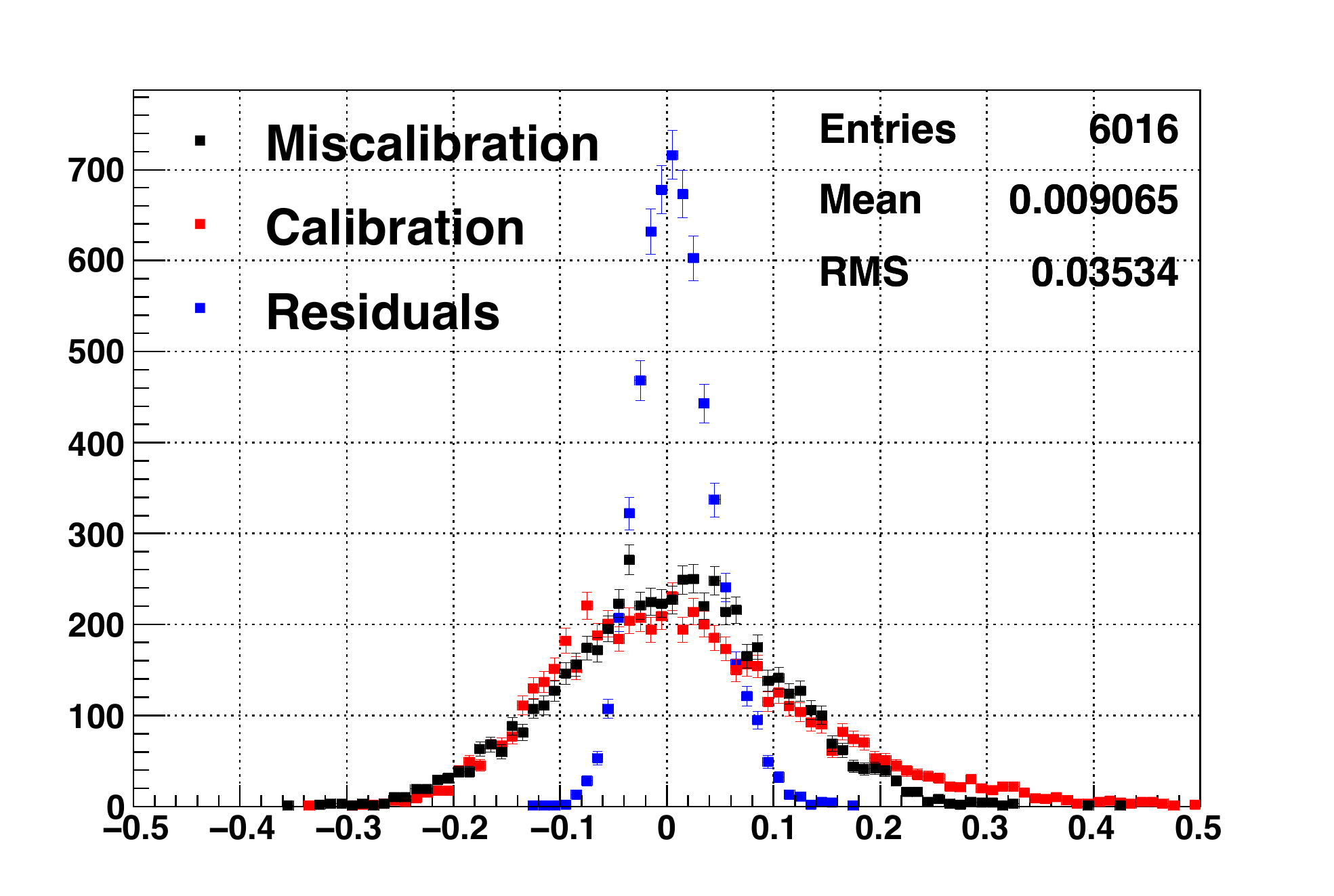}
\end{minipage}
\begin{minipage}{0.49\textwidth}
\includegraphics[width=1.0\linewidth]{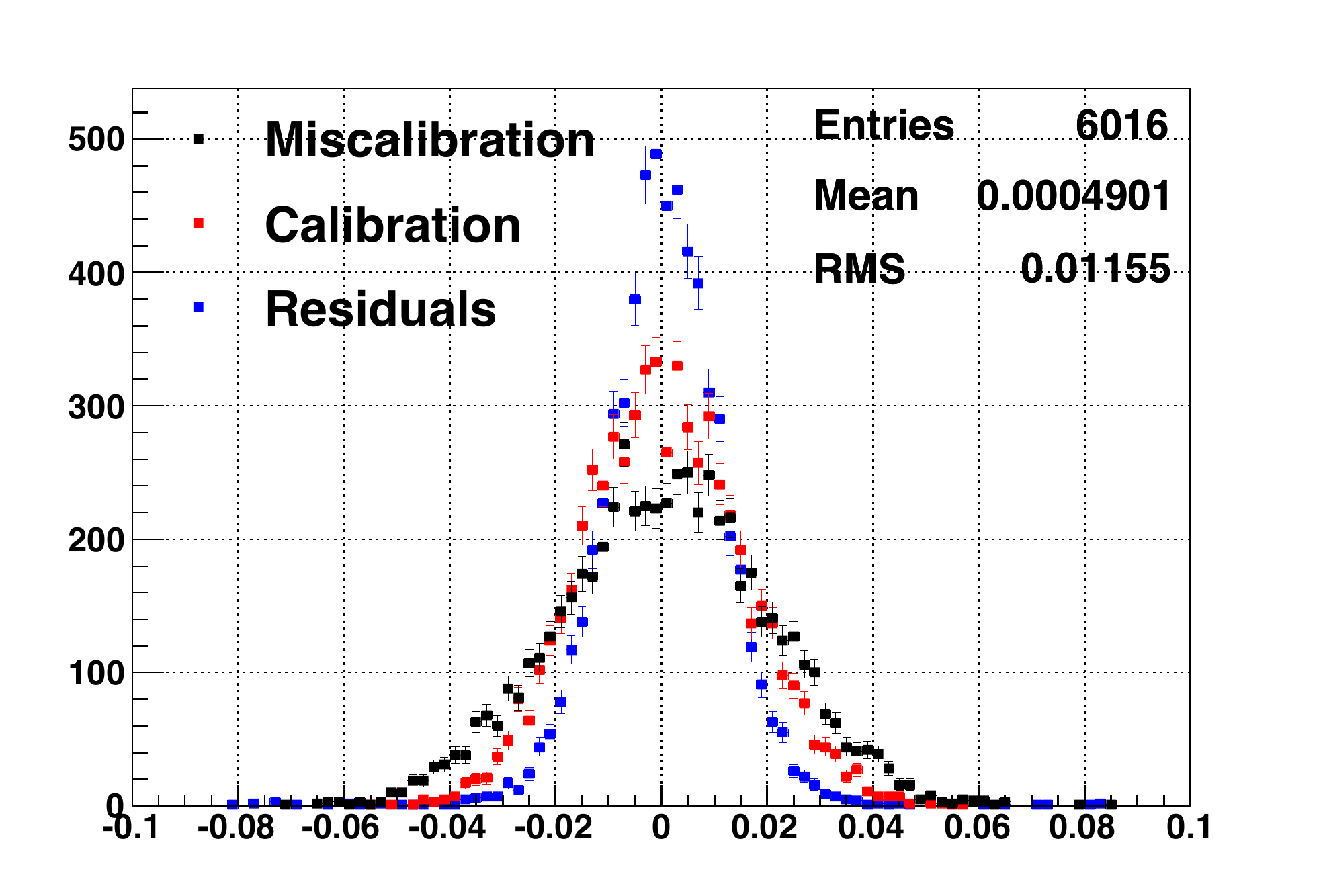}
\end{minipage}
\caption{Miscalibration (black), calibration (red) and residuals (blue) of the energy flow calibration procedure for an initial mis-calibration of 10\% (left) and 2\% (right), obtained on simulated samples. The left (right) plot corresponds to the situation after the initial (final) calibration. In both cases a significant improvement of the calibration is obtained.}
\label{ecal_miscal_eflow}
\end{figure}

After this first adjustment, an energy flow method based on the continuity of the cumulative energy deposit was used to equalise the gain in each of the 3 \ecal zones. Fluctuations of the flux among neighbouring cells due to initial mis-calibration are smoothed by averaging the energy flow over clusters of $3\times 3$ cells and by exploiting the symmetry of the energy flow over the calorimeter surface~\cite{Martens:1391730}. This procedure allows to equalise the calibration and give calibration coefficients, but for the absolute calibration the \piz mass response is used. Simulation samples, simulated with a known mis-calibration, were used to determine the sensitivity of the method. The calibration, as shown in Figure~\ref{ecal_miscal_eflow}, is improved by approximately a factor three (two) when the initial calibration precision is of 10\% (2\%). These two exercises correspond typically to situations where the energy flow is used after the initial calibration or the final calibration. The method has also been used to cross-check the precision of the \presh and \hcal calibrations. 

\subsubsection{Pile-up}

Pile-up effects are present in \lhcb since the number of collisions occurring for one selected event can be larger than 1. It is a special concern in calorimeters and more specifically in the \ecal where most of the e.m. energy is deposited. To estimate the effect of the pile-up, the following method has been designed: 
\begin{itemize}
\item for each 3x3 energy cluster $E_{cluster}$ obtained in \ecal at (x,y) a 3x3 "mirror" energy $E_{mirror}$ is measured at (-x,-y),
\item if $E_{mirror} < E_{cluster}/2$, $E_{mirror}$ is the pile-up energy $E_{PU}$ for this cluster; otherwise, $E_{cluster}$ and $E_{mirror}$ are swapped,
\item SpdMult being the \spd multiplicity, $<E_{PU}> = <E_{mirror}>(i+1) - <E_{mirror}>(i)$ for SpdMult $\ge$ 200 and i being bins of 50 SpdMult,
\item the linearity of the pile-up with SpdMult is obtained when SpdMult is corrected for the pile-up in the \spd.
\end{itemize}
This method permits to subtract the pile-up from the measured energy. This improves the quality of the measurements (\piz). The pile-up energy can be as high as 4-5 \adc channel per 50 \spd hits in the inner bending plane.

\subsubsection{Run 1 \ecal ageing corrections}

\begin{figure}[!ht]
\centering
\begin{minipage}{0.49\textwidth}
\includegraphics[width=1.0\textwidth]{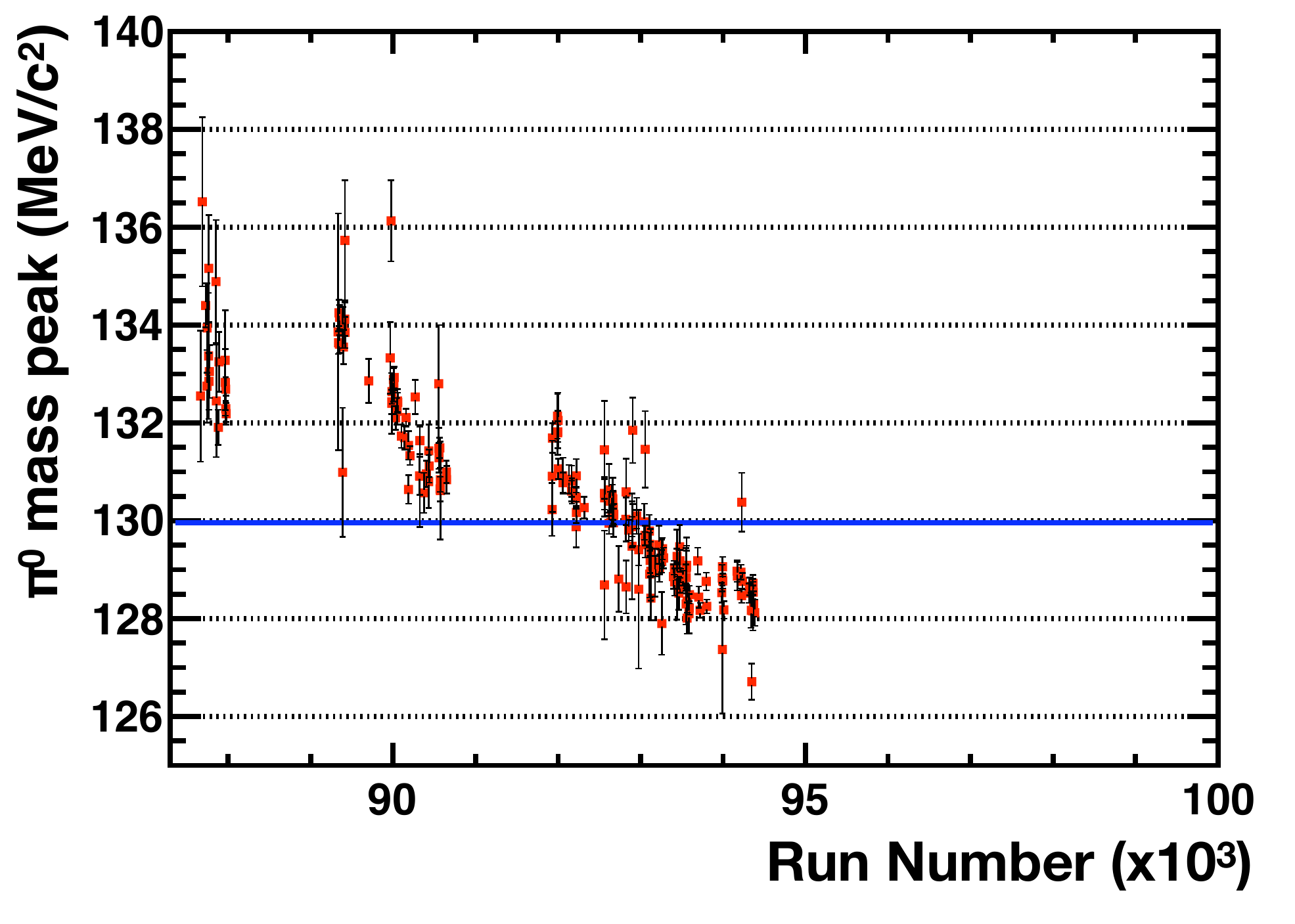}
\end{minipage}
\begin{minipage}{0.49\textwidth}
\includegraphics[width=1.0\linewidth]{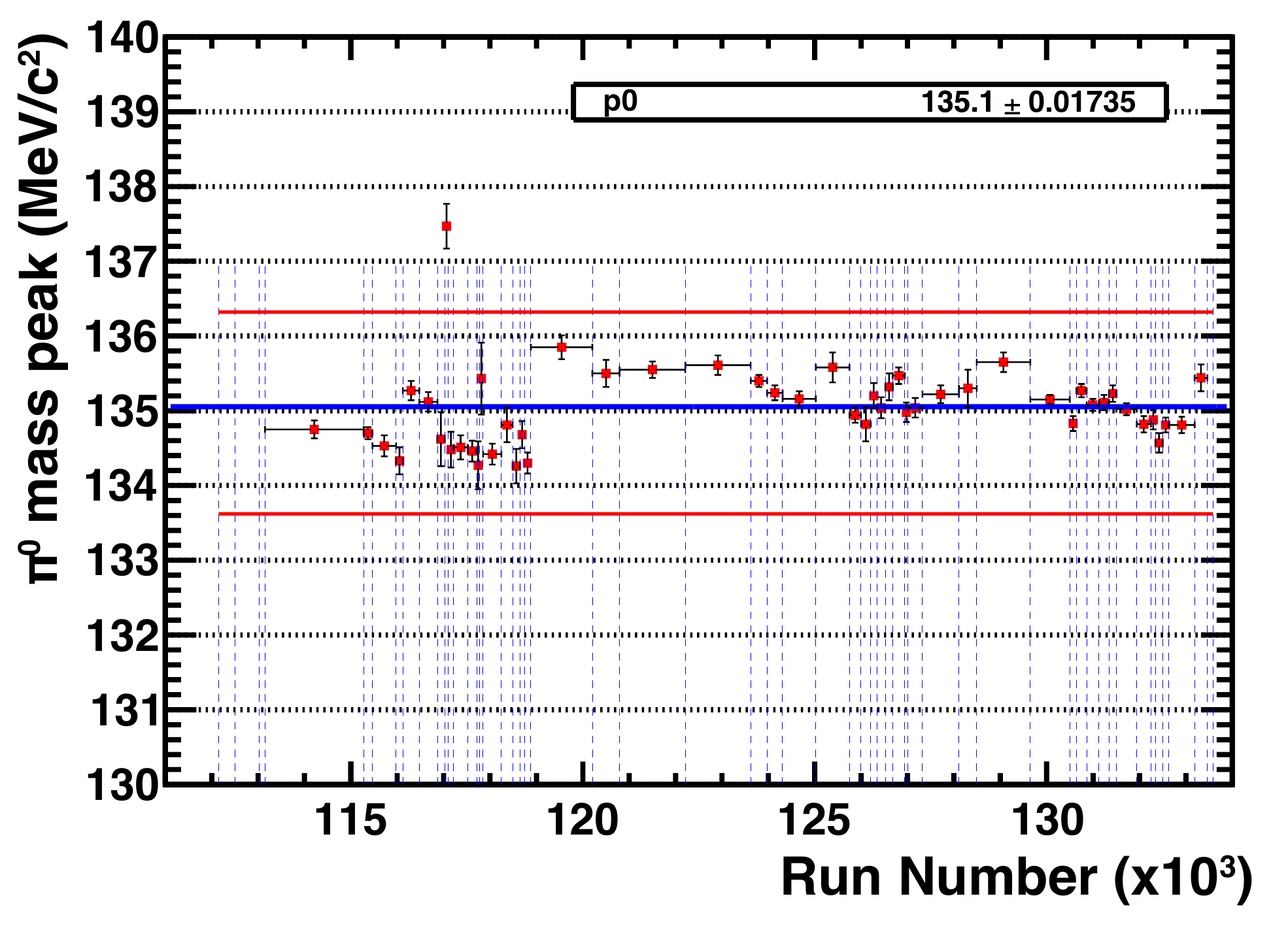}
\end{minipage}
\caption{Value of the fitted \piz mass as a function of time (run number, 2011 data). The decrease observed is due to the \ecal ageing (left). \piz mass fit value as a function of time (run number) after applying the gain correction (right).}
\label{pi0_mass_ageing} 
\end{figure}

\ecal ageing effects can be seen from the variation over time of the \piz mass fit results obtained for the calibration procedure described before (Figure~\ref{pi0_mass_ageing} (left)). To address this problem, the calibration of the \ecal is realised in two steps. First,  the fine calibration of each \ecal cell is applied, using the \piz mass fit method described below, and data taken during a short period. This procedure is applied monthly. This corresponds to about $200\,{\rm pb}^{-1}$ of recorded data and gives a monthly fine calibration reference. These calibrations have been used to update regularly the \hv of the \ecal PMTs during the entire data taking period in order to mitigate the effect of ageing on the \lone trigger performances. The measured \piz mass after these corrections is shown in Figure~\ref{pi0_mass_ageing} (right).

\begin{figure}[!ht]
\centering
\begin{minipage}{0.49\textwidth}
\includegraphics[width=1.0\textwidth]{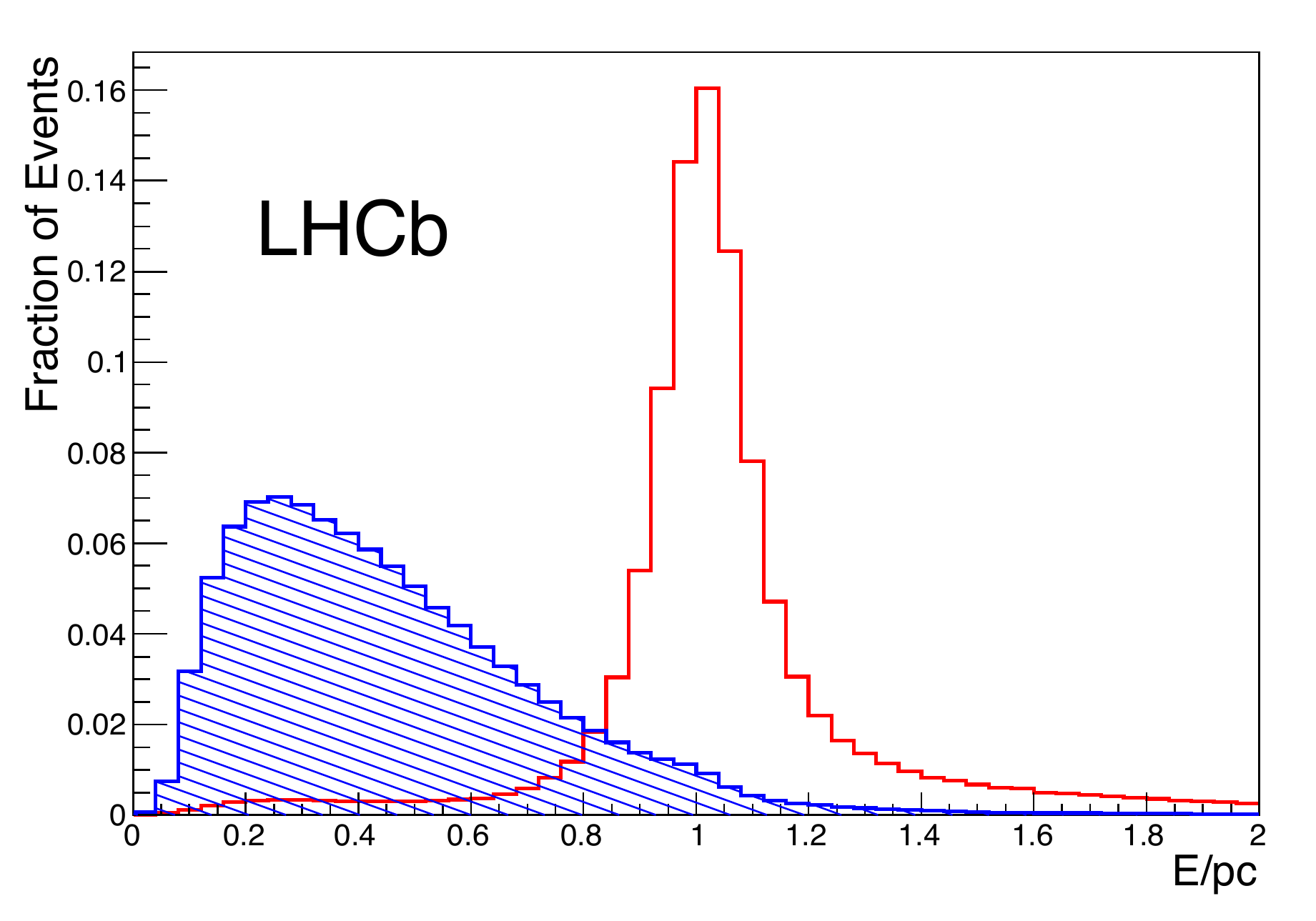}
\end{minipage}
\begin{minipage}{0.39\textwidth}
\includegraphics[width=1.0\linewidth]{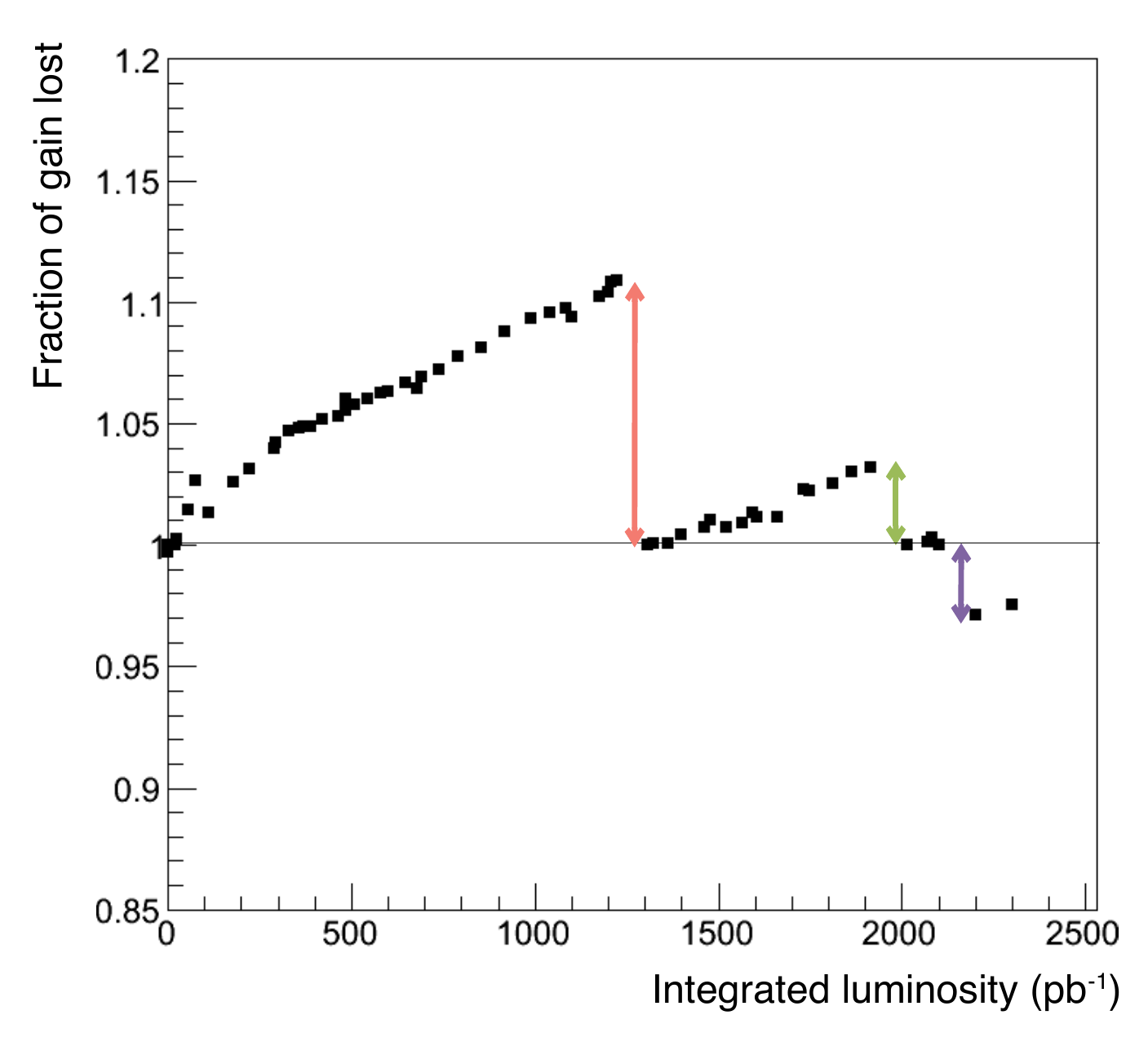}
\end{minipage}
\caption{Distribution of $E/p$ in the \ecal for electrons (red) and hadrons (blue) using 2011 data (left). Fraction of gain lost for electrons in the \ecal as a function of the integrated luminosity (right). The two first steps correspond to the \hv changes done during the data taking period, the third one is due to annealing during one month without beam.}
\label{ecal_e_over_p}
\end{figure}

The ageing trend can be measured from electrons in a shorter time scale by using converted photons. In this case, stringent selection requirements are applied on one of the electrons from the conversion and the study is performed on the other one (tag and probe method). The $E/p$ distribution for electrons and hadrons (where $E$ is the energy measured by \ecal and $p$ the momentum measured by the \lhcb tracking system) is represented in Figure~\ref{ecal_e_over_p} (left), showing that an electron sample with large purity can be obtained. From the untagged electron $E/p$ values, the trend of ageing in 2012 is determined, in the different calorimeter regions and shown in Figure~\ref{ecal_e_over_p} (right). The gains used in the offline analysis of the individual cells are updated every $40\,{\rm pb}^{-1}$ of recorded luminosity, using the convolution of the reference calibration with the measured ageing trend defined per detector zone. 

The ageing effect is understood as coming from two main sources. 
\begin{itemize}
\item The high \pmt current induces a gain degradation of the \pmt, part of which is expected to be recovered after long shut down periods.
\item Radiation damage to the scintillator tiles and the fibres is also present. An annealing effect is expected during long stops.
\end{itemize}

\subsubsection{Run 1 \ecal calibration}

\begin{figure}[!ht]
\centering
\includegraphics[width=0.7\textwidth]{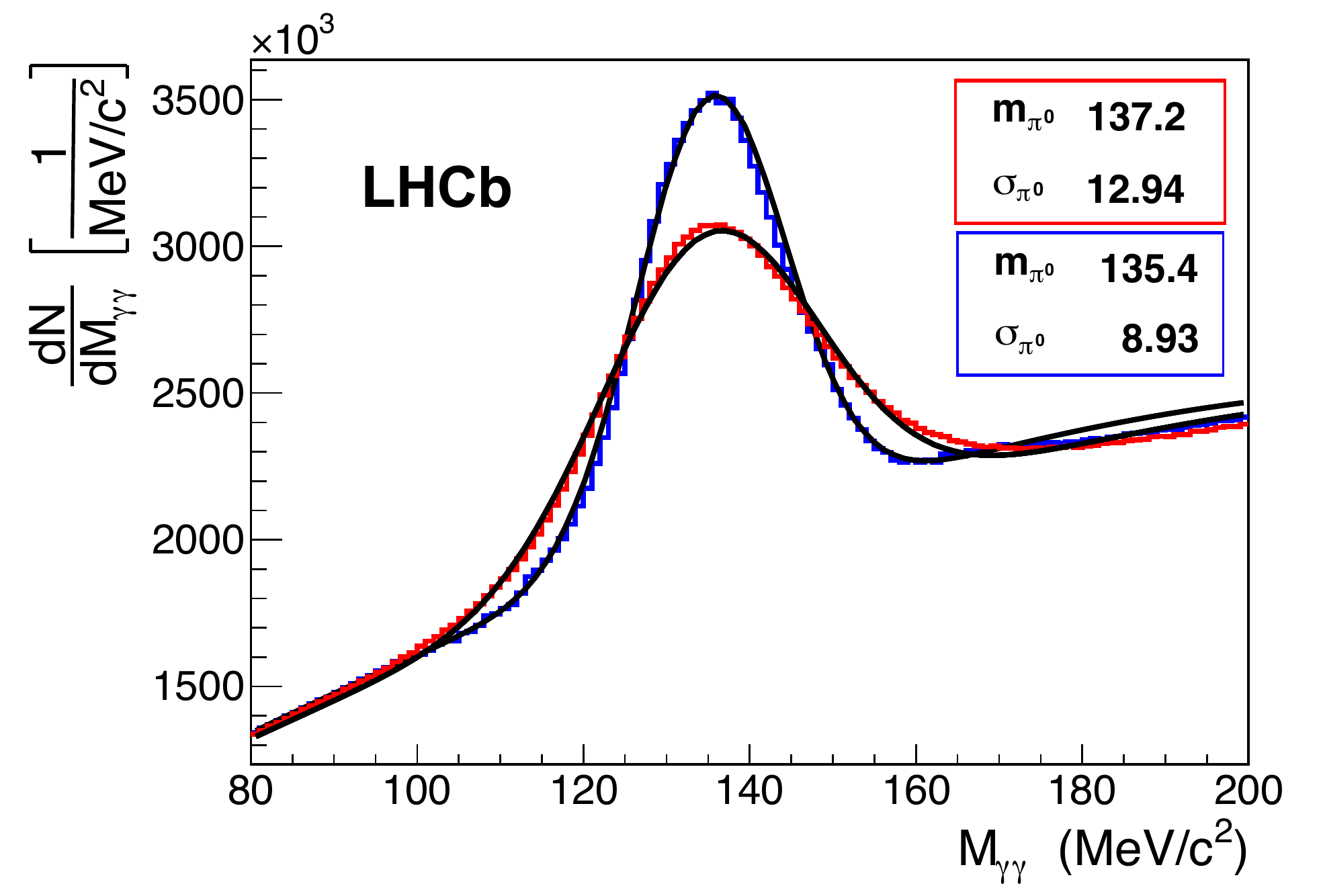}
\caption{Invariant mass distribution for $\piz \to \gamma\gamma$ candidates upon which the fine calibration algorithm is applied. The red curve corresponds to the distribution before applying the method, while the blue curve is the final one. Values in the red (blue) box are the mean and sigma of the signal peak distribution in \mevcc before (after) applying the fine calibration method.}
\label{pi0_mass_calib}
\end{figure}

The fine calibration approach uses a fitting iterative method of the \piz mass distribution from two photon combinations~\cite{Belyaev:2011zz,Belyaev:2014zga}. In this procedure from minimum bias events, photons are defined from $3\times 3$ neutral clusters consistent with single photon signals, in which the cell with the highest energy deposit is called the seed. A selection is imposed to use only photons leaving a small energy deposit in the \presh. The selected \piz candidates must belong to low multiplicity events in order to limit pile-up effects. For each cell, the di-photon invariant mass is fitted. When enough statistics is reached, the energy of each seed is corrected to match the \piz nominal mass~\cite{PDG2012} and the procedure is repeated until the obtained calibration is stable. In Figure~\ref{pi0_mass_calib}, the effect of the fine calibration algorithm is reported. The estimated precision of the calibration reached by this method is of the order of 2\%, estimated by applying the method on a mis-calibrated simulation sample. In Run 1 the foreseen scheme to correct ageing, based upon gain corrections monitored by \led response was not able to be applied, as the \ecal \led clear fibres showed strong and non-uniform radiation ageing. The convolution of ageing monitored by electrons each 40\invpb and full calibration performed on 200 millions minimum bias data, was applied during Run 1 before the annual reprocessing of the data.

\subsubsection{Run 2 \ecal calibration}

\begin{figure}[!ht]
\centering
\includegraphics[width=0.7\textwidth]{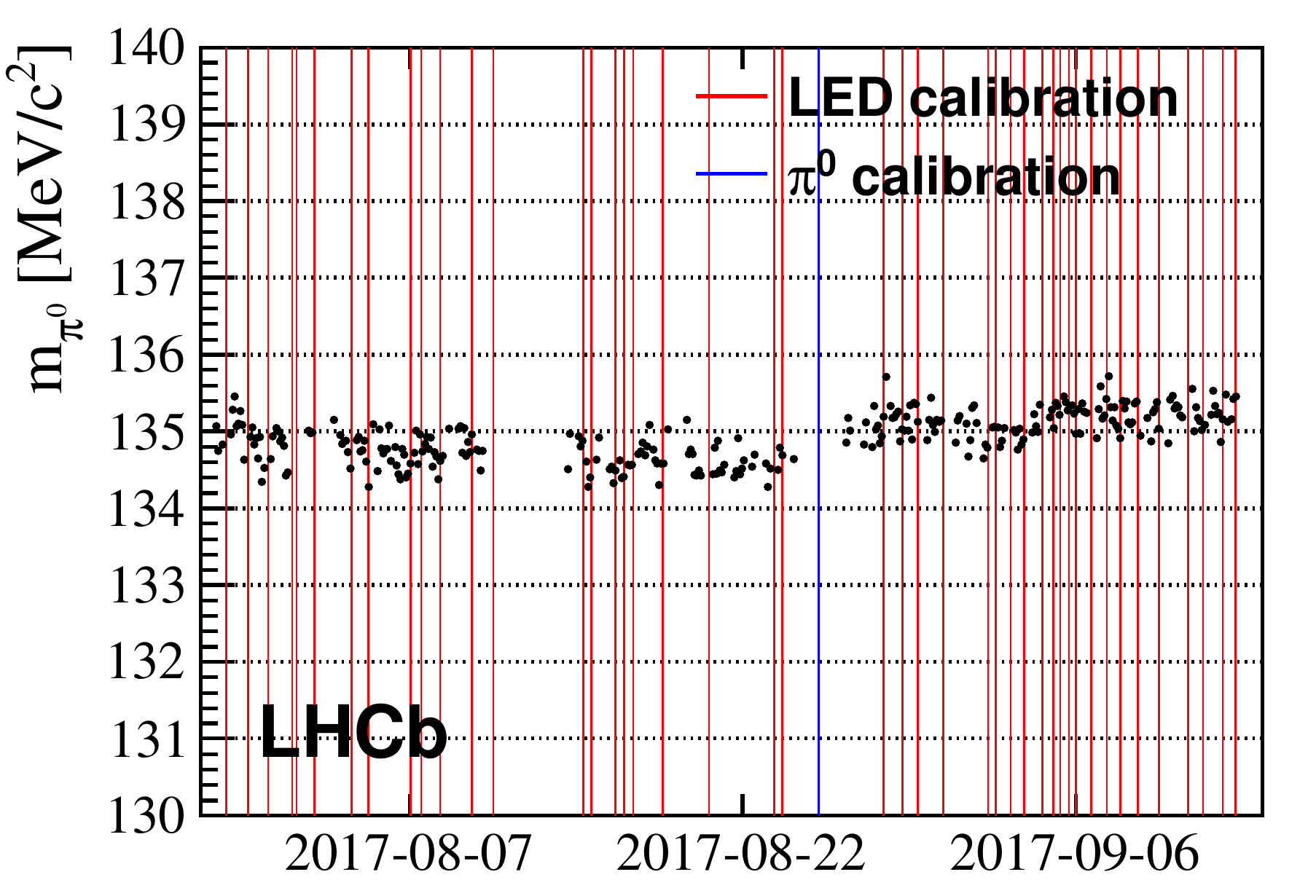}
\caption{\piz mass as a function of time in summer 2017. The red (blue) lines correspond to \hv changes from the \led system (\piz calibration).}
\label{pi0_mass_2017}
\end{figure}

During Run 1 the clear fibres bringing \led signal to \pmt were damaged by radiation preventing their use for calibration purpose. In order to use the designed monitoring system, during the shutdown period between Run 1 and Run 2 all \ecal clear fibres have been changed for high radiation resistant ones. The second concern was to avoid reprocessing and to follow in line the calibration based on \pmt gain adjustments. From 2015 onwards the calibration scheme applied to \ecal is the following:
\begin{itemize}
\item \pmt ageing is performed by the \led system, the corresponding gain changes being calculated online and validated by operator decision in 2015 and 2016. Later on (2017-2018), the procedure has been fully automatised. A reference file is produced after the first fine calibration of the year using the \led signals for each of the 6016 cells. The high voltage of the PMTs are adjusted after each fill so that the \led signal of the current fill matches the \led reference value.
\item Fine calibration based upon \piz mass calculation for each cell, with an iterative procedure, is performed from stream reconstructed online from minimum bias events. The rate of this calibration is once per running month (6 times per year). Minimum bias events are collected at a rate of 400 \hz and saved on the \hlt farm. Reconstructed data are automatically produced at the end of each fill and are accumulated. The fine calibration based upon \piz mass calculation is launched as soon as the total number of events reaches 300 millions. The calculations run on the \hlt farm at the end of the fill just after the production of reconstructed data. The full procedure requires 1 to 2 hours and is stopped if the \lhc is filled again before completion of the procedure. In this case, the calibration procedure is re-scheduled to start at the end of the following fill; it thus benefits from a larger number of events. The fine calibration is an iterative procedure consisting of 7 iterations running on the initial data, followed by a full reconstruction of the events using the newly computed constants\footnote{$\lambda_i = m_\piz(PDG)/m_\piz(fit)$} $\lambda_1$ for each cell. This full reconstruction leads to a new set of data on which 7 iterations are performed, producing a new set of constants $\lambda_2$. The final set of constants is $\lambda_1*\lambda_2$ for each cell and is directly applied to the reference used to obtain the new \hv values for the PMTs from the \led system. Thus it is automatically taken into account. 
\end{itemize}
As a consequence, the stability of the reconstructed \piz mass is greatly improved as can be seen in Figure~\ref{pi0_mass_2017}.

\subsection{\hcal}

Similar to \ecal, \hcal calibration is chosen such as to provide a uniform \et response, $E_{\rm max}=E_{\rm Tmax}/\sin\theta$, where $E_{\rm max}$ and $E_{\rm Tmax}$ are the maximal values of the \hcal dynamic range where $\theta$ is the angle between the beam axis and a line from the interaction point to an \hcal cell. The value of $E_{\rm Tmax}$ is equal to 15\gev for the 2010 and 2011 running periods. In 2012, the gains of all the \hcal PMTs were reduced by factor of 2, meaning $E_{\rm Tmax}=30\gev$ 
because of the increased instantaneous luminosity at the \lhcb collision point, in order to avoid too high \pmt anode currents. 

\subsubsection{Calibration with Caesium source}

The main \hcal calibration tool is its built-in $^{137}{\rm Cs}$ source system~\cite{LHCb-DP-2008-001}. Two $10\,{\rm mCi}$ $^{137}{\rm Cs}$ sources, one per detector half, are driven by an hydraulic system and traverse each scintillator tile. The \pmt response to the source passage through the corresponding \hcal cell, \ie the anode current,
is measured with the help of a dedicated system of current integrators. Due to the uniform structure of the \hcal, this current is expected to be proportional to the \hcal cell sensitivity to hadron shower energy\footnote{This can be defined e.g. as the \pmt pulse charge divided by the energy deposited in the cell.}, with a coefficient depending on the activity of the source. Deviations from proportionality can arise from spreads in the gains of the FEBs (which are not included in the source calibration chain), as well as from various mechanical tolerances of the \hcal modules assembly.

The relation between the \pmt anode currents measured during the Cs source calibration runs and the cell sensitivities was studied during a test beam campaign with several \hcal modules before they were installed in the \lhcb detector cavern. The tests consisted in Cs calibration runs for each module and characterisations of the module energy response with 50 to 100\gev pion beams performed right afterwards.

It was shown that the proportionality holds with a 2-3\% precision, which is by far enough for the \hcal operation in the \lhcb experiment. However the proportionality coefficient itself was determined with a 10\% uncertainty, which corresponds to the uncertainty on the knowledge of the activity of the Cs sources (the sources used at the beam test and in \lhcb are different). 

The final tuning of the absolute scale was performed at the beginning of the \lhcb operation, from data-simulation comparison using the $E/p$ method, as explained in more detail below. The Cs scans are performed regularly out of data taking, normally during \lhc technical stops. Then the \hv values of the \hcal PMTs are updated such as to restore the nominal sensitivity of the cells. Figure~\ref{hcal_css_gain_corr} (left) shows the values of the gain corrections applied to the \hcal after a typical $^{137}{\rm Cs}$ calibration run. 

\subsubsection{\hcal \pmt gain monitoring}

\begin{figure}[!ht]
\centering
\begin{minipage}{0.49\textwidth}
\includegraphics[width=1.0\textwidth]{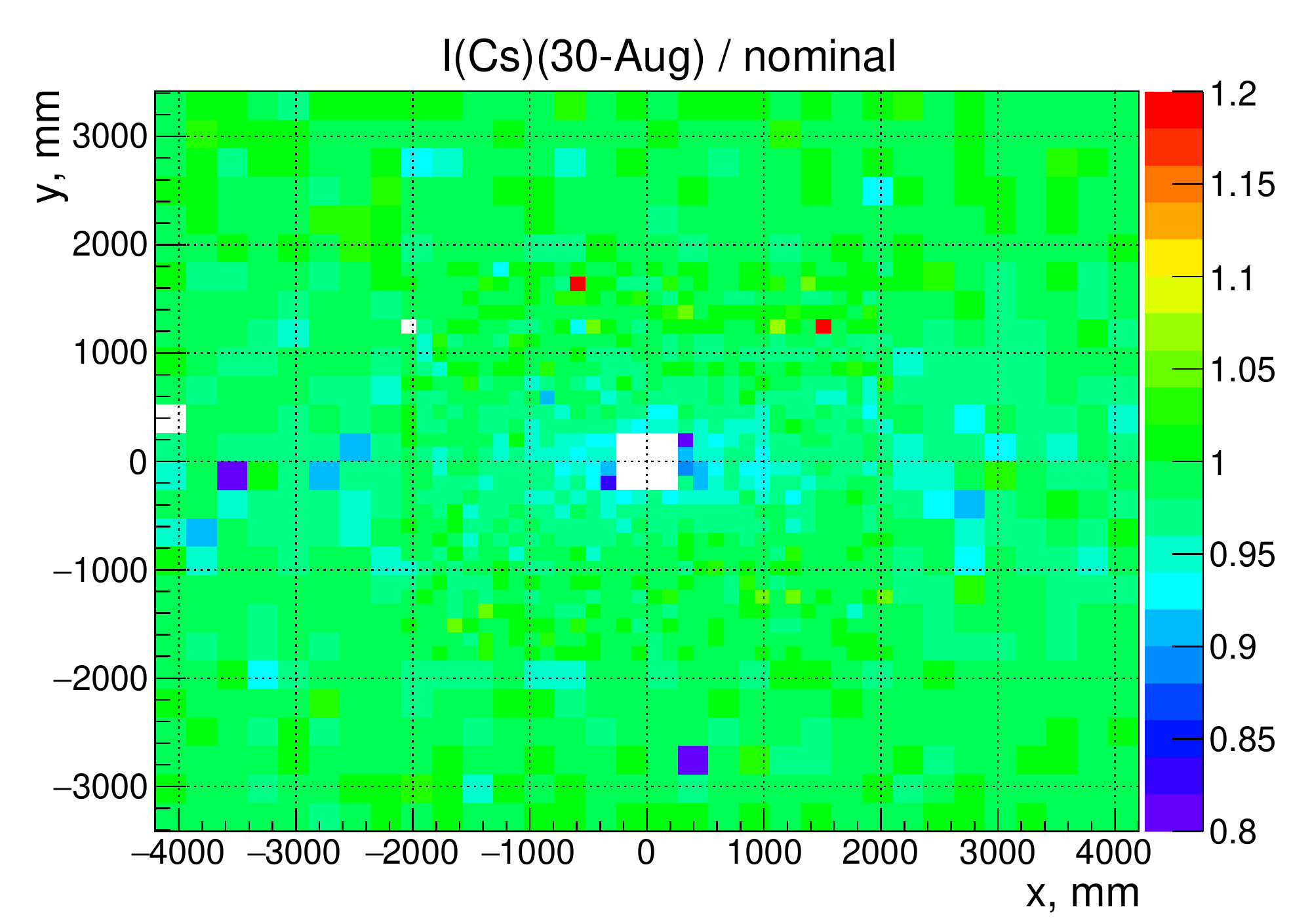}
\end{minipage}
\begin{minipage}{0.49\textwidth}
\includegraphics[width=1.0\textwidth]{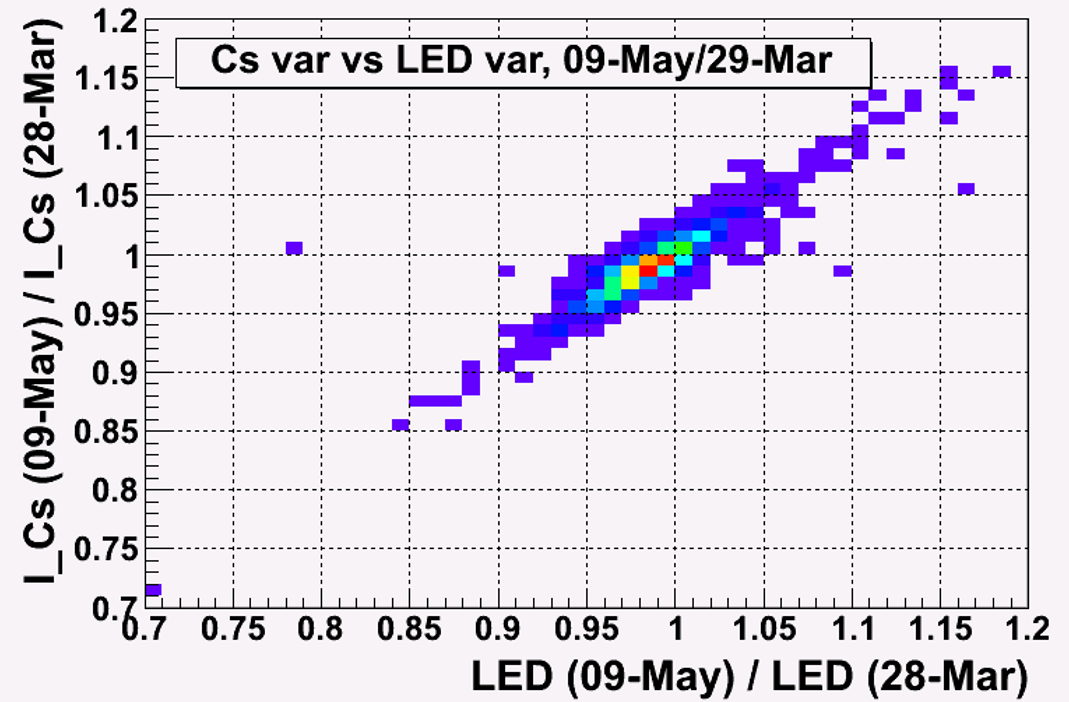}
\end{minipage}
\caption{Values of \pmt gain corrections, obtained from a typical $^{137}{\rm Cs}$ calibration run (left). Relative variations in gain obtained from two measurements with the $^{137}{\rm Cs}$ source with respect to the \led measurement variation (right).}
\label{hcal_css_gain_corr}
\end{figure}

During normal data taking, the behaviour of the \hcal is continuously monitored using the \led system. By construction, since the \led light is distributed directly to the \pmt entrance windows with clear fibres, this system can follow only the \pmt gain variations but not variations of the light yield of the \hcal cells. This mainly arises from radiation damage or ageing of the scintillator tiles and of the WLS fibres and are taken into account by the Cs source calibrations described in the previous paragraph. Figure~\ref{hcal_css_gain_corr} (right) shows the linear relation between the ratio of the \pmt anode currents obtained from two Cs calibration runs and the ratio of the \led signal amplitudes taken after these source calibrations. 

\subsubsection{\hcal ageing correction}

The \hcal information is mainly used by the \lhcb \lone trigger, for fast selection of high-\et hadrons. It is therefore important, in order to keep the \lone trigger efficiency constant, to closely follow the variations of the sensitivity of the \hcal cells and to perform corrections regularly. Out of several possible ways of correction, the chosen one is to change the \pmt \hv such as to compensate the change in sensitivity.  

\begin{figure}[!ht]
\centering
\includegraphics[width=0.7\textwidth]{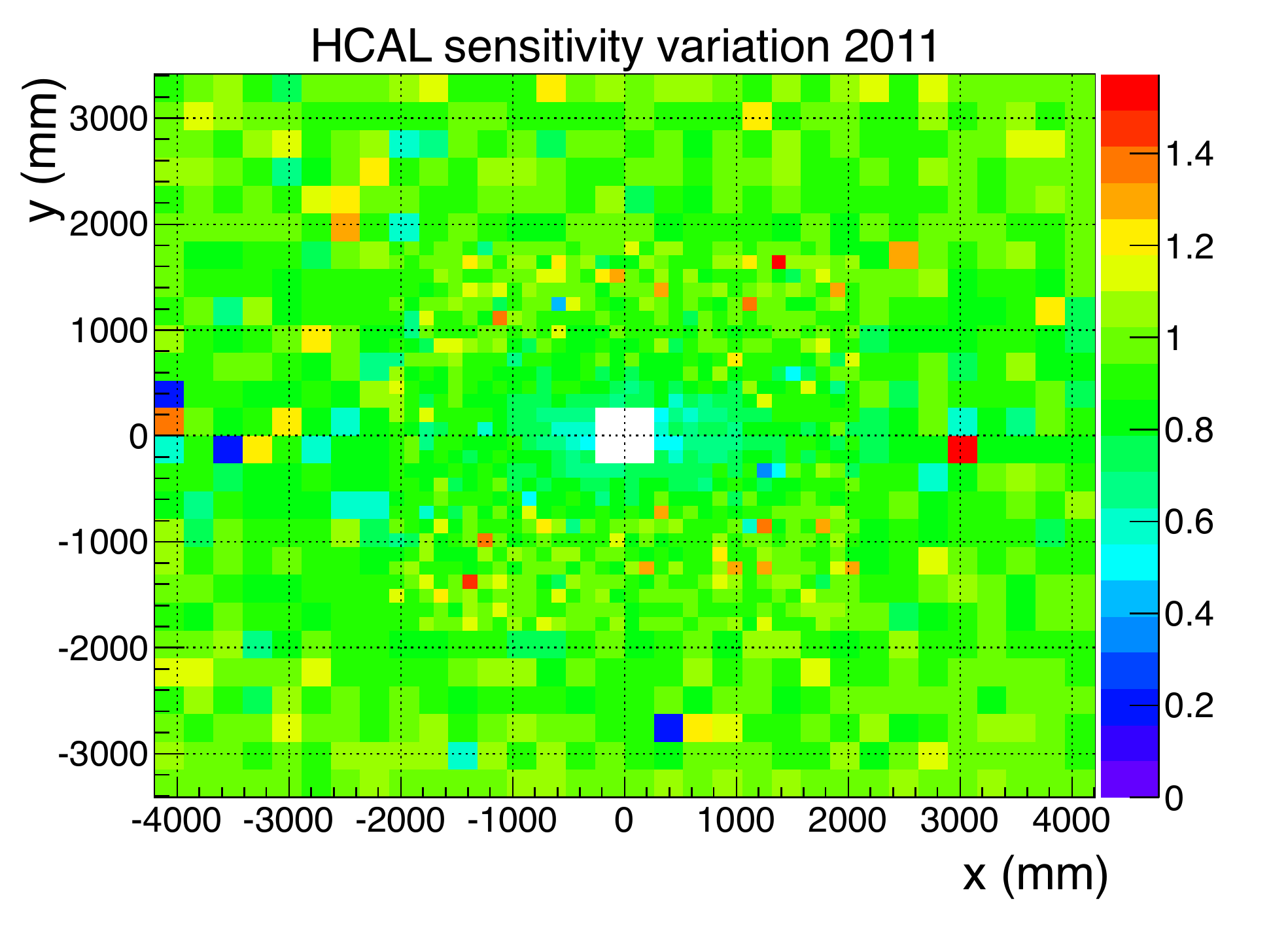}
\caption{\label{hcal_gain_var_2011} Variation of the sensitivities of the \hcal cells in 2011.}
\end{figure}

The total variation of sensitivities of each \hcal cell in 2011 which would occur without ageing corrections, is shown in Figure~\ref{hcal_gain_var_2011}. It is calculated as the inverse of all applied gain corrections in 2011. One can see a systematic 30-40\% drop in the central cells which have the highest occupancy, as well as sizeable variations in individual cells. 

\begin{figure}[!ht]
\centering
\begin{minipage}{0.49\textwidth}
\includegraphics[width=1.0\textwidth]{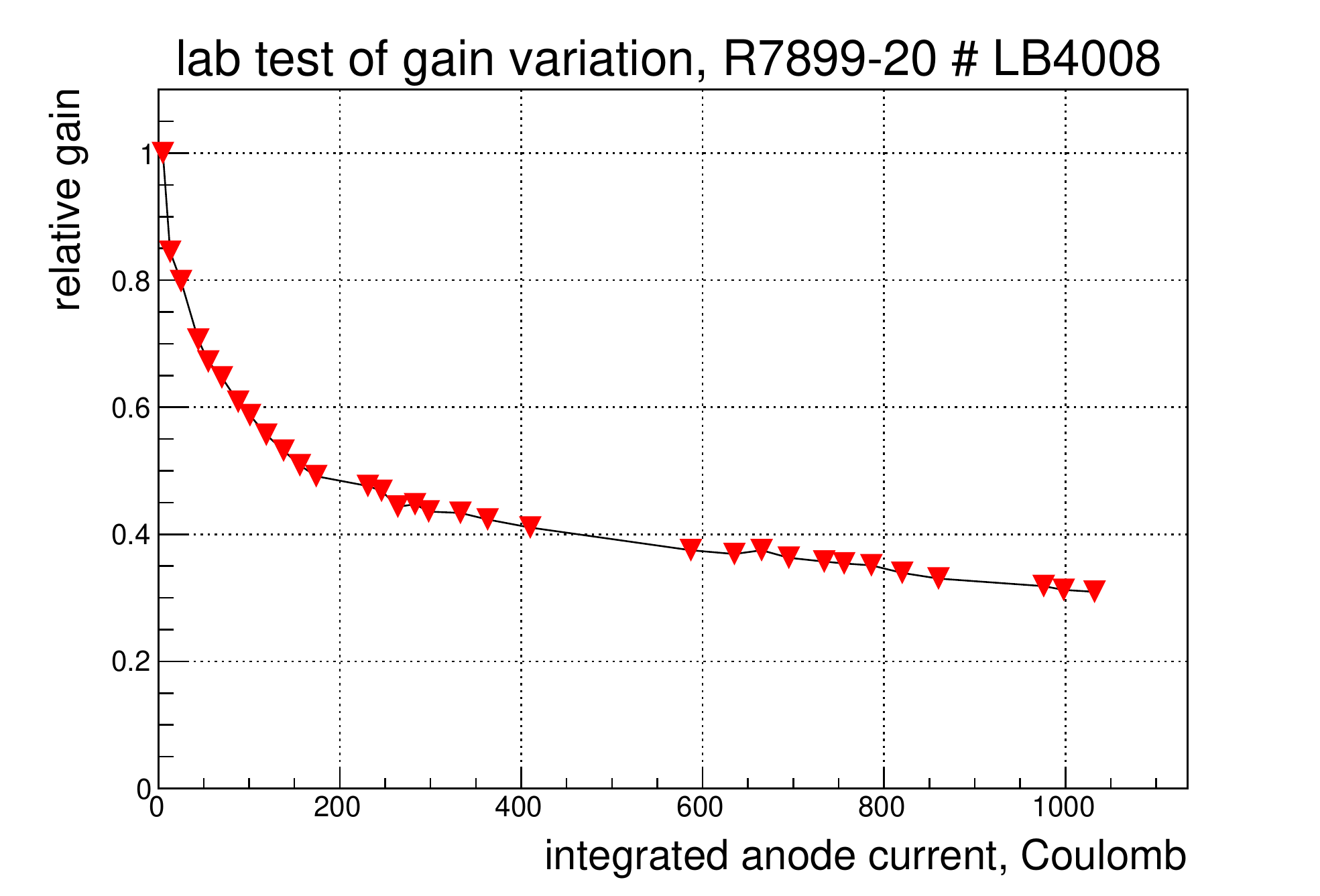}
\end{minipage}
\begin{minipage}{0.49\textwidth}
\includegraphics[width=1.0\textwidth]{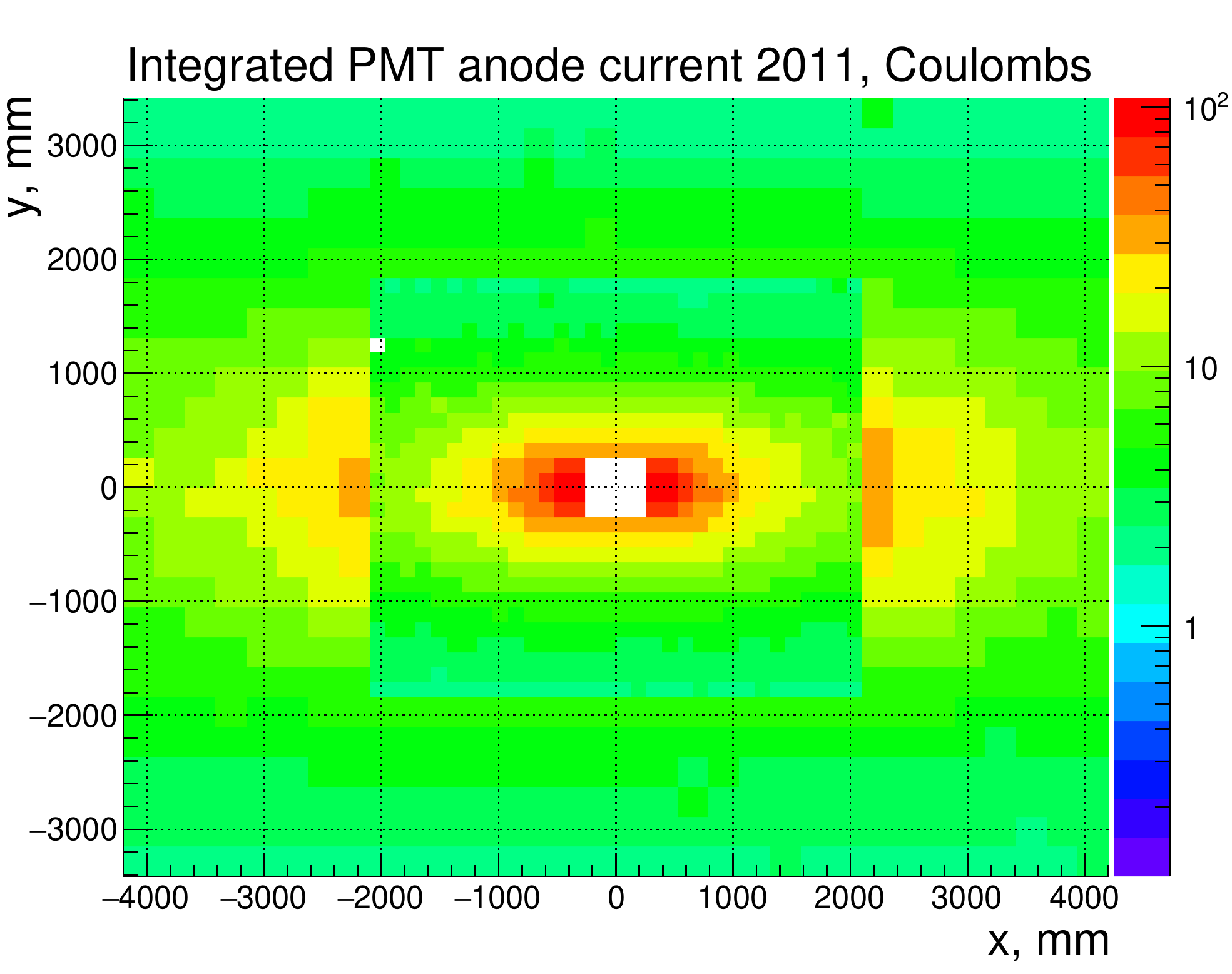}
\end{minipage}
\caption{\pmt gain degradation as a function of the integrated anode current measured at lab (left). Integrated \pmt anode currents accumulated during 2011 (right).}
\label{hcal_pmt}
\end{figure}

The sensitivity of the \hcal cells may change because of the following reasons: 
\begin{itemize}
\item \pmt gain variation, which can, in turn, be subdivided into:
\begin{itemize}
\item Degradation of the \pmt dynode system at high integrated anode current. 
      The size of the effect was measured on test bench with a single \pmt, as seen in Figure~\ref{hcal_pmt} (left). 
      The integrated anode currents in the \hcal PMTs accumulated during 2011 are shown in Figure~\ref{hcal_pmt} (right). 
      One can conclude that the 30-40\% drop in the centre seen in Figure~\ref{hcal_gain_var_2011} is compatible with the 
      degradation of the \pmt dynode system. 
\item Instability, intrinsic to \pmt and not related to anode current may occur at a very low rate.
\end{itemize}
\item Light yield degradation of \hcal cells arising from radiation damage of the scintillator tiles and of the WLS fibres. 
\end{itemize}

\begin{figure}[!ht]
\centering
\includegraphics[width=0.7\textwidth]{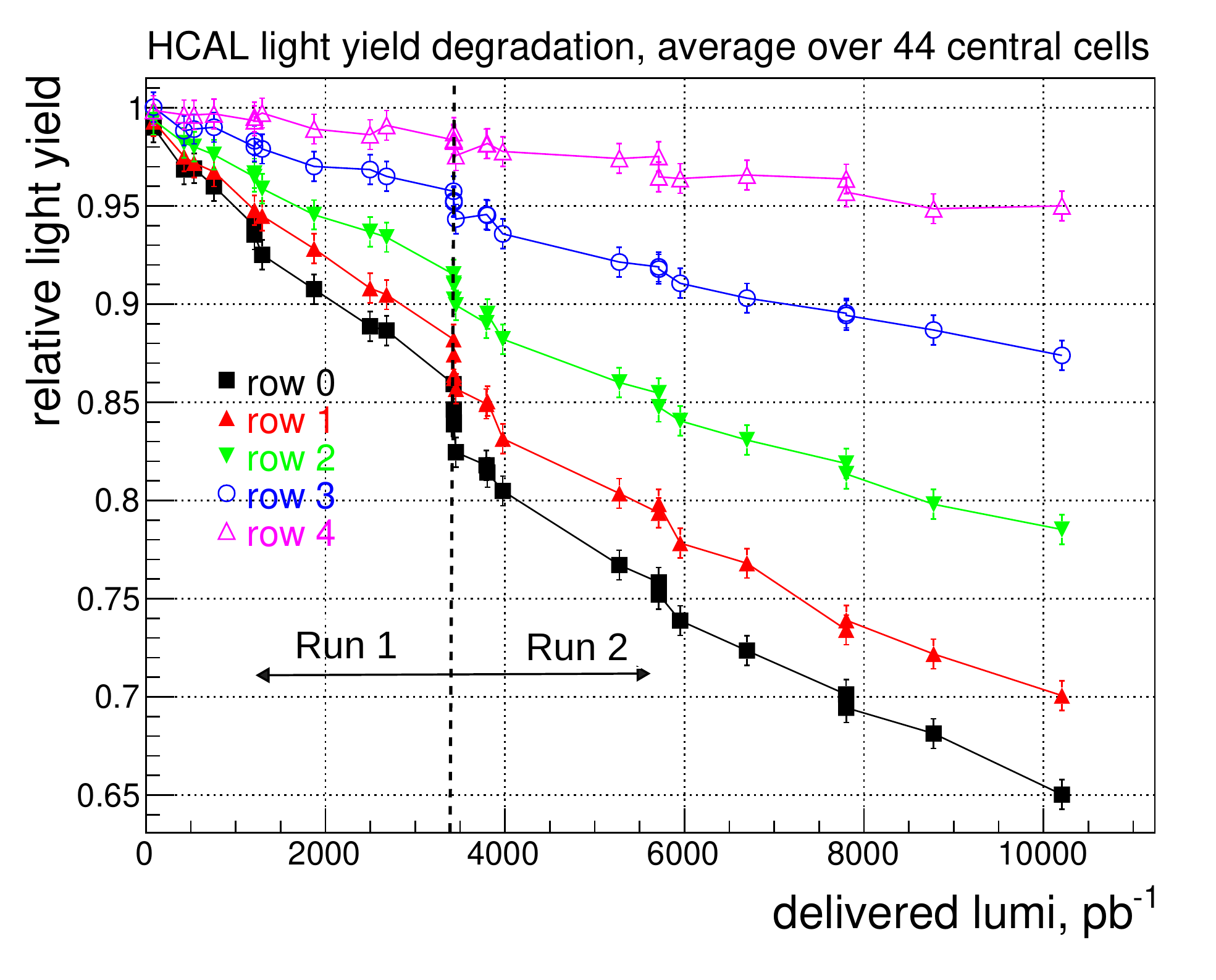}
\caption{Light yield $Y^r_i$ averaged over the 44 \hcal most central cells as a function of the delivered luminosity.}
\label{hcal_light_center}
\end{figure}

In the \hcal, the main radiation dose is deposited in the two front rows (0 and 1) of tiles that can be seen in Figure~\ref{hcal_light_center} since this corresponds to the position of the maximum development of the hadronic showers. The radiation dose received by the two rear rows (4 and 5) is much lower. 

Considering that the fifth row does not suffer ageing, one can determine the light yield degradation of each \hcal cell from Cs source calibration data, in a way independent from the study of \pmt gain variations. The light yield degradation $D_i^r$ in a tile row $r$ belonging to a cell $i$, at the time of a $^{137}{\rm Cs}$ source calibration run performed after a delivered luminosity ${\cal L}$, can be determined as the change in the ratio of the anode currents measured at passages of the Cs source in rows $r$ and 5, $I_i^r/I_i^5$, with respect to the reference value taken before the \lhc start,
\begin{equation}
D_i^r=\frac{Y^r_i({\cal L})}{Y^r_i(0)} = \frac{(I^r_i/I^5_i)({\cal L})}{(I^r_i/I^5_i)(0)},
\end{equation} 
where $Y^r_i({\cal L})$ is the light yield as a function of the integrated luminosity.

The light yield degradation from 2011 (Run 1)  to 2018 (Run 2) for each individual \hcal cell, averaged over the six tile rows, is shown in Figure~\ref{hcal_light_center}. From the comparison of Figure~\ref{hcal_light_center} and Figure~\ref{hcal_gain_var_2011}, one can conclude that the light yield degradation in scintillator tiles and WLS fibres amounts to 5-35\% of the overall sensitivity variation. Therefore, a two-step strategy is followed to compensate for the detector ageing.
\begin{itemize}
\item \led based corrections, which take into account only \pmt gain drift effects, are performed automatically after every long fill in 2018.
\item Corrections based on Cs calibration scans performed during \lhc technical stops, are computed every 2 month. 
They take into account residual  \pmt ageing and detector effects.
\end{itemize}

\subsubsection{\hcal offline calibration}

The absolute scale of the \hcal calibration is tuned using a $E/p$-based method. This method allows also to account for the spread of sensitivities of the FEB channels, which is not the case with the Cs calibration. 

Similarly to the \ecal, the $E/p$ method for the \hcal consists in comparing the momentum of a charged hadron measured with \lhcb tracking detectors and its energy deposition in the \hcal. The energy deposition of a hadron is determined as the energy of the \hcal cluster build around the cell crossed by the projection of the particle's trajectory to the \hcal plane.

\begin{figure}[!ht]
\centering
\includegraphics[width=1.0\textwidth]{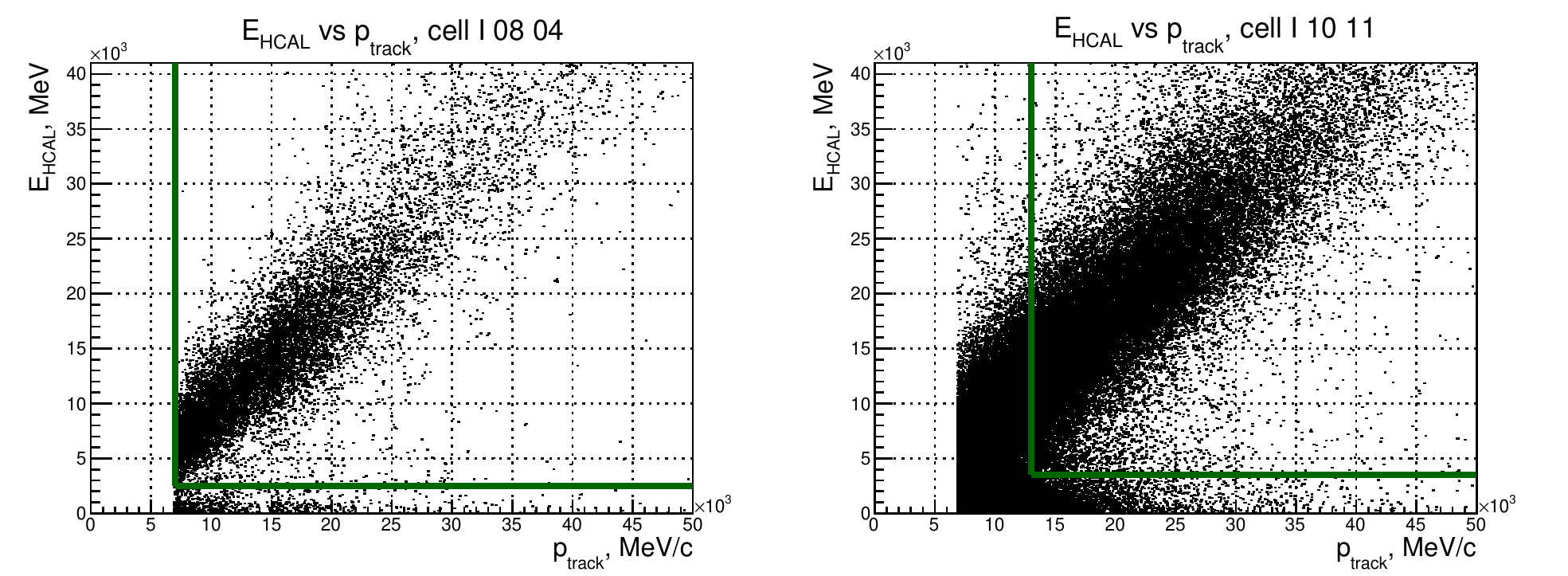}
\caption{ $E_{\rm \hcal}$ and $p_{\rm track}$ distributions for track selected in different cells, for the \hcal calibration based on $E/p$. The lines show the cuts applied for the $E/p$ calculations.}
\label{hcal_e_vs_p_cell}
\end{figure}

Tracks are selected out of all reconstructed tracks in an event according to the following criteria:
\begin{itemize}
\item The track deposits energy in the \hcal: $E_{\rm \hcal}>0$.
\item The \hcal cells associated with the track are not associated with any other track.
\item The track is reconstructed with hits in all tracking devices and the first hit's position is within $-200<z<400\mm$.
\item The track momentum is more than 7\gevc.
\end{itemize}
Figure~\ref{hcal_e_vs_p_cell} shows the $E_{\rm \hcal}$ and $p_{\rm track}$ distributions for the tracks selected with the selection described above in two different \hcal cells. For the $E/p$ analysis, the hadrons must meet the following additional criteria: 
\begin{itemize}
\item The particle is not identified as a muon in the muon system. Muons do not produce hadronic showers and their energy deposition in the \hcal is almost independent of their momentum.
\item The particle is not identified as a proton (anti-proton) in the RICH detectors. Protons and anti-protons with equal momenta give different energy depositions in the \hcal. This can give an asymmetric bias in the selection.
\item $E_{\rm \ecal}<0.8\gev$ and $E_{\rm \presh}<20\mev$, in order to reject hadronic showers starting before the \hcal.
\item $E_{\rm \hcal}>2.5\gev$ for most of the cells (the threshold is somewhat higher for the innermost ones).
\end{itemize}
The average  $E/p$ obtained from simulation studies~\cite{LHCb-PROC-2011-006} can be used for the offline \hcal calibration. The calibration procedure is iterative and 5 or 6 iterations are required to converge. 

\begin{figure}[!ht]
\centering
\begin{minipage}{0.49\textwidth}
\includegraphics[width=1.0\textwidth]{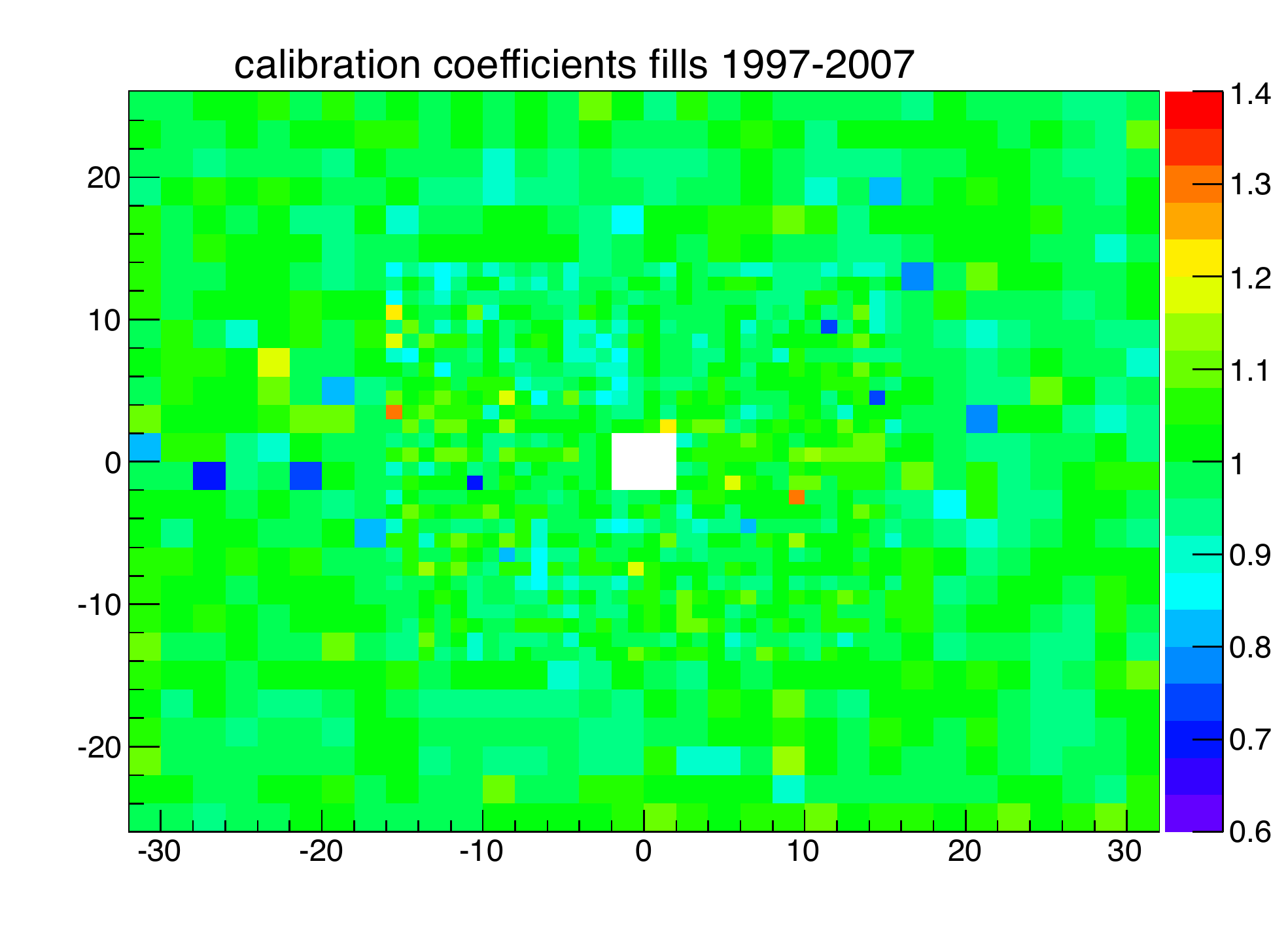}
\end{minipage}
\begin{minipage}{0.49\textwidth}
\includegraphics[width=1.0\textwidth]{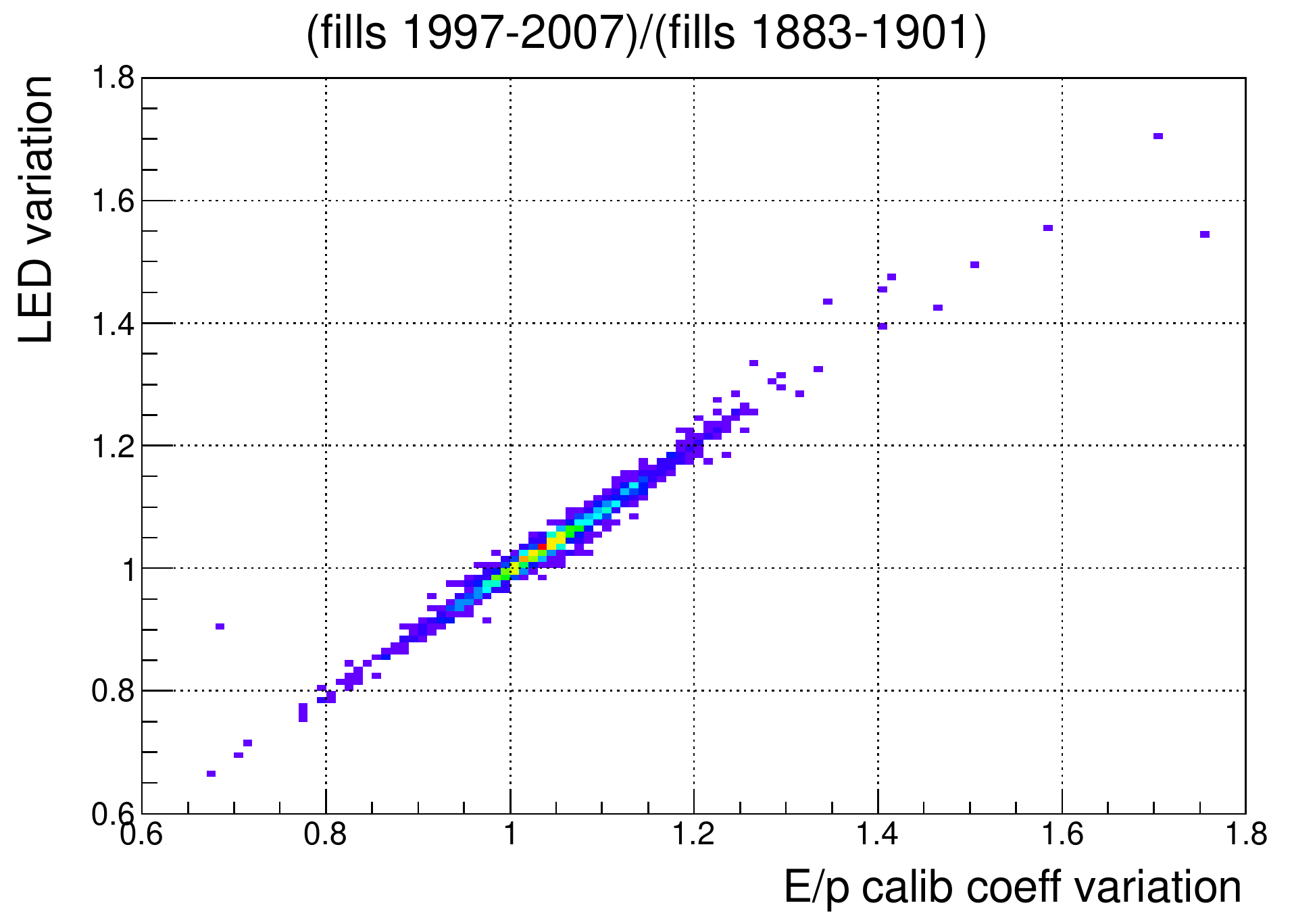}
\end{minipage}
\caption{Averaged $E_{\rm \hcal}/p_{\rm track}$ values in the \hcal cells, right after a $^{137}$Cs calibration (left). Ratio between $E/p$-based calibration coefficients in a 5 week time period and the corresponding ratio of the \led corrections for the same period (right).}
\label{hcal_e_over_p}
\end{figure}

The calibration coefficients obtained are shown in Figure~\ref{hcal_e_over_p} (left). The cells with \pmt having stability problems are clearly seen. Figure~\ref{hcal_e_over_p} (right) shows the correlation between the ratio of $E/p$-based calibration coefficients for two different periods (one week long intervals, with 5 weeks in between) and the ratio of \led based corrections for the same periods. This validates the use of the \led corrections for short time scales.

Another important characteristics of the \hcal is its response to muons. Muons do not produce a shower; the energy deposition of a muon in the \hcal represents ionisation losses of a charged particle in the scintillator tiles; it is expected to follow a Landau distribution and to be almost independent of the muon energy. However, because of the orientation of the \hcal structure along the beam axis, the response to muons is not uniform over the calorimeter surface. 

\begin{figure}[!ht]
\centering
\begin{minipage}{0.49\textwidth}
\includegraphics[width=1.0\textwidth]{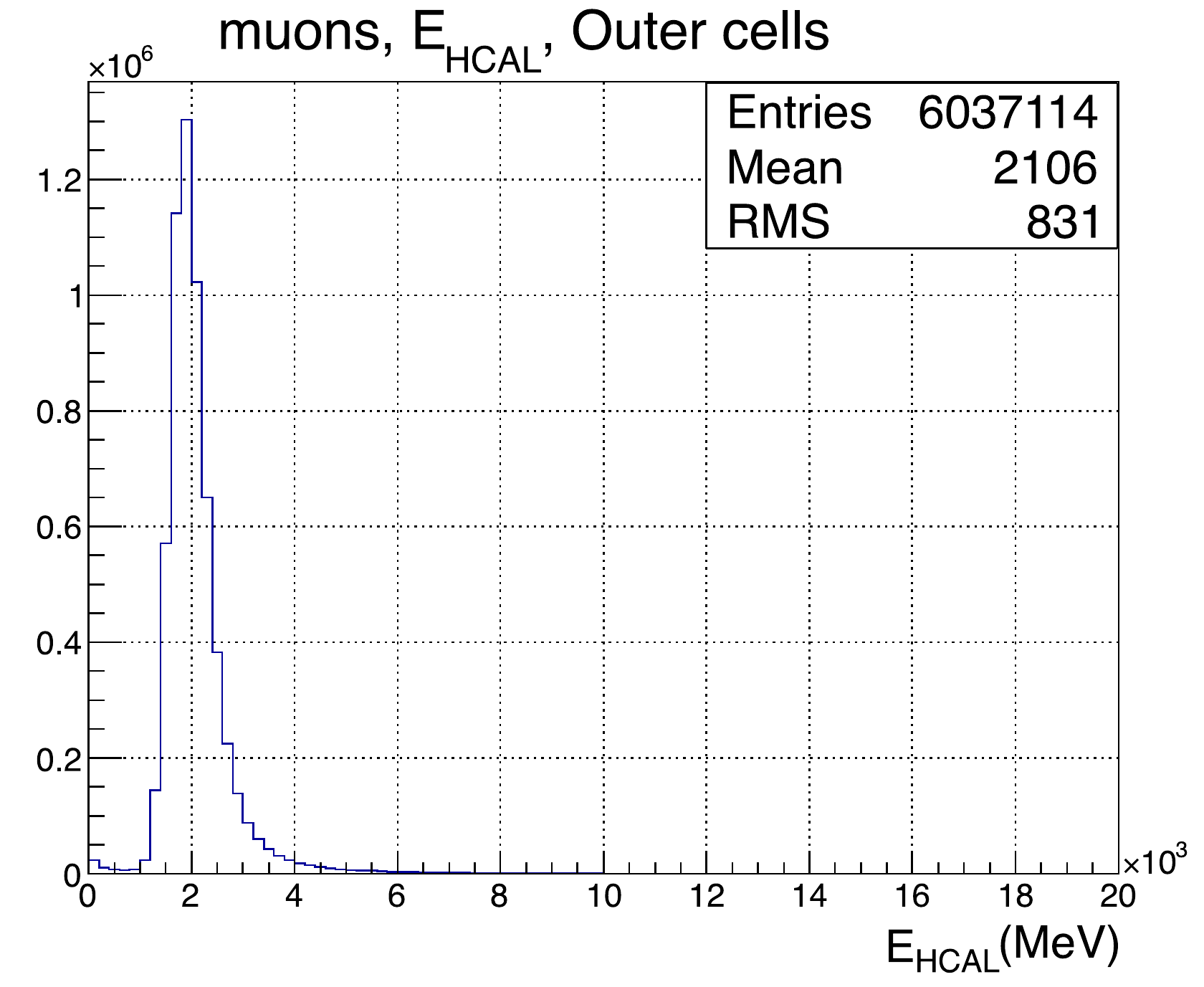}
\end{minipage}
\begin{minipage}{0.49\textwidth}
\includegraphics[width=1.0\textwidth]{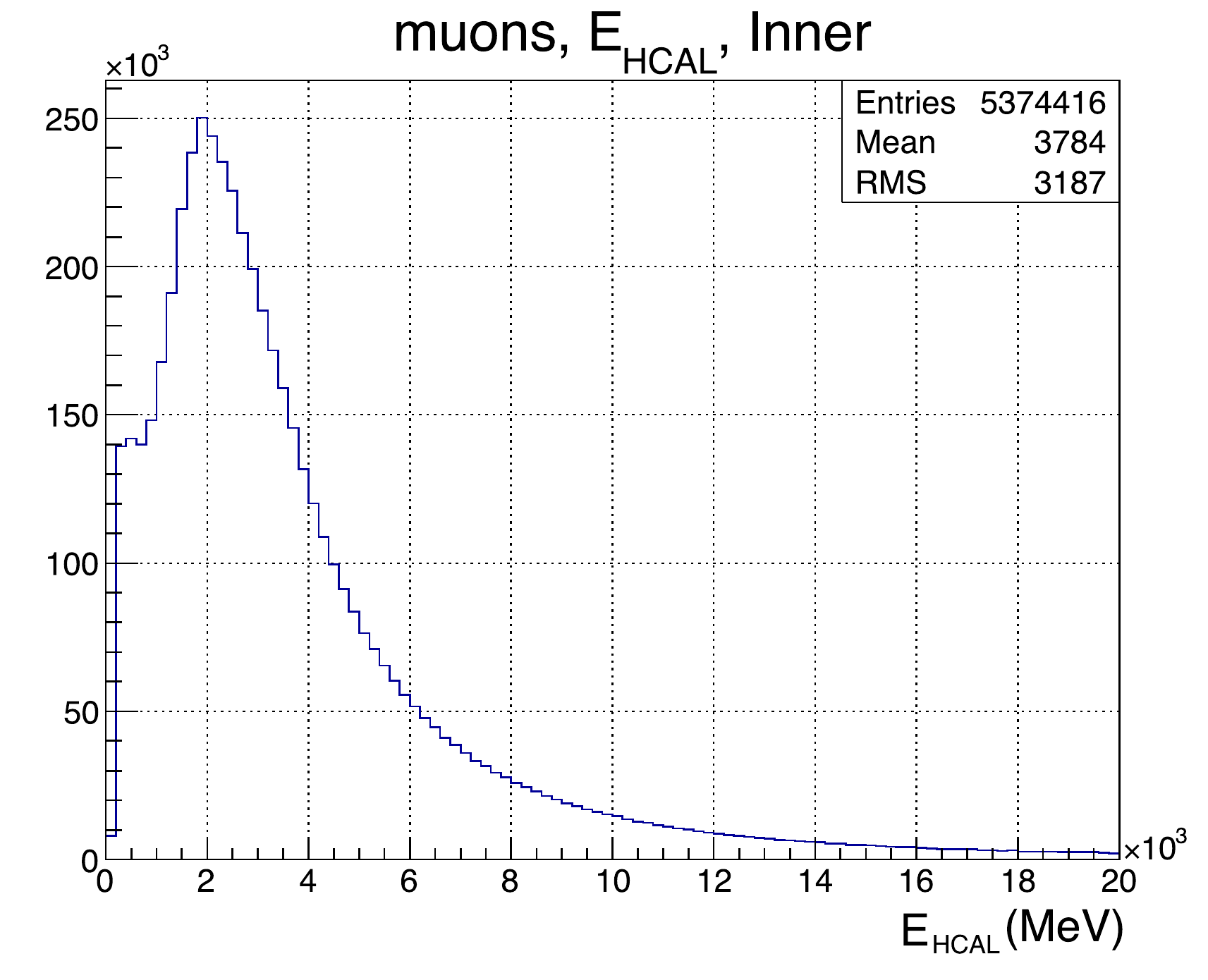}
\end{minipage}
\caption{Distribution of the muon energy deposition in the outer cells (left) and in the most central cells (right) of the \hcal.}
\label{hcal_muon}
\end{figure}

In the inner part of the \hcal, muon tracks cross the calorimeter almost parallel to the iron and scintillator planes, so that the path length (and therefore the energy deposition) in the scintillator has large fluctuations. In the outer part of the \hcal, muons cross the structure at larger angles, and the response is much less fluctuating. The \hcal signal distribution for muons identified with the muon detectors, crossing the outer part of the \hcal, is shown in Figure~\ref{hcal_muon} (left). The similar distribution for the 122 most central cells is shown in Figure~\ref{hcal_muon} (right). 

Unlike the \ecal, no major change in monitoring and calibration methods occurred for Run 2.

\section{Performances of the \lhcb Calorimeters}

\subsection{Neutral particle reconstruction}

Energy deposits in \ecal cells are clusterised applying a $3\times 3$ cell pattern around local maxima, i.e the calorimeter cells with a larger energy deposition compared to their adjacent cells. As a direct consequence of these formal definitions, the centres of the reconstructed clusters are always separated by at least one cell. In case one cell is shared between several reconstructed clusters, the energy of the cell is redistributed between the clusters proportionally to the total cluster energy. The process is iterative and converges rapidly due to the relatively small ratio between the Moliere radius and the cell size\footnote{The ratios between the Moliere radius, 3.5 cm, and the cell side length are approximately 0.87, 0.58 and 0.29 for inner, middle and outer regions respectively.}. After the redistribution of energy of shared cells, the energy-weighted cluster moments up to the order 2 are evaluated to provide the following hypothesis-independent cluster parameters:
\begin{itemize}
\item energy: $\epsilon_{\rm cl} = \Sigma_i \epsilon_i$,
\item transverse barycentre: $\vec p_{\rm cl} = \frac{1}{\epsilon_{\rm cl}} (\Sigma_i \epsilon_i x_i, \Sigma_i \epsilon_i y_i) = (x_b,y_b)$,
\item transverse dispersion: ${\cal S}_{\rm cl} = \frac{1}{\epsilon_{\rm cl}} \begin{pmatrix}
\Sigma_i \epsilon_i (x_i - x_b)^2 & \Sigma_i\epsilon_i (x_i - x_b)(y_i -  y_b) \\
\Sigma_i\epsilon_i (x_i - x_b)(y_i -  y_b) & \Sigma_i \epsilon_i (y_i - y_b)^2
\end{pmatrix}$,
\end{itemize}
where $\epsilon_i$ is the local energy deposit measured in cell $i$ of the cluster and $x_i$ and $y_i$ give the position of the center of the cell.

The selection of neutral clusters is  performed using a matching technique with reconstructed tracks. All reconstructed tracks in the event are extrapolated to the calorimeter reference plane and then a one-to-one matching with the reconstructed clusters is performed. The matching $\chi^2_{\rm 2D}$ between a cluster and a track is defined as
\begin{equation}\label{eq:chideuxecal}
\chi^2_{\rm 2D}(\vec p) = (\vec p_{\rm tr} - \vec p)^T {\cal C}_{\rm tr}^{-1}(\vec p_{\rm tr} - \vec p)+(\vec p_{\rm cl} - \vec p)^T {\cal S}_{\rm cl}^{-1}(\vec p_{\rm cl} - \vec p),
\end{equation}
where $\vec p_{\rm tr}$ is the extrapolated track impact 2D-point to the calorimeter reference plane, ${\cal C}_{\rm tr}$ is the $2 \times 2$ covariance matrix of $\vec p_{\rm tr}$, $\vec p_{\rm cl}$ is the cluster barycentre 2D position and ${\cal S}_{\rm cl}$, the $2 \times 2$ cluster covariance matrix (transverse dispersion as defined above). The $\chi^2_{\rm 2D}$ is minimised with respect to $\vec p$ and the minimal value of $\chi^2_{\rm 2D}$ is used to select neutral clusters. The clusters with a minimal value of the $\chi^2_{\rm 2D}$ estimator in excess of 4 are selected as neutral clusters or photon candidates. This cut rejects the clusters due to electrons and significantly suppresses clusters due to other charged particles keeping a high efficiency for clusters due to photons.

\begin{figure}[!ht]
\centering
\includegraphics[width=0.7\textwidth]{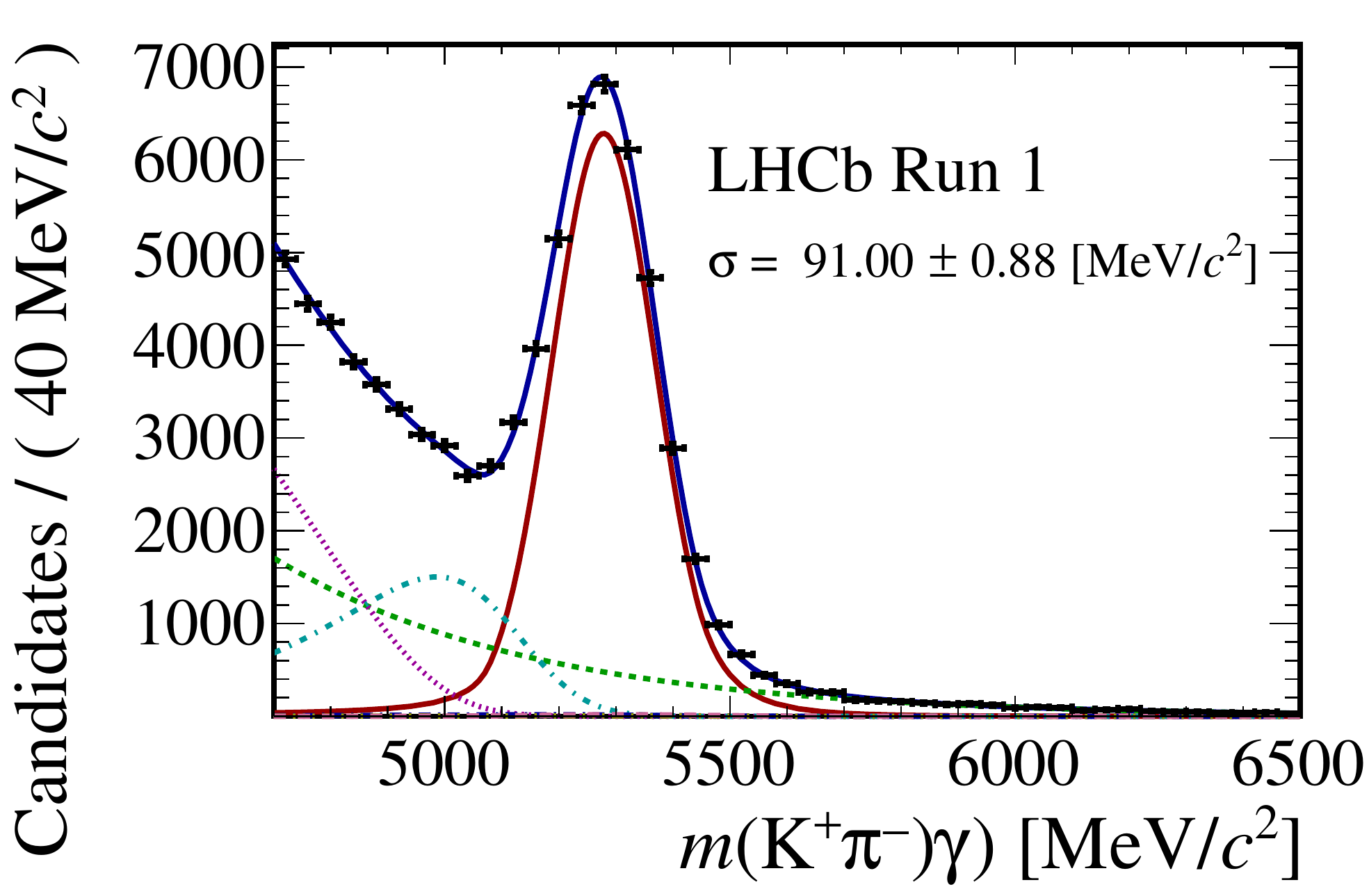}
\caption{Invariant mass distribution of $\Bz\to \Kstarz(\Kp \pim)\gamma$ candidates in Run 1. The blue curve corresponds to the mass fit. 
The $\Kstarz\gamma$ signal component of the fit function (red line) and the various background contaminations are shown.}
\label{perf_kstargamma_run1}
\end{figure}

The photon energy, $E_c$, is evaluated from the total \ecal cluster energy, $\epsilon_{\rm cl}$, corrected for leakage according to the following relation
\begin{equation}\label{eq:totalecal}
E_c = \alpha \times \epsilon_{\rm cl} + \beta \times \epsilon_{\rm \presh} + \gamma,
\end{equation}
where $\epsilon_{\rm \presh}$ is the \presh cluster energy, and the parameters $\alpha$ and $\beta$ account for energy leakages in the \ecal and \presh, respectively~\cite{Deschamps:691634,Terrier:691743}. The parameter $\gamma$ was introduced to make the $\piz$ mass more stable. The values of the parameters $\alpha$ and $\beta$ depend on the position of the cluster on the calorimeter surface. The photon direction is derived from the position of the barycentre of the photon shower. The transversal position $(x_c,y_c)$ of the barycentre, perpendicular to the beam axis, is obtained from the energy-weighted barycentre position $(x_b,y_b)$, corrected for the non linear transversal profile of the shower shape. The parametrisation of the longitudinal barycentre position depends on both the photon energy and its incidence angle on the calorimeter plane. For high mass radiative decays, some of the corrections described above might not be perfectly appropriate since these corrections have been obtained from low mass particles ($\piz$). Therefore a post-calibration of the energy of the photons is performed in each region of the calorimeters to correct for residual bias. The performance of high energy photon reconstruction, after corrections and post-calibration, is illustrated by the reconstructed $\Bz\to \Kstarz\gamma$ mass distribution shown in Figure~\ref{perf_kstargamma_run1}. The mass resolution obtained for this radiative decay, around 90\mevcc, is dominated by the \ecal energy resolution~\cite{LHCb-DP-2019-001}.  

\begin{figure}[!ht]
\centering
\includegraphics[width=0.7\textwidth]{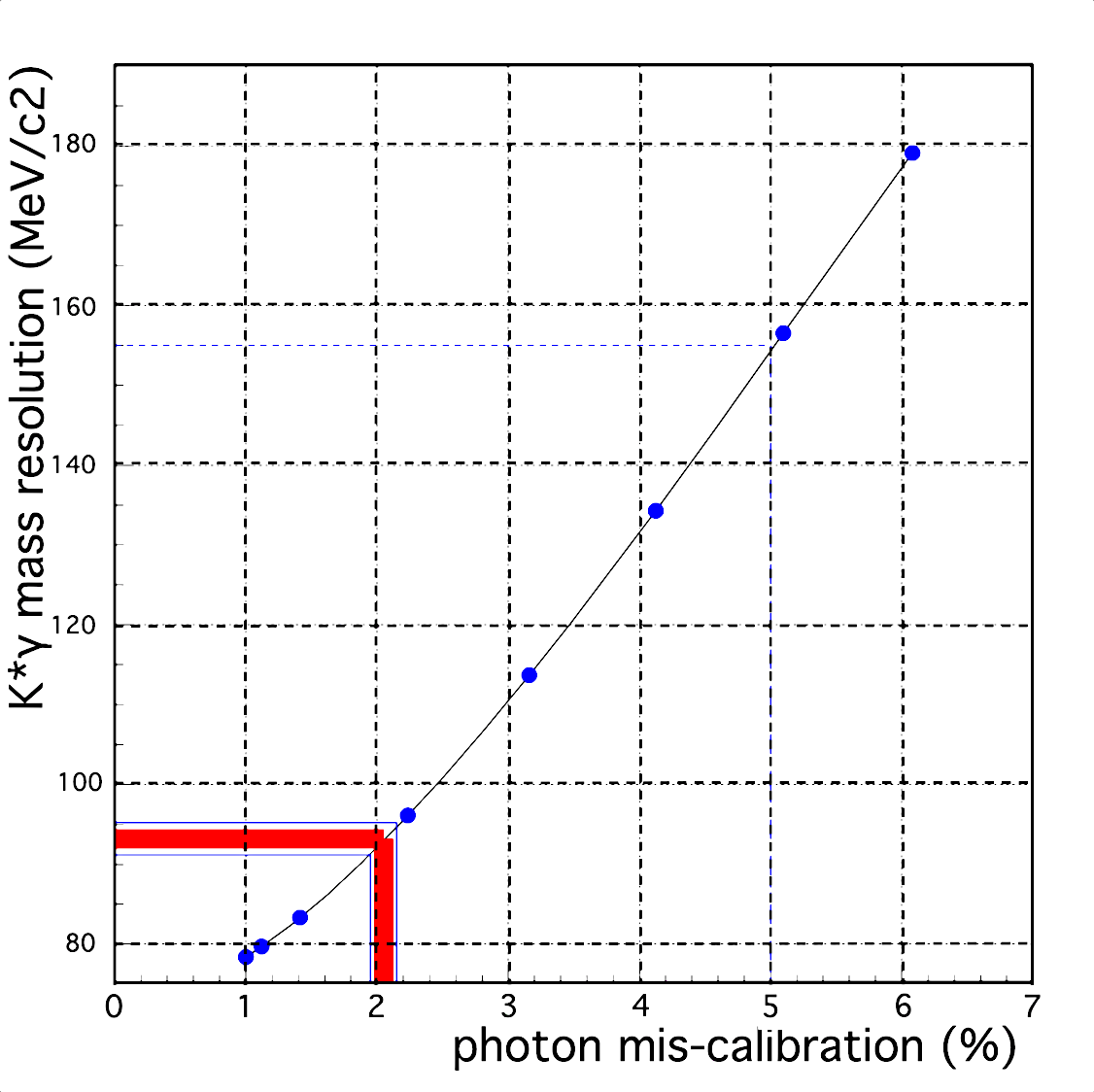}
\caption{$B\to K^* \gamma$ mass resolution in terms of photon mis-calibration from simulation for high energy photons. The comparison with the actual value suggests an intrinsic resolution in the photon energy of $2$\%.}
\label{perf_kstargam_photon_miscal}
\end{figure}

In order to have an estimation of the intrinsic resolution from data for high energy photons, one may compare the resolution of the peak of the decay $\Bz\to \Kstarz \gamma$ on both data and simulation samples. Both data and simulation meson candidates are combined with photon candidates whose energy is larger than 2.6\gev. The simulation uses a 1.2 \adc counts noise from the electronics in \et and a stochastic term of $10$\%, which is compatible with the data. The constant gain term is 1\%. To reproduce the resolution of the data with the simulation samples, the intrinsic resolution of the photon is around $2$\%, as shown in Figure~\ref{perf_kstargam_photon_miscal}. Also a study measuring the relative yields of $\Bu \to \jpsi\Kp$ and $\Bu \to \jpsi\Kstarp$ shows that corrections to the $\piz$ and $\gamma$ reconstruction efficiencies are compatible with one~\cite{Govorkova:2015vqa}.

\begin{figure}[!ht]
\centering
\begin{minipage}{0.49\textwidth}
\includegraphics[width = 1.0\textwidth]{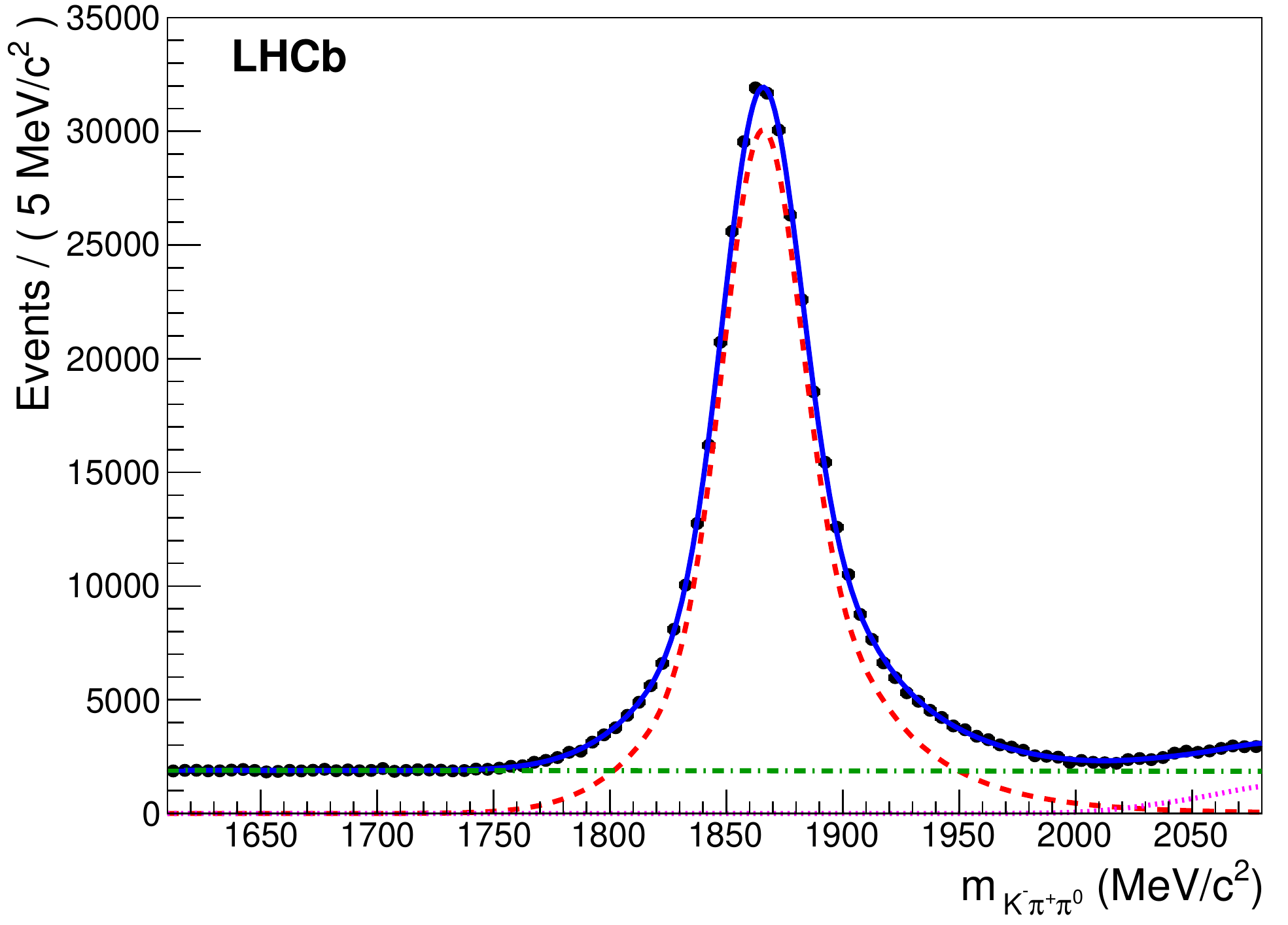}
\end{minipage}
\begin{minipage}{0.49\textwidth}
\includegraphics[width = 1.0\textwidth]{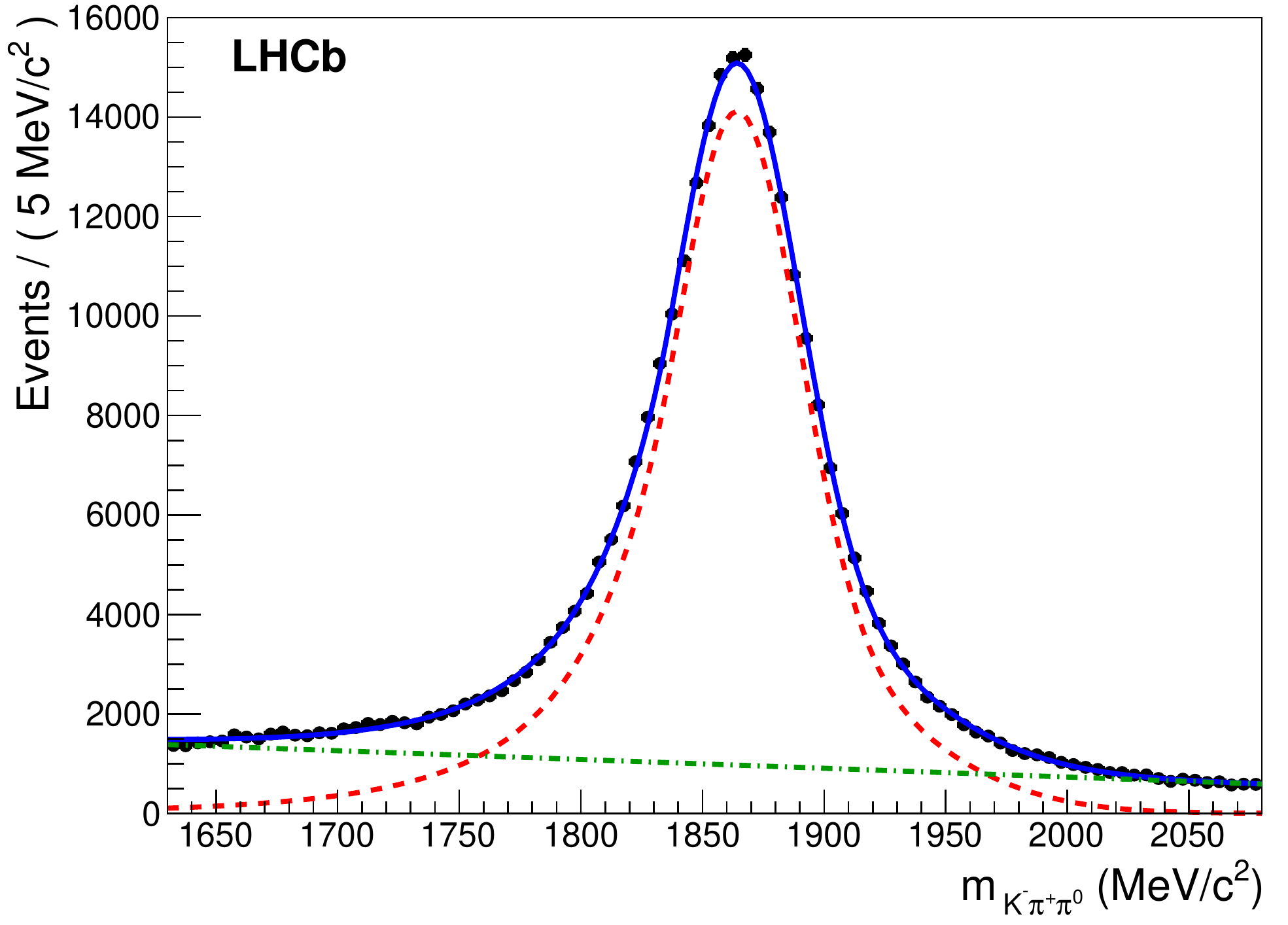}
\end{minipage}
\caption{Mass distribution of the reconstructed $\Dz\to \Km\pip \piz$ candidates with resolved \piz (left) and merged \piz (right) obtained from the 2011 data sample. The blue curve corresponds to a fit. The signal component of the fit function (red dashed line) and the background (green dash-dotted line) contributions are shown.}
\label{perf_d0_resolved_merged}
\end{figure}

Neutral pions with low transverse energy are mostly reconstructed as a resolved pair of well separated photons from their decays. A mass resolution of 8\mevcc is obtained for such neutral pions. On the opposite, a large fraction of pairs of photons coming from the decay of high energy \piz cannot be separated as a pair of clusters within the \ecal granularity. Such a configuration, hereafter referred as "merged", essentially appears for \piz with transverse momentum above 2\gevc. To reconstruct such merged \piz, a procedure has been designed to disentangle a potential pair of photons merged into the single clusters. The algorithm consists in splitting each single \ecal clusters into two interleaved $3\times 3$ sub-clusters built around the two highest deposits of the original cluster. The energy of the common cells is then shared among the two sub-clusters according to an iterative procedure based on the expected transverse shape of photon showers. The sharing of the energy depends on the barycentre position of each sub-cluster which is a function of the energy sharing. The procedure is repeated over a few iterations and is quickly converging. A mass resolution of 20\mevcc is obtained for such merged configuration. The performance of neutral pions reconstruction is illustrated in Figure~\ref{perf_d0_resolved_merged} that displays the reconstructed $\Dz\to \Km\pip\piz$ mass  distribution for resolved \piz (left) and merged \piz (right).

\subsection{Electron identification}

The electron identification from the calorimeter system is based on the information collected from the \presh, \ecal and \hcal. The procedure combines these different sources of information and builds signal and background reference samples. Samples of electrons/positrons are obtained from converted photon in reconstructed events. In order to avoid any influence from the calorimeter trigger, only events triggered by the muon detectors are used. These samples are the "signal" samples. Samples of kaons and pions are obtained from $\Dz \to \Km \pip$ decays. These samples are the "background" samples.

For each sub-detector (\presh, \ecal, \hcal), an estimator is built in the form of a signal/background log-likelihood as follows: a 2 dimensional correlation histogram is filled with the track momentum and a variable coming from the sub-detector measurement (energy for example); one histogram is filled with signal data, another histogram is filled with background data; the content of each slice in momentum is normalised to 1 and the logarithm of the content is taken; the difference between the signal and background of such histograms is therefore a 2-dimensional difference of log-likelihoods of signal and background hypotheses, labelled as $\Delta \log {\mathcal L}$. These reference histograms have been produced using the first $340\,{\rm pb}^{-1}$ recorded in 2011.

\begin{figure}[!ht]
\centering
\begin{minipage}{0.48\textwidth}
\includegraphics[width=1.0\textwidth]{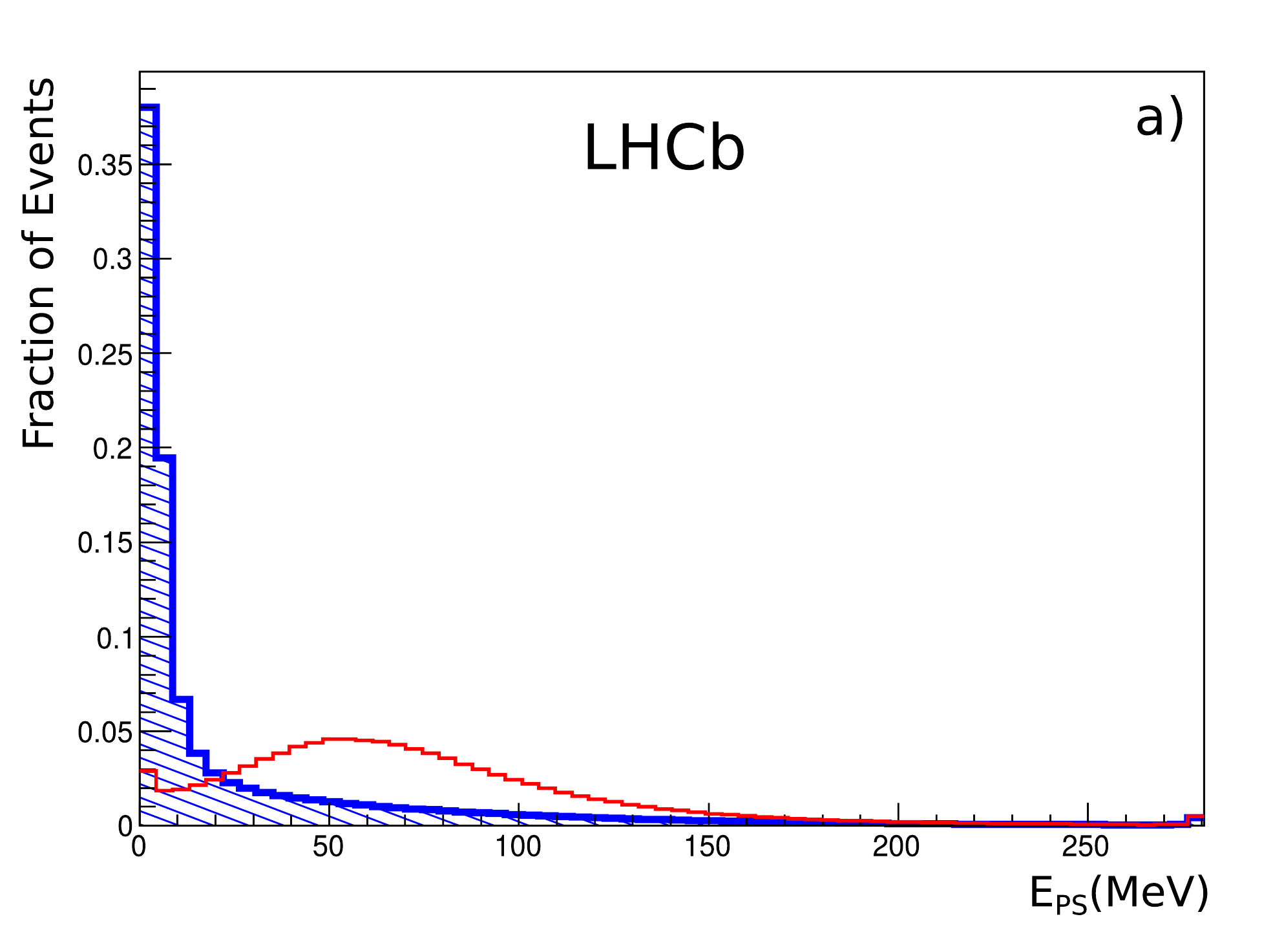}
\end{minipage}
\begin{minipage}{0.50\textwidth}
\includegraphics[width=1.0\textwidth]{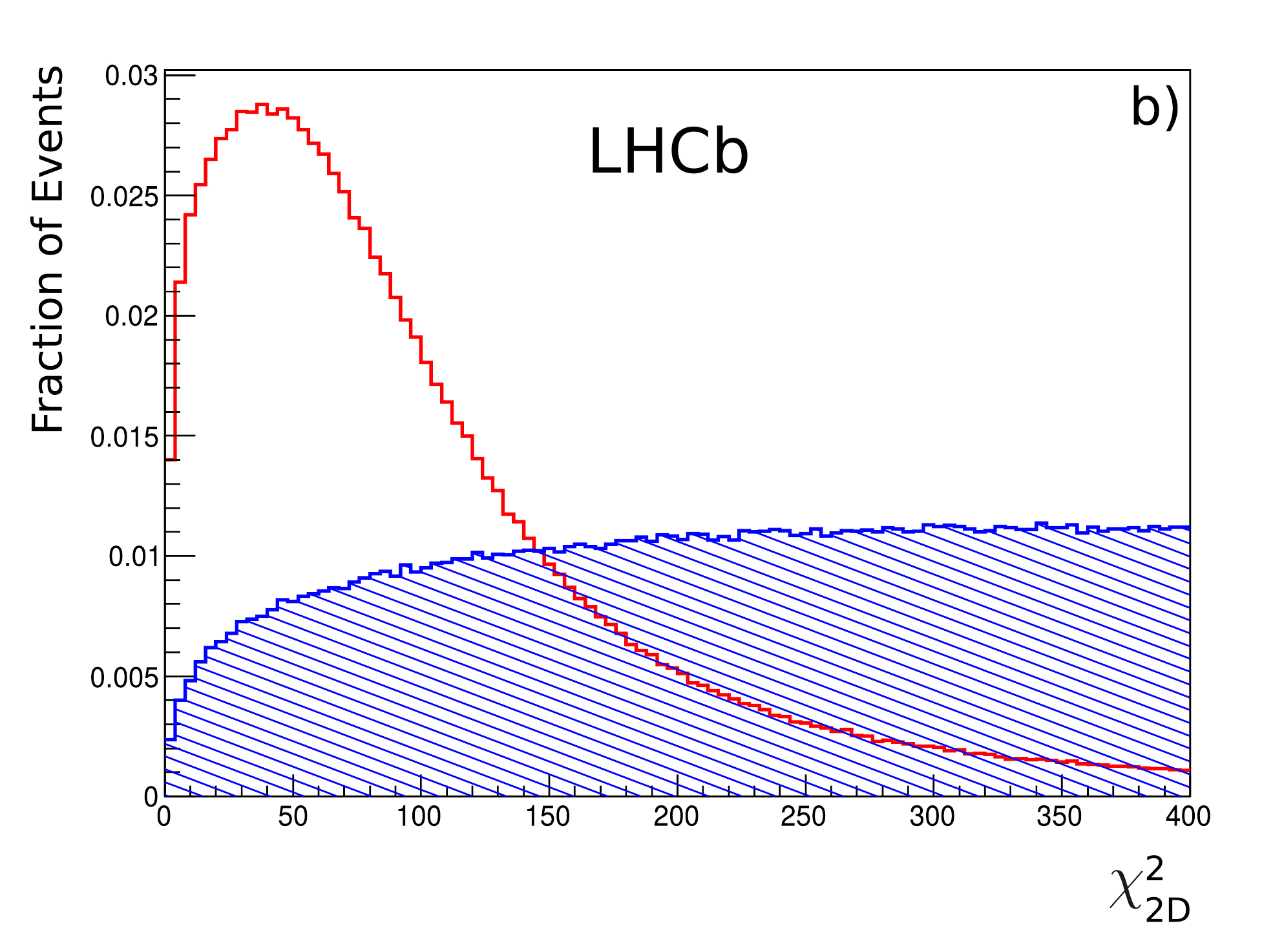}
\end{minipage} 
\includegraphics[width = 0.49\textwidth]{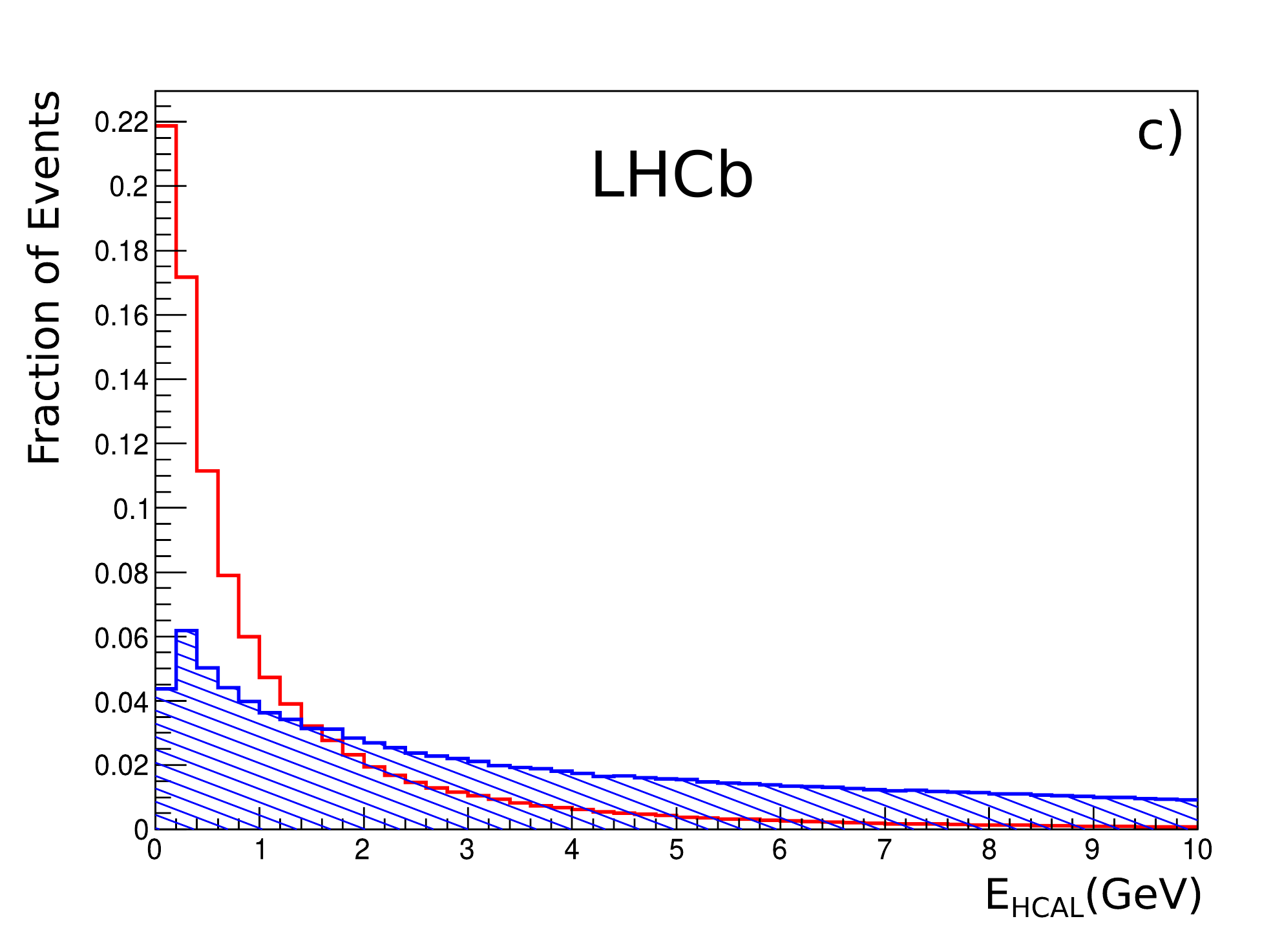}
\caption{Electron identification estimators (open histogram for electrons, shaded histogram for hadrons): a) energy deposited in the \presh,  b) value of the $\chi^2_{\rm 2D}$ estimator in \ecal,  c) energy deposited in the \hcal. Data taken in 2011.}
\label{perf_e_estimators}
\end{figure}

In the \presh, the electromagnetic particles start to develop a shower, while the depth is still small in terms of nuclear interaction length ($\lambda_I$), therefore electrons are expected to produce a larger signal than hadrons as illustrated in Figure~\ref{perf_e_estimators}~(a). The log-likelihood difference for electron and hadron hypotheses for the \presh, $\Delta \log {\mathcal L} ^{\rm \presh}_{e/h}$, is computed using the distribution of $\tanh(E_{\rm \presh}/100)$ versus $\tanh(p/100)$, where $E_{\rm \presh}$ is expressed in \mev and $p$ in \mevc. The \ecal-based estimator $\chi^2_{\rm 2D}$ defined in Eq.~\ref{eq:chideuxecal} is represented in Figure~\ref{perf_e_estimators}~(b). The $\Delta \log {\mathcal L}^{\rm \ecal}_{e/h}$ is computed from the 2D distribution of $\tanh(\chi^2_{\rm 2D}/2500)$ versus $\tanh (p/100)$, where $p$ is expressed in \gevc. For the \hcal, the distribution of $E_{\rm \hcal}$ is shown in Figure~\ref{perf_e_estimators}~(c). The effects of multiple scattering in both the \ecal and \hcal are neglected.  Due to the thickness of \ecal (25~\Xrad), the leakage of electromagnetic showers into the \hcal is expected to be extremely small. The  $\Delta  \log {\mathcal L} ^{\rm \hcal}_{e/h}$ is computed from the distribution of $\tanh(E_{\rm \hcal}/5)$ versus $\tanh(p/100)$, where $E_{\rm \hcal}$ is expressed in \gev and $p$ in \gevc.

\subsection{Photon and merged \texorpdfstring{\piz}{pi0} identification}

The first photon identification algorithm consisted in evaluating a photon hypothesis likelihood from the signal and background probability density functions (pdf) of several variables. Three variables were used:
\begin{itemize}
\item the \presh energy deposited in the cells facing the 9 \ecal cluster cells,
\item the $\chi^2_{\rm 2D}$ defined in formula~\ref{eq:chideuxecal}, which still contains a valuable information on the candidate after the cut at 4,
\item the ratio of the energy in the central cell to the energy of the cluster.
\end{itemize} 
Probability density functions for the signal and background hypothesis were built for the 3 zones of the \ecal and for several bins in energy using simulated samples. 

Later, a new classifier for photon identification was developed resulting in a significant improvement in performance, as documented in~\cite{Hoballah:2026848}. The current algorithm consists on a Neural Network (specifically, a Multi-Layer Perceptron) discriminant of the TMVA~\cite{Hocker:2007ht} tool. Two independent estimators, known as IsNotH and IsNotE, are built to establish the photon hypothesis: the former is aimed to distinguish the photon from non-electromagnetic deposits associated to hadrons while the later allows to separate between photons and electrons. The input variables of the Neural Network are the three variables used by the previous likelihood estimator, besides the following ones:
\begin{itemize}
\item the ratio of the energy in the \presh cell in front of the seed cell and the total energy measured in the \presh,
\item the ratio of the energy in the \hcal, in the projective area matching the cluster, and the cluster energy in the \ecal,
\item the energy deposited in the $2\times2$ matrix of \presh cells with the highest energy deposit in front of the reconstructed cluster,
\item the lateral shower shape variables, 
\begin{equation*}
{\cal S}_c = \frac{1}{\epsilon_{\rm cl}}
\begin{pmatrix}
\sum_i \epsilon_i(x_i-x_c)^2   & \sum_i \epsilon_i(x_i-x_c)(y_i-y_c) \\
\sum_i \epsilon_i(x_i-x_c)(y_i-y_c) & \sum_i \epsilon_i(y_i-y_c)^2
\end{pmatrix},
\end{equation*}
where ($x_c$, $y_c$) is the position of the transverse barycentre of the photon,
\item the spread (second lateral cluster moment) of the shower defined as 
\begin{equation*}
    <r^2> = S_{c,xx}+S_{c,yy} = \frac{1}{\epsilon_{\rm cl}} \sum_i \epsilon_i \left((x_i-x_c)^2+(y_i-y_c)^2 \right),
\end{equation*}
\item the number of \presh cells in front of the cluster with non-zero energy deposit, and
\item the multiplicity of hits in the $3\times3$ \spd cells matrix in front of the reconstructed cluster.
\end{itemize}

\begin{figure}[!ht]
\centering
\begin{minipage}{0.49\textwidth}
\includegraphics[width=1.0\textwidth]{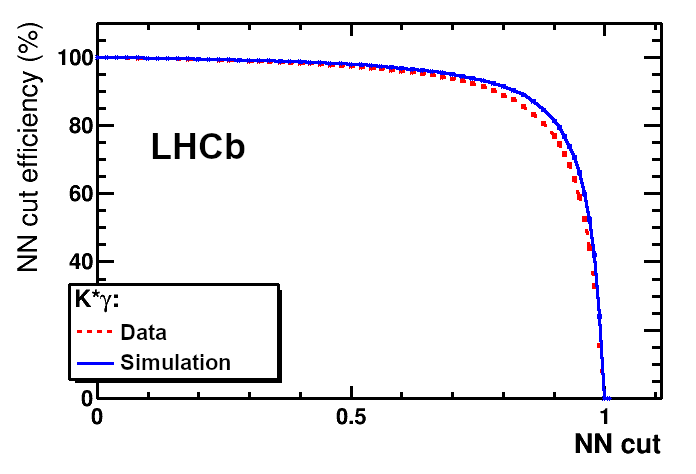}
\end{minipage}
\begin{minipage}{0.49\textwidth}
\includegraphics[width=1.0\textwidth]{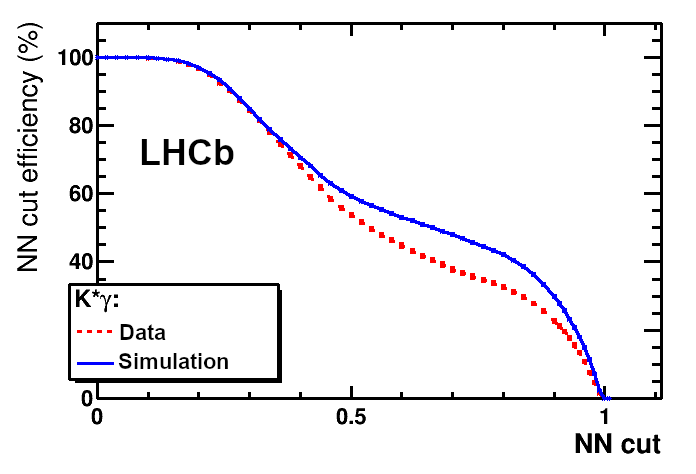}
\end{minipage}
\caption{Photon identification efficiencies as a function of the applied cut for IsNotH (hadron rejection) (left) and IsNotE (electron rejection) (right) with $\Bz\to\Kstarz\gamma$ data (red dashed line) and simulation (blue solid line).}
\label{perf_IsNotHE}
\end{figure}

The efficiencies as a function of the IsNotH (IsNotE) applied cut (NN in the plot) are shown on the left (right) side of Figure~\ref{perf_IsNotHE} for $\Bz\to\Kstarz\gamma$ data (dashed red line) and simulation (solid blue line). In both plots the data are background subtracted using the \sPlot method~\cite{Pivk:2004ty} and a fit to the B invariant mass with a dedicated model for the background. As can be noticed, for low values of the IsNotH cut, data and simulation exhibit a good agreement.

Sometimes, a high \et merged \piz, reconstructed as a single \ecal cluster, can be mis-identified as a single photon. A good $\gamma - \piz$ separation is an important pre-requisite for the study of radiative decays in order to reduce physical background containing \piz and it can also be useful in the opposite way, in the study of $B$ decays with a \piz in the final state. An algorithm based on simulation has been developed to distinguish between these two neutral particles. The tool, known as IsPhoton, is based on the fact that electromagnetic clusters produced by merged \piz have a broader and more elliptical shape with respect to the one produced by a single photon.

A few discrimination variables, here referred as shape variables, are constructed from the information of the energy deposited in the cells of the cluster. These provide information about the size of the cluster, the importance of the tails and the squashiness of the cluster. Similar information from the $3\times3$ \presh cell matrix in front of the seed of the electromagnetic cluster is used, as well as the hit multiplicity in these \presh cells with different requirements of the minimum energy deposited in the cells. The detailed description of the variables can be found in~\cite{CalvoGomez:2042173}. All the variables are taken as input in a Multi Layer Perceptron (MLP) of the TMVA~\cite{Hocker:2007ht} tool. True photons from $\Bz\to \Kstarz\gamma$ simulated samples are used as signal sample and a realistic background is obtained from a mixture of $B$ decays containing \piz in the final state, by requiring true merged \piz being reconstructed and selected as photons using the same $\Bz\to \Kstarz\gamma$ pre-selection as used for the signal sample. The three regions of the calorimeter are studied separately, because of the different size of the cells. 

\begin{figure}[!ht]
\centering
\includegraphics[width=1.0\textwidth]{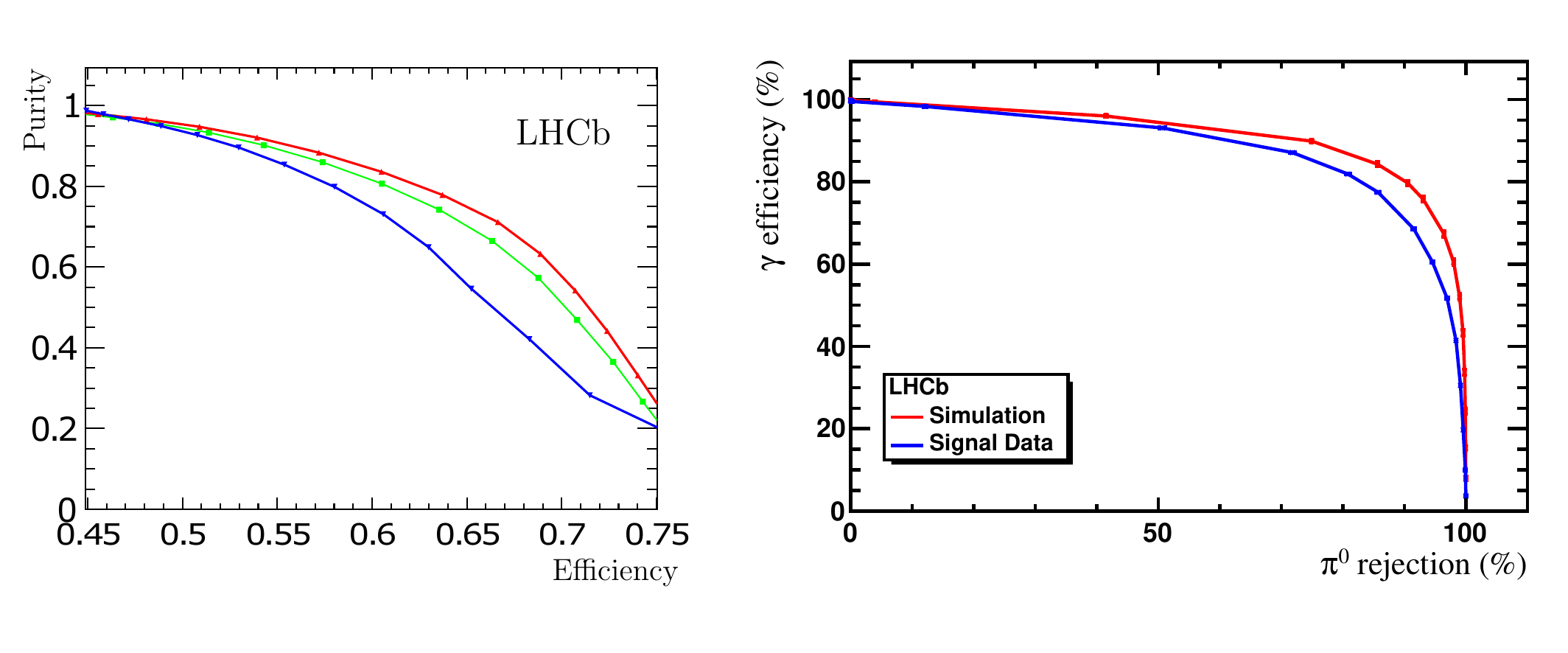}
\caption{Purity as a function of efficiency for (green) the full photon candidate sample, (blue) converted candidates according to the \spd information and (red) non-converted candidates (left). Photon identification efficiency as a function of \piz rejection efficiency for the $\gamma - \piz$  separation tool for simulation (red curve), and data (blue curve) (right).}
\label{perf_pi0_gam_eff}
\end{figure}

As shown in Figure~\ref{perf_pi0_gam_eff}, applying a cut at 0.6 on the output of the IsPhoton tool, a high signal efficiency of 95\% can be obtained with a rejection of 50\% of the \piz background reconstructed as photon. The use of the \presh variables in addition to the \ecal ones helps to decrease by 5\% the background contamination for high signal efficiencies.

A data driven calibration of the performance of this $\gamma-\piz$ separation tool has been performed, necessary due to some data/simulation discrepancies. The real photon efficiency and purity for different values of the discriminating cut and $p_{\rm T}$ ranges are evaluated using clean samples of $\Dz\to \Km \pip \piz$ and $\Bz \to\Kstarz\gamma$ decays.

\subsection{Converted photon reconstruction}

Photons converted before the magnet are reconstructed from opposite sign electron/positron tracks. The electrons are selected on the basis of their electron PID variables, electron confidence level, $p_{\rm T}$  and $E/p$ range;  these parameters can be adjusted in the conversion algorithm. The algorithm combines only electron/positron pairs having their vertical position within a short relative distance (3-200\mm). The electron pairs are selected on the basis of their transverse momentum ($p_{\rm T}>500\mevc$), their di-electron mass and their reconstructed vertex position, including bremsstrahlung photons measured in the calorimeters giving a photon pair candidate. As in most case bremsstrahlung photons are compatible with both electrons and positrons tracks, the bremsstrahlung photon is added to only one of the electron/positron track to avoid double counting through the following steps:
\begin{itemize}
\item the  photon momentum is added to the electron and the pair energy is calculated with the electron mass,
\item the electron direction is kept and the photon energy is re-evaluated taking into account the electron mass,
\item the bremsstrahlung origin allows to estimate the electron direction and the converted photon energy is re-evaluated.
\end{itemize}

\begin{figure}[!ht]
\centering
\includegraphics[width=0.7\textwidth]{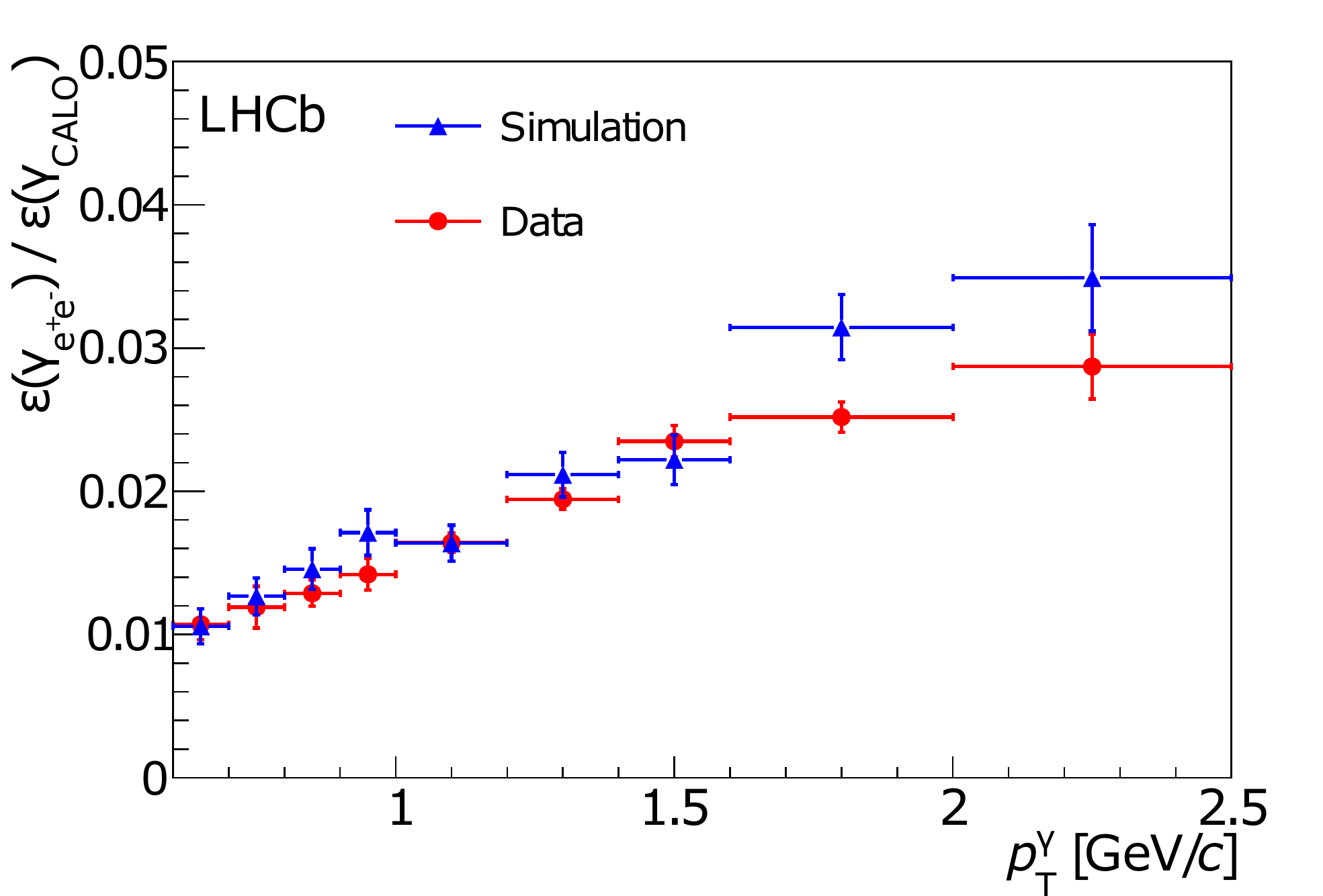}
\caption{Ratio of photon detection efficiencies $\epsilon (\gamma \to ee)/\epsilon(\gamma_{CALO})$ from the decay of \piz mesons in data (red) and simulations (blue). }
\label{perf_pi0_convert_photons}
\end{figure}

\begin{figure}[!ht]
\centering
\includegraphics[width=0.7\textwidth]{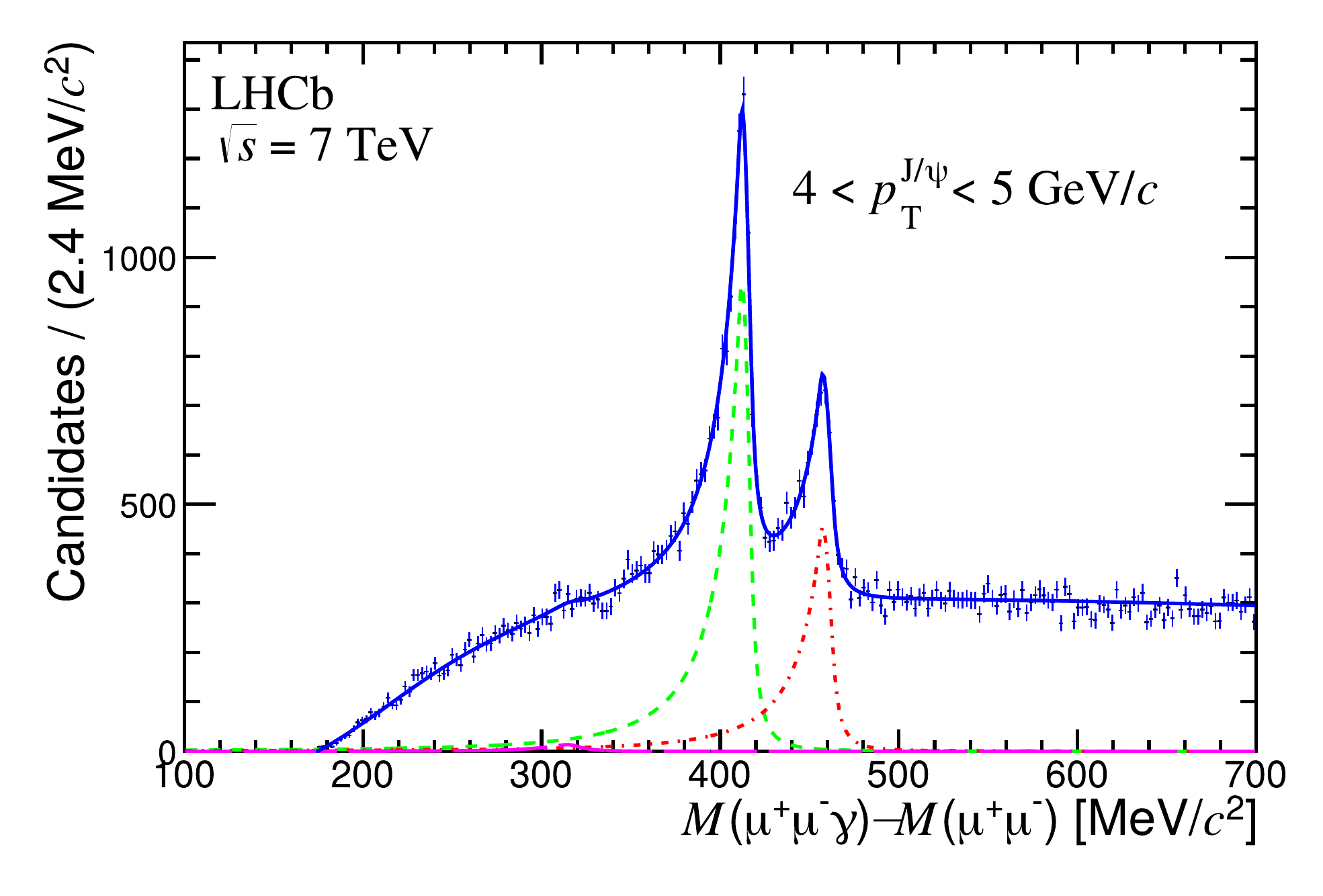}
\caption{Distribution of $\Delta M = M(\mup\mun\gamma)-M(\mup\mun)$ for $p_{\jpsi}$ in the \pt-range 4-5\gevc. Fits are also shown, with the total fitted function (blue solid curve), the \chicone signal (green dashed curve), the \chictwo signal (red dot-dashed curve) and the \chiczero signal (purple long-dashed curve).}
\label{perf_chic1_chic2}
\end{figure}

The photon reconstruction efficiency is well reproduced in the simulation, implying not only a good description of the detector material where the conversions occur as represented in Figure~\ref{perf_pi0_convert_photons}, but also good performances of the reconstruction algorithms. Some physics channels are being analysed benefiting from the good resolution obtained with converted photons, such as  $\chic \to \jpsi\gamma$ or $\chi_b  \to \Upsilonres \gamma$. For example in the case of  the \chic, the resolution on the mass difference  $\Delta M = M(\mup \mun\gamma) - M(\mup \mun)$ of 4 to 5\mevcc allows to disentangle the \chiczero, \chicone and \chictwo states as shown in Figure~\ref{perf_chic1_chic2}. More details are provided in Ref.~\cite{LHCb-PAPER-2013-028}.

\subsection{Combined performance of electron identification}

\begin{figure}[!ht]
\centering
\includegraphics[width = 0.7\textwidth]{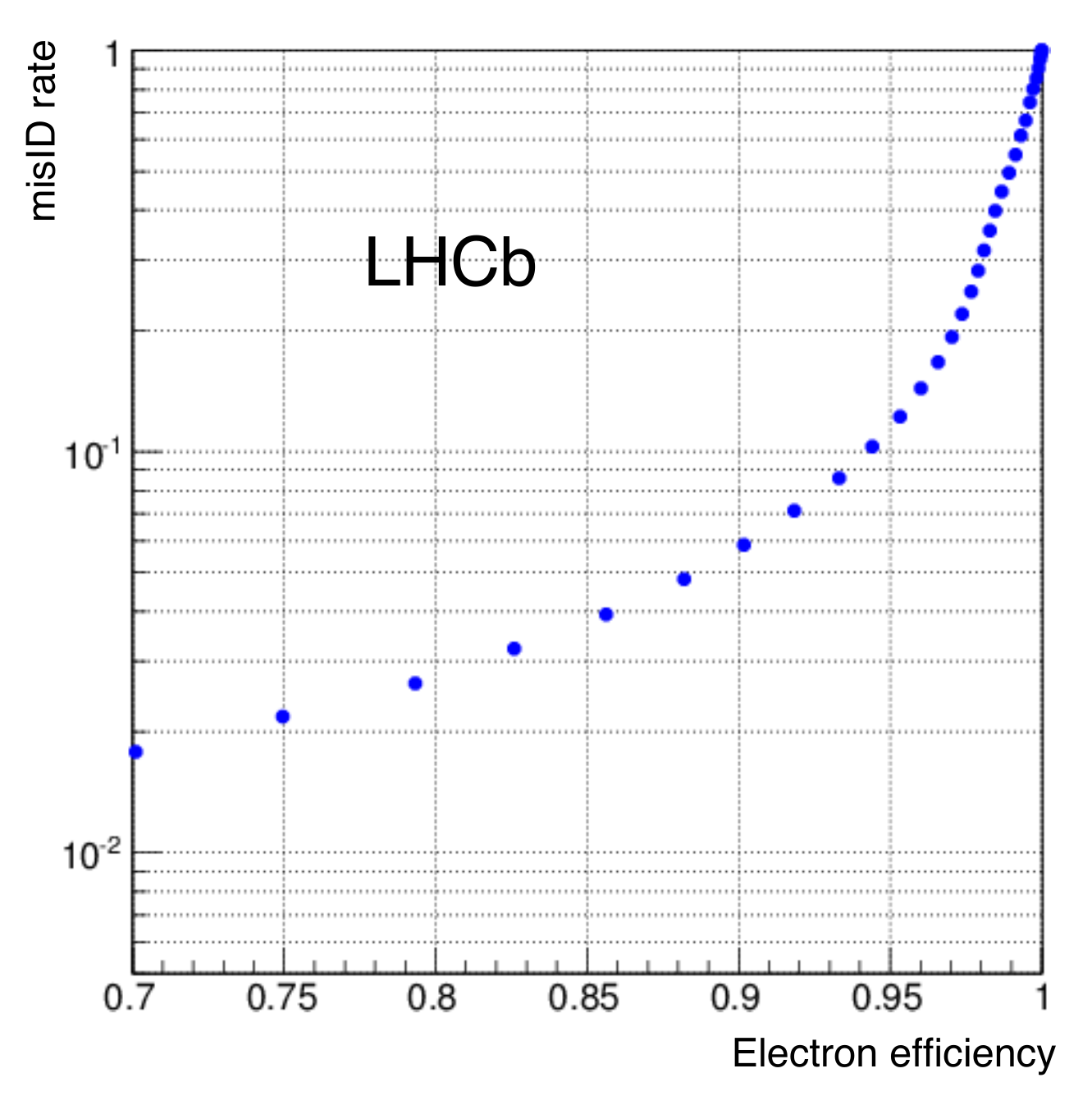}
\caption{Electron identification efficiency versus mis-identification rate. }
\label{perf_e_dll_misid}
\end{figure}

Since there is more than one electron identification estimator available from the \lhcb calorimeter system, a combined one can be simply obtained
by taking the sum of the individual estimators from the \presh, the \ecal and the \hcal, 
\begin{equation}
\Delta \log {\mathcal L}^{\rm CALO}_{e/h} = 
\Delta \log {\mathcal L}^{\rm \ecal}_{e/h} +
\Delta \log {\mathcal L}^{\rm \hcal}_{e/h}  +
\Delta \log {\mathcal L}^{\rm \presh}_{e/h}.
\end{equation}
Figure~\ref{perf_e_dll_misid} shows the combined electron identification efficiency for the whole calorimeter system defined above versus the misidentification rate obtained by varying the selection criteria applied to the log-likelihood difference. With an electron identification efficiency of 90\%, the mis-identification rate is 5.5\%.

The electron identification performance is evaluated using the data acquired in 2011. The available statistics allows to measure the electron identification efficiency with a tag-and-probe method. The method is applied using $\Bpm \to \jpsi(\ep\en) \Kpm$ candidates, where one of the electrons is required to be identified by the electron identification algorithm ($e_{\rm tag}$) while the second electron is selected without using any information from the calorimeter system ($e_{\rm probe}$). With this second electron, the efficiency of the electron identification is estimated. The use of \jpsi from $B$ decays has the advantage of getting rid of most of the combinatorial background originating from the tracks coming from the primary vertex. Both electron tracks must have a transverse momentum above 500\mevc with a total momentum above 3\gevc, a good quality of the track fit $\chisqndf < 5$, where ${\rm ndf}$ is the number of degrees of freedom and a minimal impact parameter to primary vertex \chisqip above 9. It is required that the global electron identification variable, $\Delta\log{\mathcal L}_{e/h}^{\rm CALO}$, is greater than 5 for the tag-electron $e_{\rm tag}$. For $e_{\rm tag}$, a harder cut on the transverse momentum is applied, $\pt > 1.5\gevc$, and on the total momentum $p>6\gevc$. Then the two tracks are combined into a common vertex and a cut $\chi^2_{\rm vtx}/{\rm ndf} <9$ is applied on the di-electrons vertex. The resulting di-electrons candidate is required to have a mass within 150\mevcc of the nominal \jpsi mass~\cite{PDG2012}. After a \jpsi candidate is identified, the other hadron track is combined with it to form \Bpm candidates. Hadrons are required to be identified as kaons by the particle identification system. The tracks are combined into a common vertex and a cut $\chisqvtxndf <$16 is applied on the resulting vertex. To select \Bpm,  prompt background from tracks originating directly from primary vertex is reduced by requiring the $B$ candidate lifetime ($c \tau$) to be greater than 100\mum. The electron identification efficiency is computed as $\epsilon = N_1/(N_1 + N_0)$, where $N_1$ is the number of \Bpm candidates passing a given cut on $\Delta\log{\mathcal L}^{\rm CALO}_{e/h}$ and $N_0$, the number of rejected events.

\begin{figure}[!ht]
\centering
\includegraphics[width = 1.0\textwidth]{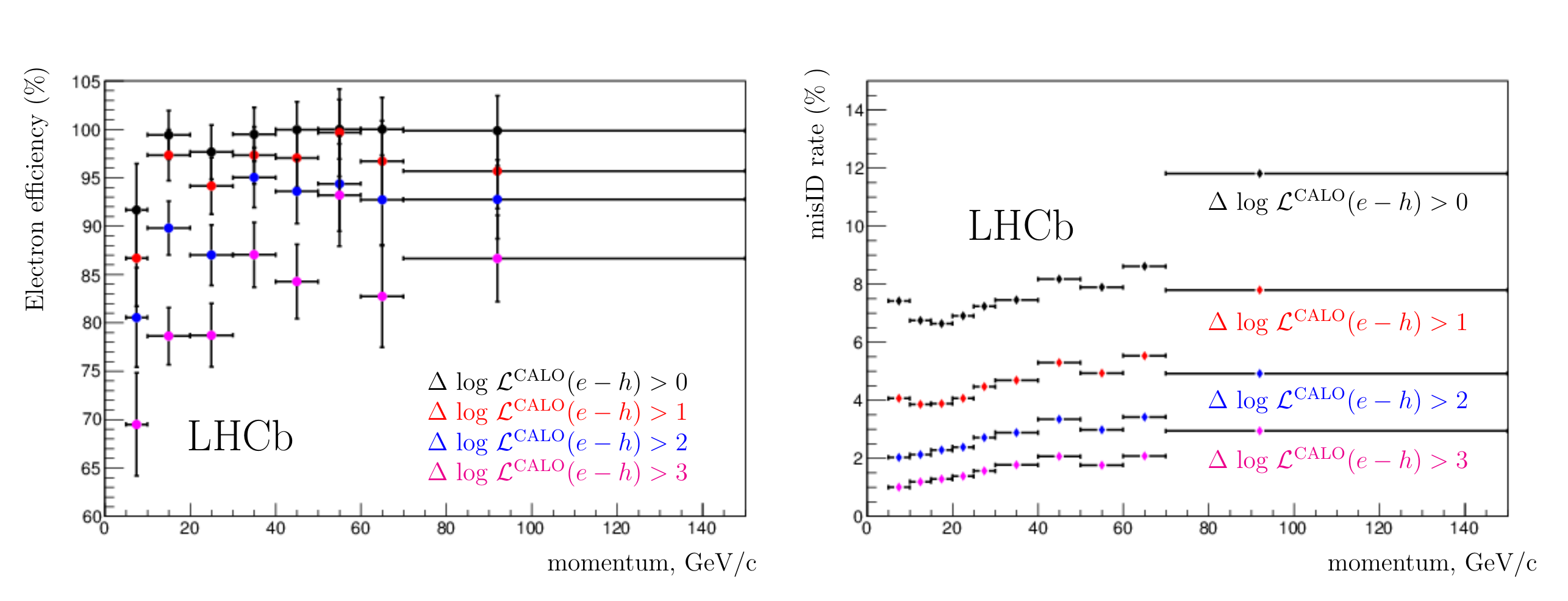}
\caption{Performance of the calorimeter for electron identification with $\Delta\log{\mathcal L}_{e/h}^{\rm CALO}$ cuts:  efficiency as a function of momentum (left) and  mis-identification rate (right) as a function of momentum.}
\label{perf_e_eff_misid_vs_mom}
\end{figure}

The efficiency and the mis-identification rate as a function of the $e_{\rm probe}$ momentum is presented in Figure~\ref{perf_e_eff_misid_vs_mom} for several cuts on $\Delta\log {\mathcal L}^{\rm CALO}_{e/h}$. For $\jpsi \to \ep\en$ from $\Bpm \to \jpsi \Kpm$  decays, the average electron identification efficiency in the complete calorimeter system is $(91.9 \pm 1.3) \%$ with a mis-identification rate of $(4.5 \pm 0.1) \%$ after requiring $\Delta\log{\mathcal L}^{\rm CALO}_{e/h} >2$. As expected, the tracks with higher momenta have higher mis-identification rates, as illustrated in Figure~\ref{perf_e_eff_misid_vs_mom}.

The combined electron identification estimator $\Delta\log{\mathcal L}_{e/h}^{\rm CALO}$ can be combined with estimators of other sub-detectors to provide a combined particle identification estimator. As a consequence, the electron mis-identification rate is improved from $\thicksim$6\% to $\thicksim$0.6\% at 90\% electron efficiency~\cite{LHCb-DP-2014-002}.

An alternative approach for electron identification, known as ProbNNe, has also been used to identify electrons. The ProbNNx (x = e, $\pi$, K, p) method~\cite{LHCb-DP-2018-001}, is based on a Neural Network using information from several sub-detectors. Simulated and background events are used to tune the algorithm. The results obtained are in general good agreement with the log-likelihood method.

\subsection{\ecal energy resolution}

\begin{figure}[!ht]
\centering
\begin{minipage}{0.49\textwidth}
\includegraphics[width=1.0\textwidth]{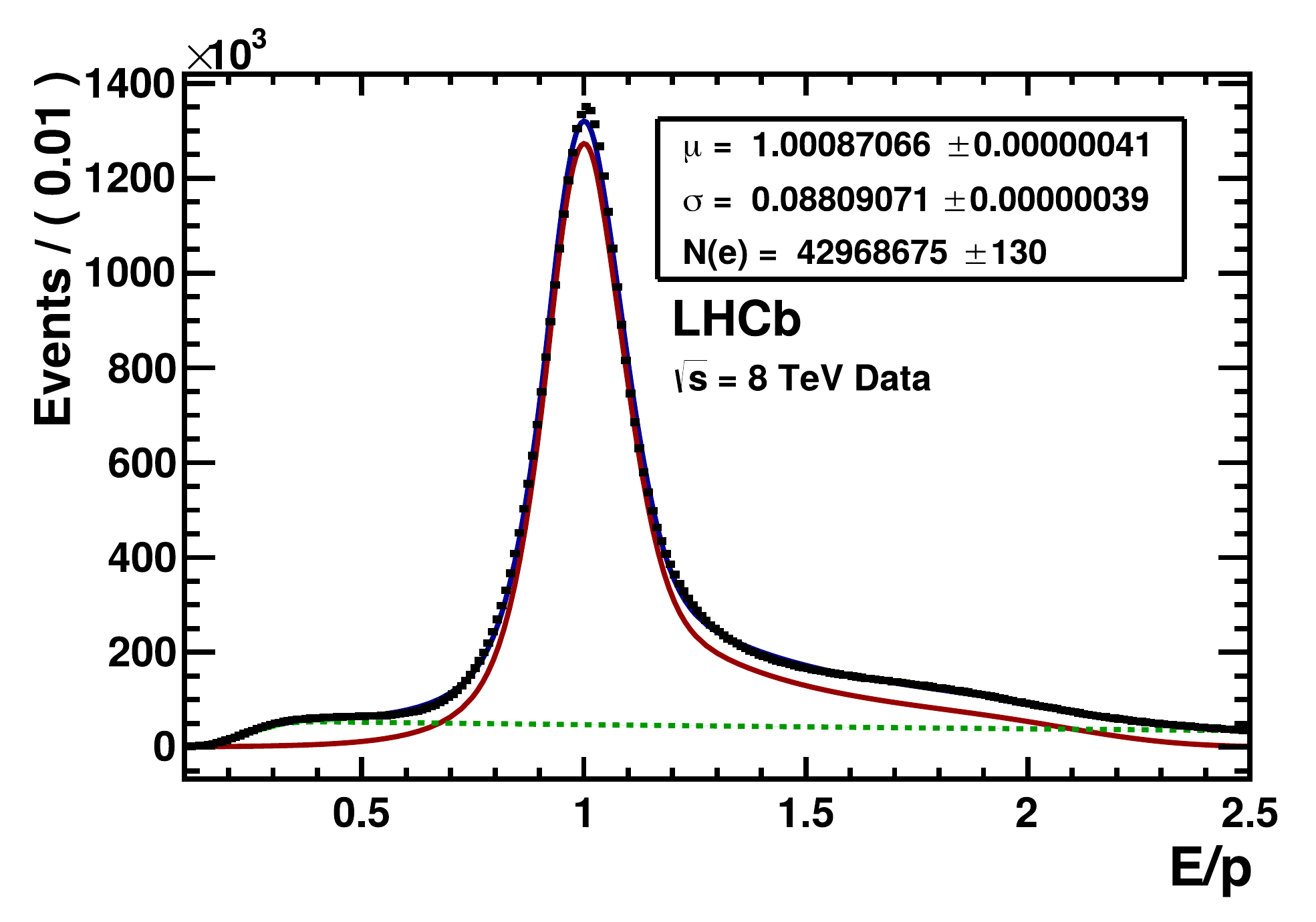}
\end{minipage}
\begin{minipage}{0.49\textwidth}
\includegraphics[width=1.0\textwidth]{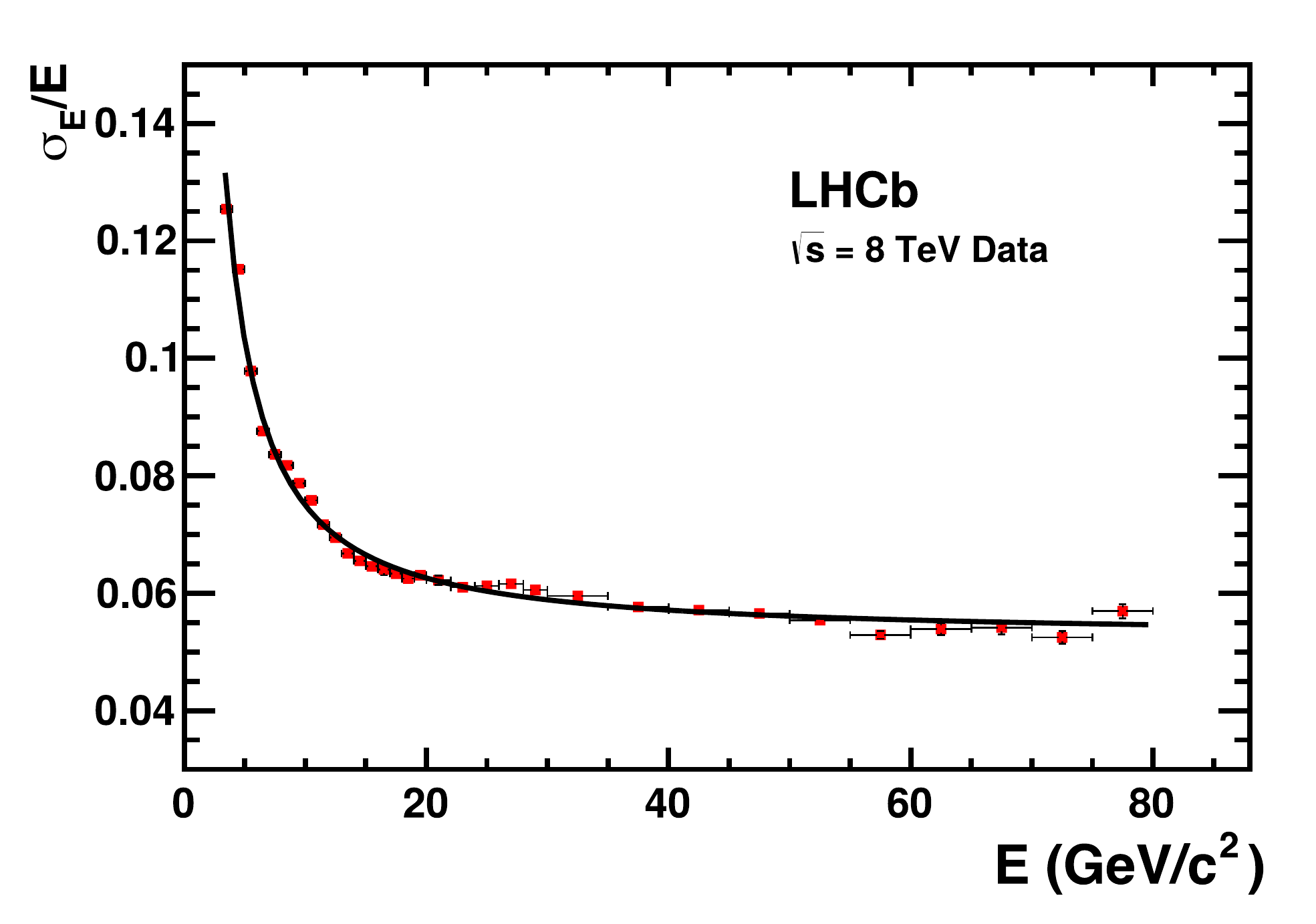}
\end{minipage}
\caption{Distribution of $E/p$ from electrons coming from converted photons with $E_{\rm \presh}>50\mev$, with at least one hit on the \spd and a track with $\pt > 200\mevc$ pointing into the \ecal cluster ($\chi^2_{2D}<25$), and a matching positron to forming invariant mass $M(ee) < 100$ \mevcc (left). Resolution for $\sigma(E/p)$ as a function of energy (right).}
\label{perf_e_over_p}
\end{figure}

In order to estimate the \ecal energy resolution from data, electrons coming from converted photons are used. These are selected by requiring the \presh energy to be $E_{\rm \presh}>50\mev$, with at least one hit on the \spd and a track with $\pt > 200\mevc$ pointing into the \ecal cluster (with a $\chi^2_{2D}<25$). In order to minimise the background, the electron is required to match a positron to form an invariant mass $M(ee) < 100$ \mevcc. Figure~\ref{perf_e_over_p} (right) shows the distribution of $E / p$ for the 2012 data. The fit uses a double-sided Crystal Ball function for the signal and an exponential for the background, with threshold functions to describe the cuts in \et/\pt.  

From Figure~\ref{perf_e_over_p} (left), the resolution $\sigma ( E/p )$ for each energy bin (Figure~\ref{perf_e_over_p} (right)), in the energy range from 3 to 80\gev is extracted. Below three \gev, threshold effects on the transverse momentum estimation of the tracks are found, while over 80\gev, saturation effects occur in the inner modules. After deconvoluting the tracking resolution, which happens to be almost negligible, and fitting the curve in Figure~\ref{perf_e_over_p} (right), the following model is obtained, 
\begin{equation}
\frac{\sigma_E}{E} = \frac{(13.5 \pm 0.7)\%}{\sqrt{E_{\rm GeV}}} \oplus (5.2 \pm 0.1)\% \oplus \frac{(320\pm 30) \mev}{E_{\rm GeV}}.
\label{eq:resolution}
\end{equation}

This model represents the effective resolution of the calorimeter, including the intrinsic resolution of the cells plus the electron reconstruction effects, the interaction with the material in front of the calorimeter and the pile-up. The actual design value, in Eq.~\ref{eq:resolution}, obtained from a test beam, did not include those effects, in particular, the material in front of the calorimeter.

\begin{figure}[!ht]
\centering
\begin{minipage}{0.49\textwidth}
\includegraphics[width=1.0\textwidth]{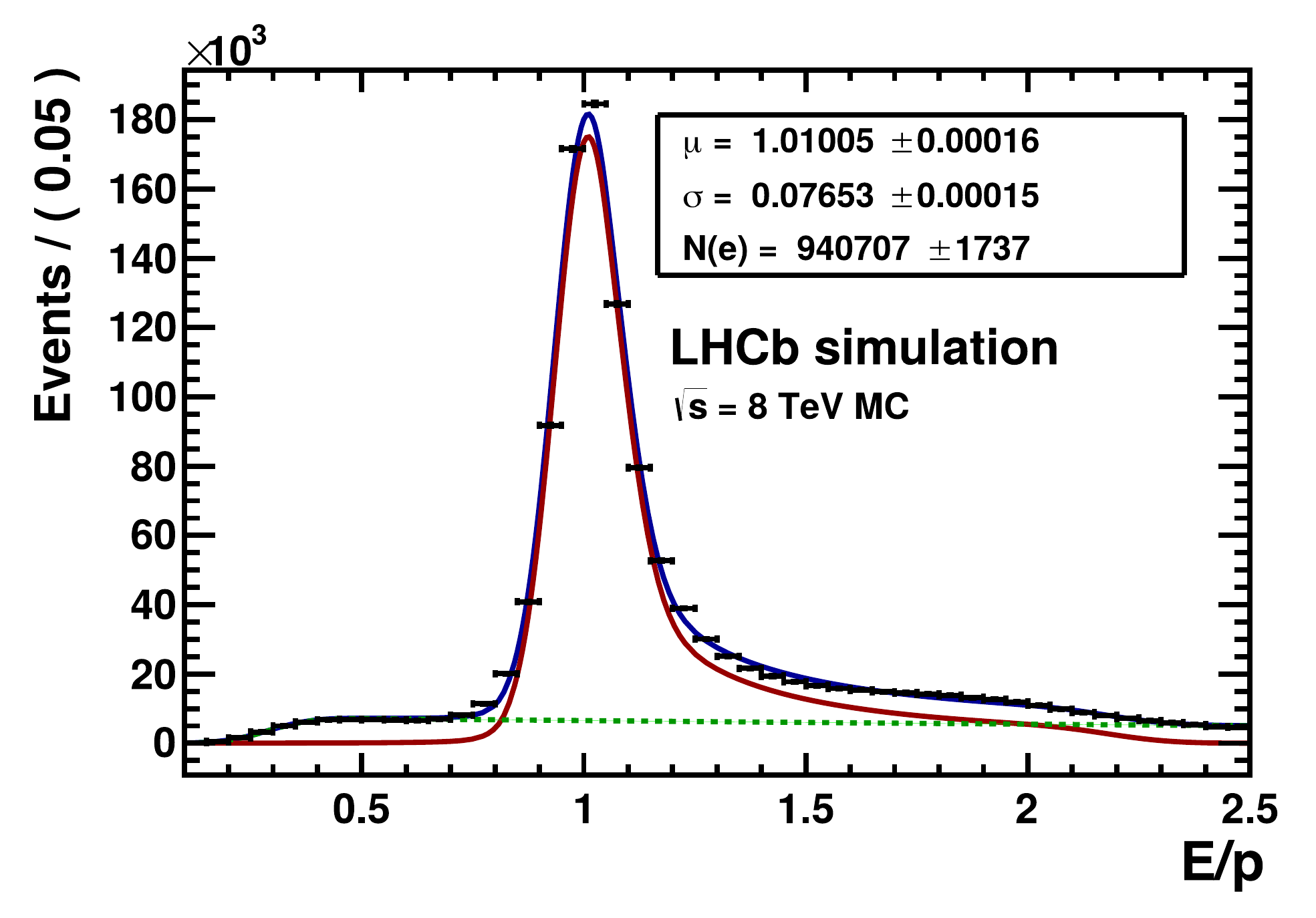}
\end{minipage}
\begin{minipage}{0.49\textwidth}
\includegraphics[width=1.0\textwidth]{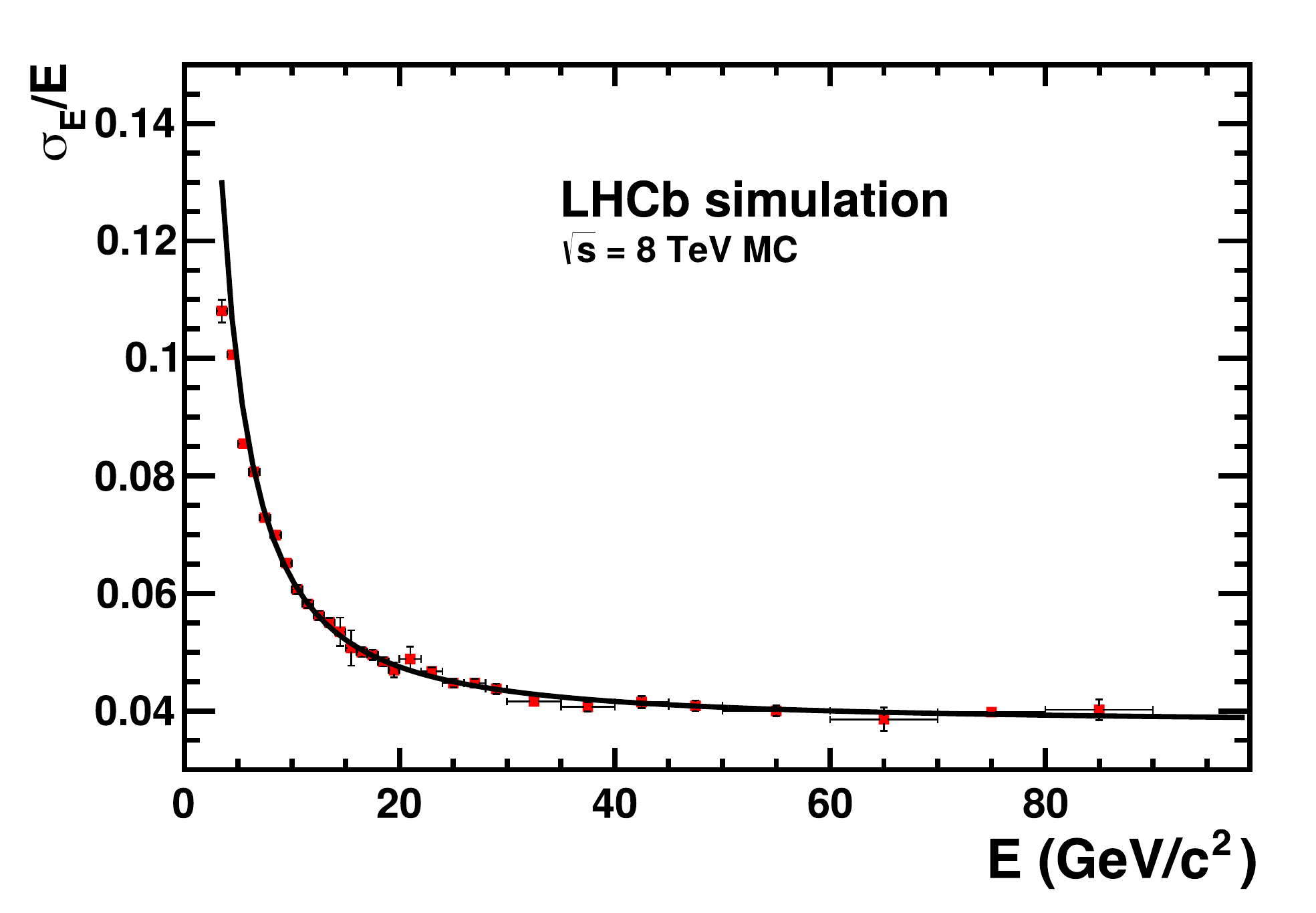}
\end{minipage}
\caption{Distribution of $E/p$ from simulations (left). Resolution $\sigma(E/p)$ as a function of energy, obtained from simulations (right).}
\label{perf_e_over_p_elec_mc}
\end{figure}

The results of the simulation study is shown in Figure~\ref{perf_e_over_p_elec_mc}, where the $E/p$ distribution (left) and its resolution (right) can be observed. From the corresponding fit, one obtains, 
\begin{equation}
\frac{\sigma_E}{E} = \frac{(9.6\pm 1.3)\%}{\sqrt{E_{\rm GeV}}} \oplus (3.7\pm0.1)\% \oplus \frac{(395\pm 30) \mev}{E_{\rm GeV}}.
\label{eq:resolutionMC}
\end{equation}

Both the background term and the stochastic terms from Eq.~\ref{eq:resolution} and~\ref{eq:resolutionMC} are of the same magnitude. Actually, \ecal is calibrated using \et so one part of the background depends on $\sin\theta$ where $\theta$ is the azimuthal angle with respect to the beam. The noise coming from the electronics is equivalent to 1.2 \adc counts (incoherent) plus 0.3 \adc counts (coherent), which corresponds to an average noise of about 6 \adc counts, equivalent to 15\mev in terms of \et over one cluster. This amounts to a background range from 45\mev to 450\mev depending on the region. The actual value, 350\mev, is an average over the full calorimeter which includes the effects of the material in front. The constant term value, of about $5\%$, is essentially due to the pile-up effect. 

\section{Examples of results}

The performance of the \lhcb calorimeter system is within expectations; ageing issues have been addressed and corrected. Among the physics results in which the calorimeter data quality is relevant, results related to decays of $\Bz \to \Kstarz \gamma$ and  $\Bs\to \phi \gamma$, which are particularly sensitive to the \ecal calibration may be singled out. At a centre of mass energy of \sqs = 7\tev, the ratio of branching fractions of $\Bz\to \Kstarz \gamma$ and $\Bz\to \phi\gamma$ decays is measured to be~\cite{Aaij:2012ita}
\begin{equation}
\frac{{\cal B}\left(\Bz\to \Kstarz \gamma\right)}{{\cal B}\left( \Bs\to \phi\gamma\right)} =   
1.23 \pm 0.06 \mathrm{(stat)} \pm 0.04 \mathrm{(syst)} \pm 0.10 (f_s/f_d),
\end{equation}
in good agreement with the theoretical prediction of $1.0 \pm 0.2$~\cite{Ali:2007sj}. Using ${\cal B}\left(\Bz\to \Kstarz \gamma\right) = (4.33\pm 0.15)\times10^{-5}$~\cite{Asner:2010qj}, the measurement is
\begin{equation}
  {\cal B}\left( \Bs\to \phi\gamma\right) = (3.5 \pm 0.4 )\times 10^{-5},
\end{equation}
where statistical and systematic uncertainties are combined, which agrees with the previous experimental value. This is the most precise measurement of the ${\cal B}\left(\Bs\to \phi\gamma\right)$ branching fraction to date.

\begin{figure}[!ht]
\centering
\includegraphics[width=0.9\textwidth]{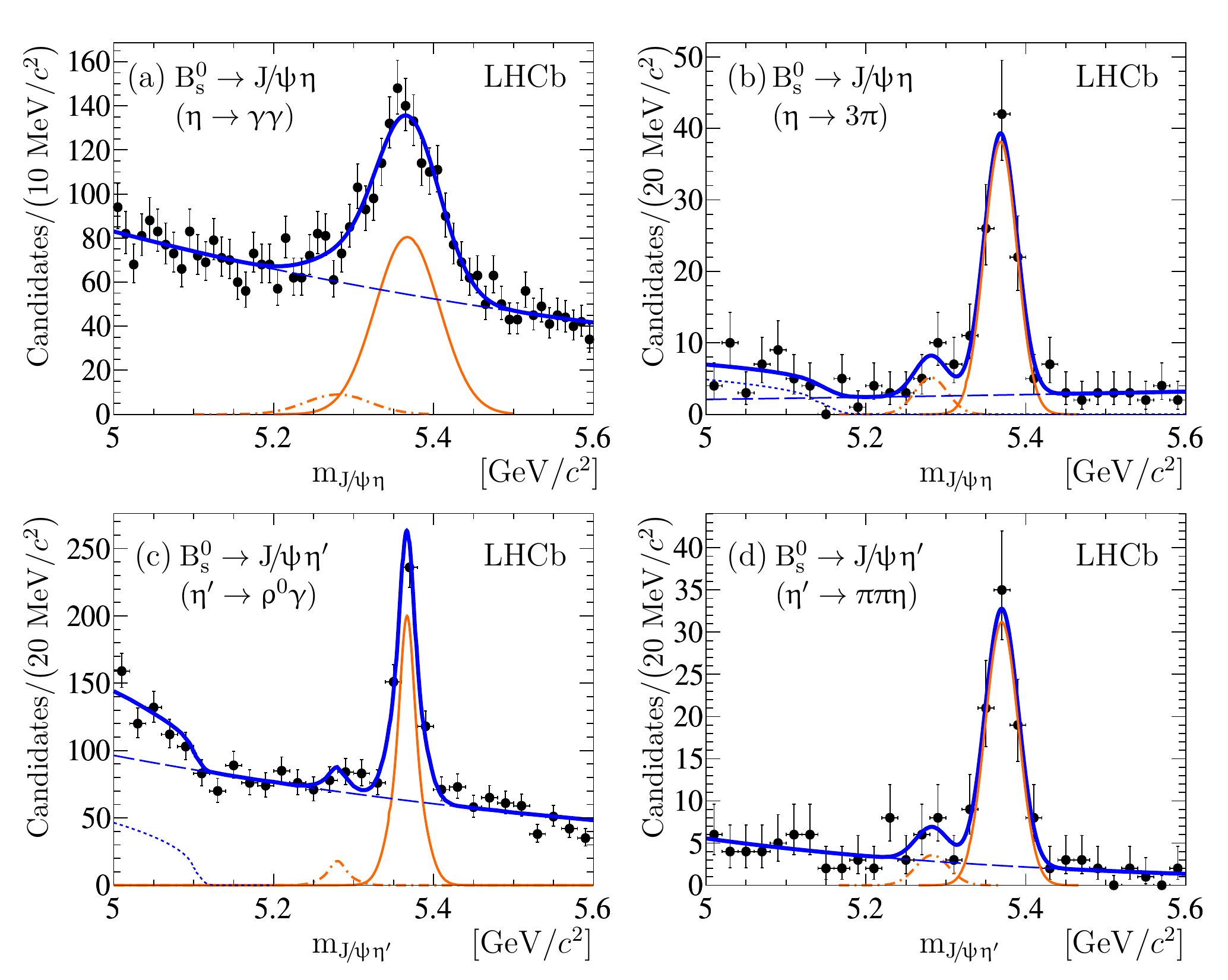}
\caption { Invariant mass distributions for selected $\Bs\to \jpsi\eta^{\left(\prime\right)}$ candidates. Black dots are the data, thin solid orange lines are the $\Bs$ contributions and orange dot-dashed lines are the $\Bz$ contributions. The blue dashed lines show the combinatorial background  contributions  and  the  dotted  blue  lines  show  the  partially  reconstructed background components.  The total fit functions are drawn as solid blue lines.}
\label{perf_bs_psi_eta_etap}
\end{figure}

The first evidence of the $\Bz\to \jpsi \omega$ decay is found, and the $\Bs\to \jpsi\eta$ and $\Bs\to \jpsi\eta^{\prime}$ decays are studied using a dataset corresponding to an integrated luminosity of 1.0\invfb (see Ref.~\cite{LHCb-PAPER-2012-022}). The ratio of the branching fractions of $\Bs\to \jpsi \eta^{\prime}$ and $\Bs\to \jpsi\eta$ 
decays is measured to be 
\begin{equation} 
  \frac{{\cal B}(\Bs\to \jpsi\eta^{\prime})}{{\cal B}(\Bs\to \jpsi \eta)} = 0.90\pm0.09\,(\mathrm{stat})\,^{+0.06}_{-0.02}\,(\mathrm{syst}).
  \end{equation} 
This result is consistent with the previous Belle measurement of
$\mathcal{R}^{\Bs,\eta^{\prime}}_{\Bs,\eta}=0.73\pm0.14$~\cite{Adachi:2009usa},
but is more precise. The invariant mass distributions for $\Bs\to \jpsi \eta^{\left(\prime\right)}$ candidates are shown in Figure~\ref{perf_bs_psi_eta_etap}.

\clearpage

\begin{figure}[!ht]
\centering
\includegraphics[width=0.7\textwidth]{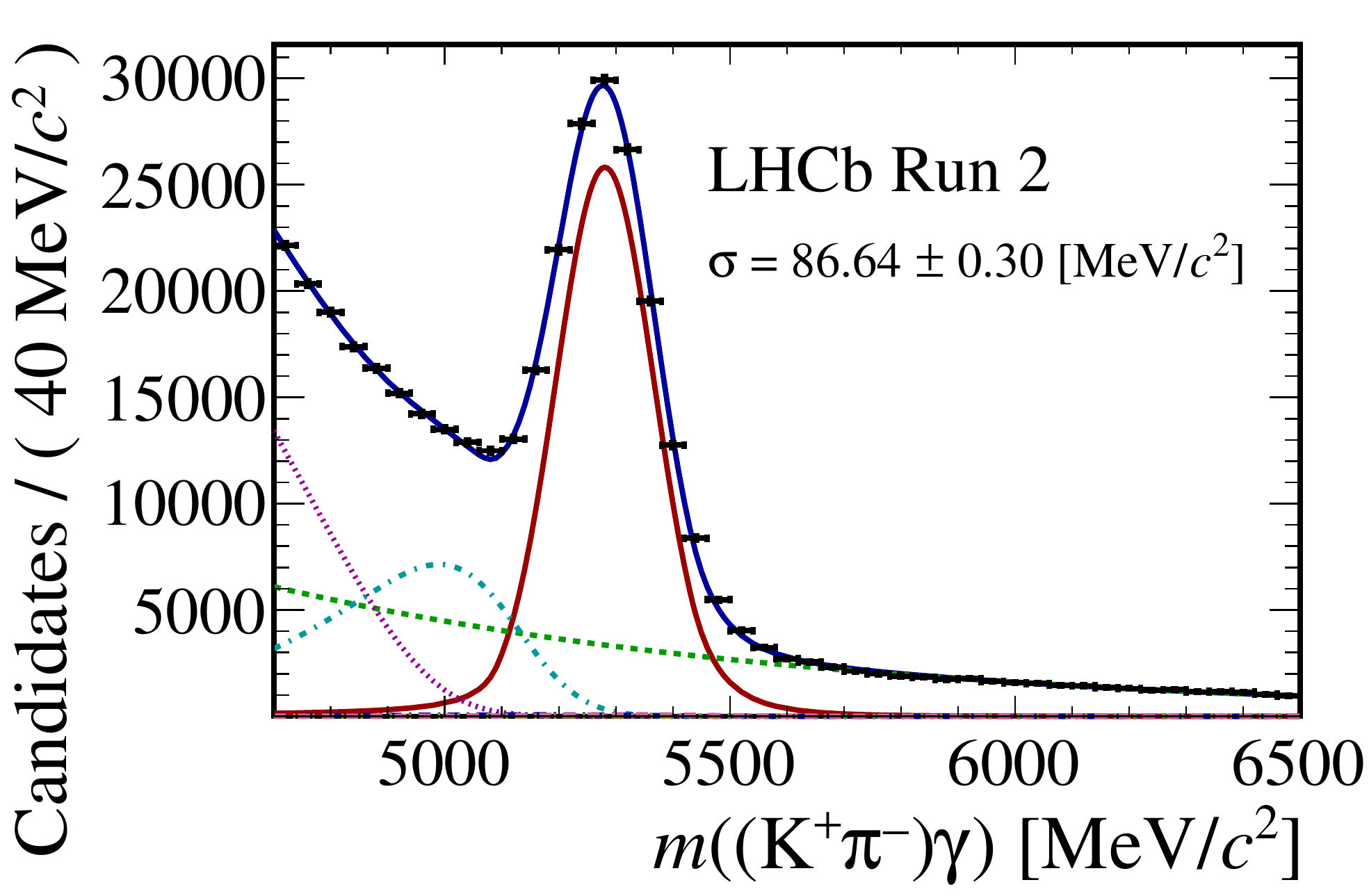}
\caption{Mass distribution of the reconstructed $\Bz\to \Kstarz(\Kp \pim)\gamma$ candidates in Run 2. The blue curve corresponds to the mass fit. The $\Kstarz\gamma$ signal component of the fit function (red line) and the various background contaminations are shown.}
\label{perf_b02kstgamma_2017_2018}
\end{figure}

\begin{figure}[!ht]
\centering
\begin{minipage}{0.49\textwidth}
\includegraphics[width=1.0\textwidth]{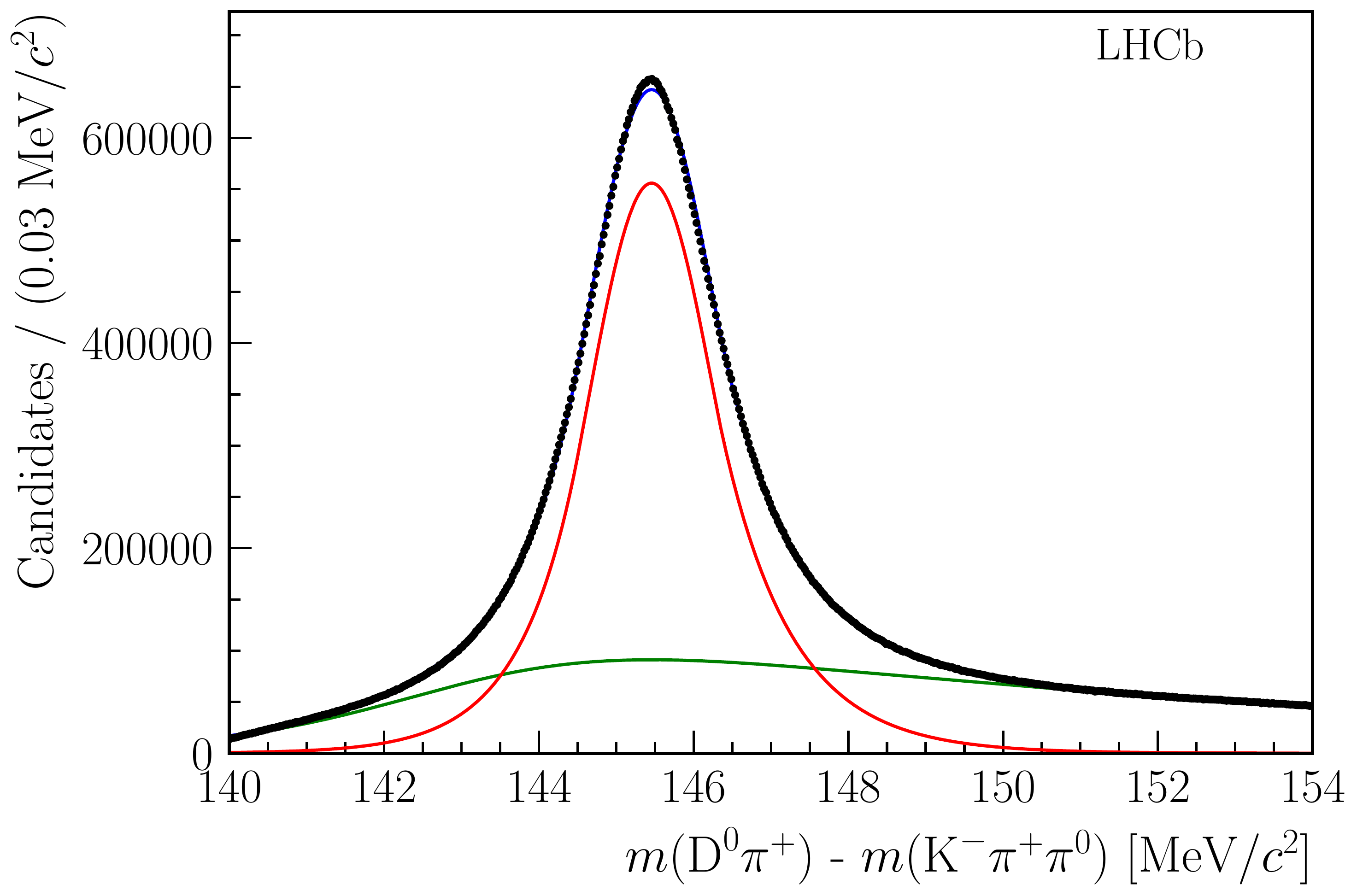}
\end{minipage}
\begin{minipage}{0.49\textwidth}
\includegraphics[width=1.0\textwidth]{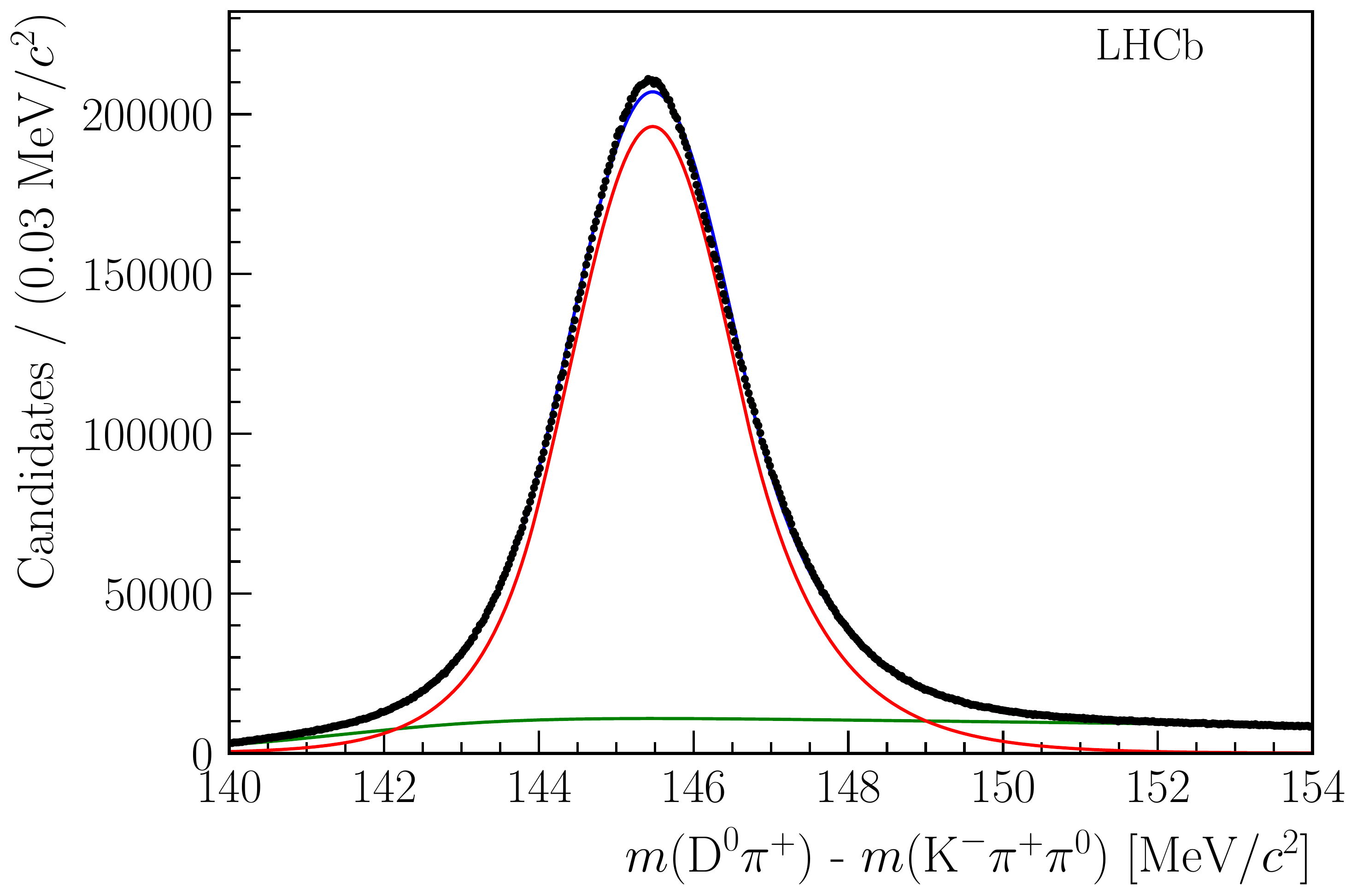}
\end{minipage}
\caption{Distribution of $\Delta M = M(\Dz\pip)-M(\Km\pip\piz)$ with resolved (left) and merged $\piz$ (right) in 2017-2018.}
\label{perf_deltam_2017_2018}
\end{figure}

Most of the results shown so far come from the Run 1 period. For Run 2 data taking, calorimetric and RICH information have been added to the second level software trigger (\hlt2), to provide offline-quality PID variables in the trigger allowing for larger calibration samples with higher signal purity~\cite{Lupton:2134057}. Below are two examples of the results obtained with Run 2 data with B decays containing photons ($\Bz\to \Kstarz\gamma$, Figure~\ref{perf_b02kstgamma_2017_2018})~\cite{LHCb-DP-2019-001} or \theDstarp decays containing \piz ($\theDstarp\to \Dz \pip, \Dz \to \Km\pip\piz$, Figure~\ref{perf_deltam_2017_2018}).

\section{Conclusion}

The Run 1 and Run 2 years of running have shown that the calorimetry was an essential piece of the \lhcb experiment. As shown here, it was used for the \lone trigger and its data have been used in many analyses. A lot of results involving calorimetry are still to come. Over the years, the calorimeters have been functioning as designed and the increase of energy from Run 1 to Run 2 was well handled. In 2019, the \spd and the \presh have been dismounted. The calorimetry, now made only of \ecal and \hcal, will still be an important part of LHCb.

\section*{Acknowledgements}
%
%
\noindent We express our gratitude to our colleagues in the CERN
accelerator departments for the excellent performance of the LHC. We
thank the technical and administrative staff at the LHCb
institutes.
We acknowledge support from CERN and from the national agencies:
CAPES, CNPq, FAPERJ and FINEP (Brazil); 
MOST and NSFC (China); 
CNRS/IN2P3 (France); 
BMBF, DFG and MPG (Germany); 
INFN (Italy); 
NWO (Netherlands); 
MNiSW and NCN (Poland); 
MEN/IFA (Romania); 
MSHE (Russia); 
MinECo (Spain); 
SNSF and SER (Switzerland); 
NASU (Ukraine); 
STFC (United Kingdom); 
DOE NP and NSF (USA).
We acknowledge the computing resources that are provided by CERN, IN2P3
(France), KIT and DESY (Germany), INFN (Italy), SURF (Netherlands),
PIC (Spain), GridPP (United Kingdom), RRCKI and Yandex
LLC (Russia), CSCS (Switzerland), IFIN-HH (Romania), CBPF (Brazil),
PL-GRID (Poland) and OSC (USA).
We are indebted to the communities behind the multiple open-source
software packages on which we depend.
Individual groups or members have received support from
AvH Foundation (Germany);
EPLANET, Marie Sk\l{}odowska-Curie Actions and ERC (European Union);
ANR, Labex P2IO and OCEVU, and R\'{e}gion Auvergne-Rh\^{o}ne-Alpes (France);
Key Research Program of Frontier Sciences of CAS, CAS PIFI, and the Thousand Talents Program (China);
RFBR, RSF and Yandex LLC (Russia);
GVA, XuntaGal and GENCAT (Spain);
the Royal Society
and the Leverhulme Trust (United Kingdom).

\clearpage
\addcontentsline{toc}{section}{References}
\setboolean{inbibliography}{true}
\bibliographystyle{LHCb}
\bibliography{standard,LHCb-PAPER,LHCb-DP,LHCb-TDR,main}

\ifx\mcitethebibliography\mciteundefinedmacro
\PackageError{LHCb.bst}{mciteplus.sty has not been loaded}
{This bibstyle requires the use of the mciteplus package.}\fi
\providecommand{\href}[2]{#2}
\begin{mcitethebibliography}{10}
\mciteSetBstSublistMode{n}
\mciteSetBstMaxWidthForm{subitem}{\alph{mcitesubitemcount})}
\mciteSetBstSublistLabelBeginEnd{\mcitemaxwidthsubitemform\space}
{\relax}{\relax}

\bibitem{LHCb-DP-2008-001}
LHCb collaboration, A.~A. Alves~Jr.\ {\em et~al.},
  \ifthenelse{\boolean{articletitles}}{\emph{{The \lhcb detector at the LHC}},
  }{}\href{https://doi.org/10.1088/1748-0221/3/08/S08005}{JINST \textbf{3}
  (2008) S08005}\relax
\mciteBstWouldAddEndPuncttrue
\mciteSetBstMidEndSepPunct{\mcitedefaultmidpunct}
{\mcitedefaultendpunct}{\mcitedefaultseppunct}\relax
\EndOfBibitem
\bibitem{LHCb-TDR-002}
LHCb collaboration, \ifthenelse{\boolean{articletitles}}{\emph{{LHCb
  calorimeters: Technical Design Report}}, }{}
  \href{http://cdsweb.cern.ch/search?p=CERN-LHCC-2000-036&f=reportnumber&action_search=Search&c=LHCb+Reports}
  {CERN-LHCC-2000-036}\relax
\mciteBstWouldAddEndPuncttrue
\mciteSetBstMidEndSepPunct{\mcitedefaultmidpunct}
{\mcitedefaultendpunct}{\mcitedefaultseppunct}\relax
\EndOfBibitem
\bibitem{LHCb-TDR-010}
LHCb collaboration, \ifthenelse{\boolean{articletitles}}{\emph{{LHCb trigger
  system: Technical Design Report}}, }{}
  \href{http://cdsweb.cern.ch/search?p=CERN-LHCC-2003-031&f=reportnumber&action_search=Search&c=LHCb+Reports}
  {CERN-LHCC-2003-031}\relax
\mciteBstWouldAddEndPuncttrue
\mciteSetBstMidEndSepPunct{\mcitedefaultmidpunct}
{\mcitedefaultendpunct}{\mcitedefaultseppunct}\relax
\EndOfBibitem
\bibitem{LHCb-DP-2012-004}
R.~Aaij {\em et~al.}, \ifthenelse{\boolean{articletitles}}{\emph{{The \lhcb
  trigger and its performance in 2011}},
  }{}\href{https://doi.org/10.1088/1748-0221/8/04/P04022}{JINST \textbf{8}
  (2013) P04022}, \href{http://arxiv.org/abs/1211.3055}{{\normalfont\ttfamily
  arXiv:1211.3055}}\relax
\mciteBstWouldAddEndPuncttrue
\mciteSetBstMidEndSepPunct{\mcitedefaultmidpunct}
{\mcitedefaultendpunct}{\mcitedefaultseppunct}\relax
\EndOfBibitem
\bibitem{Arefev:1103500}
A.~Arefev {\em et~al.}, \ifthenelse{\boolean{articletitles}}{\emph{{Beam Test
  Results of the LHCb Electromagnetic Calorimeter}}, }{}
  \href{http://cdsweb.cern.ch/search?p=LHCb-2007-149&f=reportnumber&action_search=Search&c=LHCb+Notes}
  {LHCb-2007-149}\relax
\mciteBstWouldAddEndPuncttrue
\mciteSetBstMidEndSepPunct{\mcitedefaultmidpunct}
{\mcitedefaultendpunct}{\mcitedefaultseppunct}\relax
\EndOfBibitem
\bibitem{Coca:691519}
C.~Coca {\em et~al.}, \ifthenelse{\boolean{articletitles}}{\emph{{The hadron
  calorimeter prototype beam-test results}}, }{}
  \href{http://cdsweb.cern.ch/search?p=LHCb-2000-036&f=reportnumber&action_search=Search&c=LHCb+Notes}
  {LHCb-2000-036}\relax
\mciteBstWouldAddEndPuncttrue
\mciteSetBstMidEndSepPunct{\mcitedefaultmidpunct}
{\mcitedefaultendpunct}{\mcitedefaultseppunct}\relax
\EndOfBibitem
\bibitem{AbellanBeteta:2012dd}
C.~Abellan~Beteta {\em et~al.},
  \ifthenelse{\boolean{articletitles}}{\emph{{Time alignment of the front end
  electronics of the LHCb calorimeters}},
  }{}\href{https://doi.org/10.1088/1748-0221/7/08/P08020}{JINST \textbf{7}
  (2012) P08020}\relax
\mciteBstWouldAddEndPuncttrue
\mciteSetBstMidEndSepPunct{\mcitedefaultmidpunct}
{\mcitedefaultendpunct}{\mcitedefaultseppunct}\relax
\EndOfBibitem
\bibitem{VazquezGomez:1399119}
R.~Vazquez~Gomez {\em et~al.},
  \ifthenelse{\boolean{articletitles}}{\emph{{Calibration and alignment of the
  SPD detector with cosmic rays and first LHC collisions}}, }{}
  \href{http://cdsweb.cern.ch/search?p=LHCb-PUB-2011-024&f=reportnumber&action_search=Search&c=LHCb+Notes}
  {LHCb-PUB-2011-024}\relax
\mciteBstWouldAddEndPuncttrue
\mciteSetBstMidEndSepPunct{\mcitedefaultmidpunct}
{\mcitedefaultendpunct}{\mcitedefaultseppunct}\relax
\EndOfBibitem
\bibitem{Aaij:2016rxn}
R.~Aaij {\em et~al.}, \ifthenelse{\boolean{articletitles}}{\emph{{Tesla : an
  application for real-time data analysis in High Energy Physics}},
  }{}\href{https://doi.org/10.1016/j.cpc.2016.07.022}{Comput.\ Phys.\ Commun.\
  \textbf{208} (2016) 35},
  \href{http://arxiv.org/abs/1604.05596}{{\normalfont\ttfamily
  arXiv:1604.05596}}\relax
\mciteBstWouldAddEndPuncttrue
\mciteSetBstMidEndSepPunct{\mcitedefaultmidpunct}
{\mcitedefaultendpunct}{\mcitedefaultseppunct}\relax
\EndOfBibitem
\bibitem{Martens:1391730}
A.~Martens, \ifthenelse{\boolean{articletitles}}{\emph{{Towards a measurement
  of the angle $\gamma$ of the Unitarity Triangle with the LHCb detector at the
  LHC (CERN): calibration of the calorimeters using an energy flow technique
  and first observation of the $B^0_s \to \overline{D}^0 \overline{K}^{*0}$
  decay}}, }{}
  \href{http://cdsweb.cern.ch/search?p=CERN-THESIS-2011-128&f=reportnumber&action_search=Search&c=LHCb+Theses}
  {CERN-THESIS-2011-128}\relax
\mciteBstWouldAddEndPuncttrue
\mciteSetBstMidEndSepPunct{\mcitedefaultmidpunct}
{\mcitedefaultendpunct}{\mcitedefaultseppunct}\relax
\EndOfBibitem
\bibitem{Aaij:1384386}
R.~Aaij and J.~Albrecht, \ifthenelse{\boolean{articletitles}}{\emph{{Muon
  triggers in the High Level Trigger of LHCb}}, }{}
  \href{http://cdsweb.cern.ch/search?p=LHCb-PUB-2011-017&f=reportnumber&action_search=Search&c=LHCb+Notes}
  {LHCb-PUB-2011-017}\relax
\mciteBstWouldAddEndPuncttrue
\mciteSetBstMidEndSepPunct{\mcitedefaultmidpunct}
{\mcitedefaultendpunct}{\mcitedefaultseppunct}\relax
\EndOfBibitem
\bibitem{Machikhiliyan:1557395}
I.~Machikhiliyan and V.~Lindahl,
  \ifthenelse{\boolean{articletitles}}{\emph{{Calibration of photomultipliers
  and initial adjustment of the LHCb Electromagnetic Calorimeter for pilot LHC
  runs}}, }{}
  \href{http://cdsweb.cern.ch/search?p=LHCb-PUB-2013-010&f=reportnumber&action_search=Search&c=LHCb+Notes}
  {LHCb-PUB-2013-010}\relax
\mciteBstWouldAddEndPuncttrue
\mciteSetBstMidEndSepPunct{\mcitedefaultmidpunct}
{\mcitedefaultendpunct}{\mcitedefaultseppunct}\relax
\EndOfBibitem
\bibitem{Arefev:1080559}
A.~Arefev {\em et~al.}, \ifthenelse{\boolean{articletitles}}{\emph{{Design,
  construction, quality control and performance study with cosmic rays of
  modules for the LHCb electromagnetic calorimeter}}, }{}
  \href{http://cdsweb.cern.ch/search?p=LHCb-2007-148&f=reportnumber&action_search=Search&c=LHCb+Notes}
  {LHCb-2007-148}\relax
\mciteBstWouldAddEndPuncttrue
\mciteSetBstMidEndSepPunct{\mcitedefaultmidpunct}
{\mcitedefaultendpunct}{\mcitedefaultseppunct}\relax
\EndOfBibitem
\bibitem{Belyaev:2011zz}
LHCb, I.~Belyaev, D.~Savrina, R.~Graciani, and A.~Puig,
  \ifthenelse{\boolean{articletitles}}{\emph{{Kali: The framework for fine
  calibration of the LHCb electromagnetic calorimeter}},
  }{}\href{https://doi.org/10.1088/1742-6596/331/3/032050}{J.\ Phys.\ Conf.\
  Ser.\  \textbf{331} (2011) 032050}\relax
\mciteBstWouldAddEndPuncttrue
\mciteSetBstMidEndSepPunct{\mcitedefaultmidpunct}
{\mcitedefaultendpunct}{\mcitedefaultseppunct}\relax
\EndOfBibitem
\bibitem{Belyaev:2014zga}
I.~M. Belyaev, D.~Y. Golubkov, V.~Y. Egorychev, and D.~V. Savrina,
  \ifthenelse{\boolean{articletitles}}{\emph{{Calibration of the LHCb
  electromagnetic calorimeter using the technique of neutral pion invariant
  mass reconstruction}},
  }{}\href{https://doi.org/10.1134/S0020441213060171}{Instrum.\ Exp.\ Tech.\
  \textbf{57} (2014) 33}\relax
\mciteBstWouldAddEndPuncttrue
\mciteSetBstMidEndSepPunct{\mcitedefaultmidpunct}
{\mcitedefaultendpunct}{\mcitedefaultseppunct}\relax
\EndOfBibitem
\bibitem{PDG2012}
Particle Data Group, J.~Beringer {\em et~al.},
  \ifthenelse{\boolean{articletitles}}{\emph{{\href{http://pdg.lbl.gov/}{Review
  of particle physics}}},
  }{}\href{https://doi.org/10.1103/PhysRevD.86.010001}{Phys.\ Rev.\
  \textbf{D86} (2012) 010001}, {and 2013 partial update for the 2014
  edition}\relax
\mciteBstWouldAddEndPuncttrue
\mciteSetBstMidEndSepPunct{\mcitedefaultmidpunct}
{\mcitedefaultendpunct}{\mcitedefaultseppunct}\relax
\EndOfBibitem
\bibitem{LHCb-PROC-2011-006}
M.~Clemencic {\em et~al.}, \ifthenelse{\boolean{articletitles}}{\emph{{The
  \lhcb simulation application, Gauss: Design, evolution and experience}},
  }{}\href{https://doi.org/10.1088/1742-6596/331/3/032023}{J.\ Phys.\ Conf.\
  Ser.\  \textbf{331} (2011) 032023}\relax
\mciteBstWouldAddEndPuncttrue
\mciteSetBstMidEndSepPunct{\mcitedefaultmidpunct}
{\mcitedefaultendpunct}{\mcitedefaultseppunct}\relax
\EndOfBibitem
\bibitem{Deschamps:691634}
O.~Deschamps {\em et~al.}, \ifthenelse{\boolean{articletitles}}{\emph{{Photon
  and neutral pion reconstruction}}, }{}
  \href{http://cdsweb.cern.ch/search?p=LHCb-2003-091&f=reportnumber&action_search=Search&c=LHCb+Notes}
  {LHCb-2003-091}\relax
\mciteBstWouldAddEndPuncttrue
\mciteSetBstMidEndSepPunct{\mcitedefaultmidpunct}
{\mcitedefaultendpunct}{\mcitedefaultseppunct}\relax
\EndOfBibitem
\bibitem{Terrier:691743}
H.~Terrier and I.~Belyaev, \ifthenelse{\boolean{articletitles}}{\emph{{Particle
  identification with LHCb calorimeters}}, }{}
  \href{http://cdsweb.cern.ch/search?p=LHCb-2003-092&f=reportnumber&action_search=Search&c=LHCb+Notes}
  {LHCb-2003-092}\relax
\mciteBstWouldAddEndPuncttrue
\mciteSetBstMidEndSepPunct{\mcitedefaultmidpunct}
{\mcitedefaultendpunct}{\mcitedefaultseppunct}\relax
\EndOfBibitem
\bibitem{LHCb-DP-2019-001}
R.~Aaij {\em et~al.}, \ifthenelse{\boolean{articletitles}}{\emph{{Performance
  of the LHCb trigger and full real-time reconstruction in Run 2 of the LHC}},
  }{}\href{https://doi.org/10.1088/1748-0221/14/04/P04013}{JINST \textbf{14}
  (2019) P04013}, \href{http://arxiv.org/abs/1812.10790}{{\normalfont\ttfamily
  arXiv:1812.10790}}\relax
\mciteBstWouldAddEndPuncttrue
\mciteSetBstMidEndSepPunct{\mcitedefaultmidpunct}
{\mcitedefaultendpunct}{\mcitedefaultseppunct}\relax
\EndOfBibitem
\bibitem{Govorkova:2015vqa}
E.~Govorkova, \ifthenelse{\boolean{articletitles}}{\emph{{Study of
  $\pi^0/\gamma$ efficiency using $B$ meson decays in the LHCb experiment}},
  }{}\href{https://doi.org/10.1134/S1063778816100070}{Phys.\ Atom.\ Nucl.\
  \textbf{79} (2016) 1474},
  \href{http://arxiv.org/abs/1505.02960}{{\normalfont\ttfamily
  arXiv:1505.02960}}\relax
\mciteBstWouldAddEndPuncttrue
\mciteSetBstMidEndSepPunct{\mcitedefaultmidpunct}
{\mcitedefaultendpunct}{\mcitedefaultseppunct}\relax
\EndOfBibitem
\bibitem{Hoballah:2026848}
M.~Hoballah, \ifthenelse{\boolean{articletitles}}{\emph{{Measurement of the
  photon polarization using ${B_s^0\to \phi\gamma}$ at LHCb}}, }{}
  \href{http://cdsweb.cern.ch/search?p=CERN-THESIS-2015-073&f=reportnumber&action_search=Search&c=LHCb+Theses}
  {CERN-THESIS-2015-073}\relax
\mciteBstWouldAddEndPuncttrue
\mciteSetBstMidEndSepPunct{\mcitedefaultmidpunct}
{\mcitedefaultendpunct}{\mcitedefaultseppunct}\relax
\EndOfBibitem
\bibitem{Hocker:2007ht}
H.~Voss, A.~Hoecker, J.~Stelzer, and F.~Tegenfeldt,
  \ifthenelse{\boolean{articletitles}}{\emph{{TMVA - Toolkit for Multivariate
  Data Analysis with ROOT}}, }{}\href{https://doi.org/10.22323/1.050.0040}{PoS
  \textbf{ACAT} (2007) 040}\relax
\mciteBstWouldAddEndPuncttrue
\mciteSetBstMidEndSepPunct{\mcitedefaultmidpunct}
{\mcitedefaultendpunct}{\mcitedefaultseppunct}\relax
\EndOfBibitem
\bibitem{Pivk:2004ty}
M.~Pivk and F.~R. Le~Diberder,
  \ifthenelse{\boolean{articletitles}}{\emph{{sPlot: A statistical tool to
  unfold data distributions}},
  }{}\href{https://doi.org/10.1016/j.nima.2005.08.106}{Nucl.\ Instrum.\ Meth.\
  \textbf{A555} (2005) 356},
  \href{http://arxiv.org/abs/physics/0402083}{{\normalfont\ttfamily
  arXiv:physics/0402083}}\relax
\mciteBstWouldAddEndPuncttrue
\mciteSetBstMidEndSepPunct{\mcitedefaultmidpunct}
{\mcitedefaultendpunct}{\mcitedefaultseppunct}\relax
\EndOfBibitem
\bibitem{CalvoGomez:2042173}
M.~Calvo~Gomez {\em et~al.}, \ifthenelse{\boolean{articletitles}}{\emph{{A tool
  for $\gamma/\pi^0$ separation at high energies}}, }{}
  \href{http://cdsweb.cern.ch/search?p=LHCb-PUB-2015-016&f=reportnumber&action_search=Search&c=LHCb+Notes}
  {LHCb-PUB-2015-016}\relax
\mciteBstWouldAddEndPuncttrue
\mciteSetBstMidEndSepPunct{\mcitedefaultmidpunct}
{\mcitedefaultendpunct}{\mcitedefaultseppunct}\relax
\EndOfBibitem
\bibitem{LHCb-PAPER-2013-028}
LHCb collaboration, R.~Aaij {\em et~al.},
  \ifthenelse{\boolean{articletitles}}{\emph{{Measurement of the relative rate
  of prompt $\chiczero$, $\chicone$ and $\chictwo$ production at
  \mbox{$\sqs=$7\tev}}}, }{}\href{https://doi.org/10.1007/JHEP10(2013)115}{JHEP
  \textbf{10} (2013) 115},
  \href{http://arxiv.org/abs/1307.4285}{{\normalfont\ttfamily
  arXiv:1307.4285}}\relax
\mciteBstWouldAddEndPuncttrue
\mciteSetBstMidEndSepPunct{\mcitedefaultmidpunct}
{\mcitedefaultendpunct}{\mcitedefaultseppunct}\relax
\EndOfBibitem
\bibitem{LHCb-DP-2014-002}
LHCb collaboration, R.~Aaij {\em et~al.},
  \ifthenelse{\boolean{articletitles}}{\emph{{LHCb detector performance}},
  }{}\href{https://doi.org/10.1142/S0217751X15300227}{Int.\ J.\ Mod.\ Phys.\
  \textbf{A30} (2015) 1530022},
  \href{http://arxiv.org/abs/1412.6352}{{\normalfont\ttfamily
  arXiv:1412.6352}}\relax
\mciteBstWouldAddEndPuncttrue
\mciteSetBstMidEndSepPunct{\mcitedefaultmidpunct}
{\mcitedefaultendpunct}{\mcitedefaultseppunct}\relax
\EndOfBibitem
\bibitem{LHCb-DP-2018-001}
R.~Aaij {\em et~al.}, \ifthenelse{\boolean{articletitles}}{\emph{{Selection and
  processing of calibration samples to measure the particle identification
  performance of the LHCb experiment in Run 2}},
  }{}\href{https://doi.org/10.1140/epjti/s40485-019-0050-z}{Eur.\ Phys.\ J.\
  Tech.\ Instr.\  \textbf{6} (2018) 1},
  \href{http://arxiv.org/abs/1803.00824}{{\normalfont\ttfamily
  arXiv:1803.00824}}\relax
\mciteBstWouldAddEndPuncttrue
\mciteSetBstMidEndSepPunct{\mcitedefaultmidpunct}
{\mcitedefaultendpunct}{\mcitedefaultseppunct}\relax
\EndOfBibitem
\bibitem{Aaij:2012ita}
LHCb, R.~Aaij {\em et~al.},
  \ifthenelse{\boolean{articletitles}}{\emph{{Measurement of the ratio of
  branching fractions $BR(B_0 \to K^{\star 0} \gamma)/BR(B_{s0} \to \phi
  \gamma)$ and the direct CP asymmetry in $B_0 \to K^{\star 0} \gamma$}},
  }{}\href{https://doi.org/10.1016/j.nuclphysb.2012.09.013}{Nucl.\ Phys.\
  \textbf{B867} (2013) 1},
  \href{http://arxiv.org/abs/1209.0313}{{\normalfont\ttfamily
  arXiv:1209.0313}}\relax
\mciteBstWouldAddEndPuncttrue
\mciteSetBstMidEndSepPunct{\mcitedefaultmidpunct}
{\mcitedefaultendpunct}{\mcitedefaultseppunct}\relax
\EndOfBibitem
\bibitem{Ali:2007sj}
A.~Ali, B.~D. Pecjak, and C.~Greub,
  \ifthenelse{\boolean{articletitles}}{\emph{{$B \to$ V $\gamma$ Decays at NNLO
  in SCET}}, }{}\href{https://doi.org/10.1140/epjc/s10052-008-0623-5}{Eur.\
  Phys.\ J.\ C \textbf{55} (2008) 577},
  \href{http://arxiv.org/abs/0709.4422}{{\normalfont\ttfamily
  arXiv:0709.4422}}\relax
\mciteBstWouldAddEndPuncttrue
\mciteSetBstMidEndSepPunct{\mcitedefaultmidpunct}
{\mcitedefaultendpunct}{\mcitedefaultseppunct}\relax
\EndOfBibitem
\bibitem{Asner:2010qj}
Heavy Flavor Averaging Group, D.~Asner {\em et~al.},
  \ifthenelse{\boolean{articletitles}}{\emph{{Averages of $b$-hadron,
  $c$-hadron, and $\tau$-lepton properties}},
  }{}\href{http://arxiv.org/abs/1010.1589}{{\normalfont\ttfamily
  arXiv:1010.1589}}\relax
\mciteBstWouldAddEndPuncttrue
\mciteSetBstMidEndSepPunct{\mcitedefaultmidpunct}
{\mcitedefaultendpunct}{\mcitedefaultseppunct}\relax
\EndOfBibitem
\bibitem{LHCb-PAPER-2012-022}
LHCb collaboration, R.~Aaij {\em et~al.},
  \ifthenelse{\boolean{articletitles}}{\emph{{Evidence for the decay
  \mbox{\decay{\Bz}{\jpsi\omegaz}} and measurement of the relative branching
  fractions of \Bs meson decays to $\jpsi\etaz$ and $\jpsi\etapr$}},
  }{}\href{https://doi.org/10.1016/j.nuclphysb.2012.10.021}{Nucl.\ Phys.\
  \textbf{B867} (2013) 547},
  \href{http://arxiv.org/abs/1210.2631}{{\normalfont\ttfamily
  arXiv:1210.2631}}\relax
\mciteBstWouldAddEndPuncttrue
\mciteSetBstMidEndSepPunct{\mcitedefaultmidpunct}
{\mcitedefaultendpunct}{\mcitedefaultseppunct}\relax
\EndOfBibitem
\bibitem{Adachi:2009usa}
Belle collaboration, J.~Li {\em et~al.},
  \ifthenelse{\boolean{articletitles}}{\emph{{First observation of $B_s^0\to
  J/\psi\eta$ and $B_s^0\to J/\psi\eta'$}},
  }{}\href{https://doi.org/10.1103/PhysRevLett.108.181808}{Phys.\ Rev.\ Lett.\
  \textbf{108} (2012) 181808},
  \href{http://arxiv.org/abs/1202.0103}{{\normalfont\ttfamily
  arXiv:1202.0103}}\relax
\mciteBstWouldAddEndPuncttrue
\mciteSetBstMidEndSepPunct{\mcitedefaultmidpunct}
{\mcitedefaultendpunct}{\mcitedefaultseppunct}\relax
\EndOfBibitem
\bibitem{Lupton:2134057}
O.~Lupton, L.~Anderlini, B.~Sciascia, and V.~Gligorov,
  \ifthenelse{\boolean{articletitles}}{\emph{{Calibration samples for particle
  identification at LHCb in Run 2}}, }{}
  \href{http://cdsweb.cern.ch/search?p=LHCb-PUB-2016-005&f=reportnumber&action_search=Search&c=LHCb+Notes}
  {LHCb-PUB-2016-005}\relax
\mciteBstWouldAddEndPuncttrue
\mciteSetBstMidEndSepPunct{\mcitedefaultmidpunct}
{\mcitedefaultendpunct}{\mcitedefaultseppunct}\relax
\EndOfBibitem
\end{mcitethebibliography}

\clearpage
\section*{LHCb Calorimeter group}

\begin{flushleft}
\small
C.~Abell{\'a}n~Beteta$^8$,
A.~Alfonso~Albero$^8$,
Y.~Amhis$^3$,
S.~Barsuk$^3$,
C.~Beigbeder-Beau$^3$,
I.~Belyaev$^4$, 
R.~Bonnefoy$^2$, 
D.~Breton$^3$,
O.~Callot$^3$,
M.~Calvo~Gomez$^{8,a}$, 
A.~Camboni$^{8,a}$, 
H.~Chanal$^2$, 
D.~Charlet$^3$,
M.~Chefdeville$^1$, 
V.~Coco$^9$, 
E.~Cogneras$^2$, 
A.~Comerma-Montells$^8$,
S.~Coquereau$^8$, 
O.~Deschamps$^2$, 
F.~Domingo~Bonal$^8$,
C.~Drancourt$^1$, 
O.~Duarte$^3$, 
N.~Dumont~Dayot$^1$, 
R.~Dzhelyadin$^7$, 
V.~Egorychev$^4$, 
D.~Esperante$^8$,
S.~Filippov$^6$, 
L.~Garrido$^8$, 
D.~Gascon$^8$, 
Ph.~Ghez$^1$, 
L.L.~Gioi$^2$, 
P.~Gironella~Gironell$^8$, 
D.~Golubkov$^4$, 
A.~Golutvin$^{10,11}$,
M.~Grabalosa~G{\`a}ndara$^8$,
R.~Graciani~Diaz$^8$, 
E.~Graug{\'e}s$^8$, 
E.~Gushchin$^6$,
Yu.~Guz$^{7,9}$,
C.~Hadjivasiliou$^2$, 
B.~Jean-Marie$^3$,
A.~Konoplyannikov$^4$, 
R.~Kristic$^9$, 
T.~Kvaratskheliya$^4$, 
R.~Lef{\`e}vre$^2$, 
J.~Lefran{\c c}ois$^3$,
Y.~Li$^3$, F.~Machefert$^3$,
I.~Machikhiliyan$^4$, 
J.-F.~Marchand$^1$, 
C.~Marin~Benito$^3$, 
A.~Martens$^3$,  
A.~Martin-S{\`a}nchez$^3$,  
V.~Matveev$^7$,
J.~Mauricio$^8$,
M.-N.~Minard$^1$, 
S.~Monteil$^2$, 
V.~Niess$^2$, 
V.~Obraztsov$^7$, 
D.~Pereima$^4$, 
P.~Perret$^2$, 
E.~Picatoste~Olloqui$^8$,
B.~Pietrzyk$^1$, 
A.~Puig~Navarro$^8$, 
V.~Rives-Molina$^8$,
P.~Robbe$^3$,
M.~Rosell{\'o}$^{8,a}$,
H.~Ruiz$^8$,
C.~Sanchez~Mayordomo$^8$,
D.~Savrina$^{4,5}$,
A.~Schopper$^9$, 
M.H.~Schune$^3$,
A.~Semennikov$^4$,
P.~Shatalov$^4$,
M.~Soldatov$^7$,
O.~Stenyakin$^7$, 
V.~Tisserand$^1$, 
S.~T'Jampens$^1$, 
V.~Tocut$^3$,
E.~Tournefier$^1$, 
A.~Vallier$^3$, 
R.~Vazquez~Gomez$^8$, 
B.~Viaud$^3$, 
X.~Vilasis-Cardona$^{8,a}$, 
M.~Winn$^3$, 
O.~Yushchenko$^7$,
A.~Zhokhov$^4$
\bigskip

{\footnotesize \it
$ ^{1}$Univ. Grenoble Alpes, Univ. Savoie Mont Blanc, CNRS, IN2P3-LAPP, Annecy, France\\ 
$ ^{2}$Universit{\'e} Clermont Auvergne, CNRS/IN2P3, LPC, Clermont-Ferrand, France\\ 
$ ^{3}$Universit{\'e} Paris-Saclay, CNRS/IN2P3, IJCLab, Orsay, France\\ 
$ ^{4}$Institute of Theoretical and Experimental Physics NRC Kurchatov Institute (ITEP NRC KI), Moscow, Russia, Moscow, Russia\\ 
$ ^{5}$Institute of Nuclear Physics, Moscow State University (SINP MSU), Moscow, Russia\\ 
$ ^{6}$Institute for Nuclear Research of the Russian Academy of Sciences (INR RAS), Moscow, Russia\\ 
$ ^{7}$Institute for High Energy Physics NRC Kurchatov Institute (IHEP NRC KI), Protvino, Russia, Protvino, Russia\\ 
$ ^{8}$ICCUB, Universitat de Barcelona, Barcelona, Spain\\  
$ ^{9}$European Organization for Nuclear Research (CERN), Geneva, Switzerland\\ 
$ ^{10}$Imperial College London, London, United Kingdom\\ 
$ ^{11}$National University of Science and Technology ``MISIS'', Moscow, Russia, associated to $^{4}$\\ 
\bigskip
$^{a}$DS4DS, La Salle, Universitat Ramon Llull, Barcelona, Spain 
}

\end{flushleft}

\end{document}